\documentclass[reqno]{amsart}

\usepackage{amsmath}
\usepackage{amsfonts}
\usepackage{amsthm}

\numberwithin{equation}{section}
\newcommand {\pa}{\partial}
\newcommand {\ar}{\rightarrow}

\newcommand{\supp}{\operatorname{supp}}
\newcommand{\Spec}{\operatorname{Spec}}

\newcommand{\dist}{{\operatorname{dist}}}

\newcommand{\curl}{{\operatorname{curl}\,}}
\newcommand{\Ran}{{\operatorname{Ran}\,}}

\newcommand{\Span}{{\operatorname{Span}}}

\newtheorem{theorem}{Theorem}[section]

\newtheorem{lemma}[theorem]{Lemma}
\newtheorem{proposition}[theorem]{Proposition}
\newtheorem{prop}[theorem]{Proposition}

\newtheorem{remark}[theorem]{Remark}
\newtheorem{corollary}[theorem]{Corollary}
\newtheorem{cor}[theorem]{Corollary}

\def\Og {{\mathcal O}} 
\def \rz {{\mathbb R}}

\title[Magnetic bottles]{Accurate estimates for magnetic bottles in connection with superconductivity}

\author{S. Fournais}
\author{B. Helffer}

\thanks{The two authors are supported by the European Research Network
`Postdoctoral Training Program in Mathematical Analysis of Large
Quantum Systems' with contract number HPRN-CT-2002-00277, and the ESF
Scientific Programme in Spectral Theory and Partial Differential
Equations (SPECT)} 
\address[S. Fournais]{CNRS and Laboratoire de
Math\'{e}matiques UMR CNRS 8628\\ Universit\'{e} Paris-Sud - B\^{a}t 425\\
F-91405 Orsay Cedex\\ France.
}
\email{soeren.fournais@math.u-psud.fr}

\address[B. Helffer]{Laboratoire de
Math\'{e}matiques UMR CNRS 8628\\ Universit\'{e} Paris-Sud - B\^{a}t 425\\
F-91405 Orsay Cedex\\ France.
}
\email{bernard.helffer@math.u-psud.fr}

\date{November 16, 2004}

\begin{document}

\bibliographystyle{plain}

\begin{abstract}
Motivated by the theory of superconductivity and more precisely by the problem
 of the onset of superconductivity in dimension two, many papers devoted to the analysis in a semi-classical regime 
 of the lowest eigenvalue of the Schr\"odinger operator 
with magnetic field have appeared recently. Here we would like to mention the works by 
Bernoff-Sternberg, Lu-Pan,  Del Pino-Felmer-Sternberg and
 Helffer-Morame and also Bauman-Phillips-Tang for the case of a disc.
In the present paper we settle one important part of this question
 completely by proving an asymptotic expansion to all orders for low-lying
eigenvalues for generic domains. The word `generic' means in this context that the curvature
 of the boundary of the domain has a unique non-degenerate maximum.
\end{abstract} 

\maketitle
\tableofcontents

\section{Introduction}
The object of study in this paper is a magnetic Schr\"{o}dinger
operator with Neumann boundary conditions in a smooth, bounded domain
$\Omega$. We are interested in finding an accurate description of the eigenvalues
near the bottom of the spectrum.
In particular, we will improve estimates given in \cite{HeMo3} in the
case of constant magnetic field.

Apart from its intrinsic mathematical interest, this question is
important for applications to superconductivity. Precise knowledge of
the lowest eigenvalues of this magnetic Schr\"{o}dinger operator is
crucial for a detailed description of the nucleation of
superconductivity (on the boundary) for superconductors of Type II and
for accurate estimates of the critical field $H_{C_3}$. These
applications will be the subject of further work and will be published
elsewhere. We refer the reader
to the works of Bernoff-Sternberg~\cite{BeSt}, Lu-Pan~\cite{LuPa1, LuPa2, LuPa3, LuPa5}, and Helffer-Pan~\cite{HePa} for further discussion of
this subject and to \cite{TiTi} and \cite{ SaSj} for 
 the physical motivation.

Let us fix the notations. The domain $\Omega \subset {\mathbb R}^2$ is supposed to be  smooth, bounded and simply connected. 
Points $(x_1,x_2)$ in ${\mathbb R}^2$ are denoted by $z$ or $x$. 
At each point $z$ of the boundary,  we denote by $\nu(z)$ the interior unit normal vector to the boundary of $\Omega$.
We define the magnetic Neumann operator ${\mathcal H}$ by
\begin{equation}\label{defHNeumann}
{\mathcal D}({\mathcal H})\ni u \mapsto {\mathcal H}u = {\mathcal H}_{h, \Omega}u = (-ih\nabla_{z} - A(z))^2 u(z)\;.
\end{equation}
Here $A(z) = (-x_2/2, x_1/2)$, so that $\curl A = 1$, and the domain ${\mathcal D}({\mathcal H})$ of the operator ${\mathcal H}$ is defined by
$$
{\mathcal D}({\mathcal H}) = \big \{ u \in H^2(\Omega) \, \big | \, \nu\cdot (-ih\nabla_{z} - A(z))u\big |_{\partial \Omega} = 0 \big\}\;.
$$

The case of the half-plane, $\Omega = {\mathbb R}\times{\mathbb R}_{+}$,  will be important for fixing notations. After a gauge transformation and a partial Fourier transformation we get, in this case and with $h=1$, the family of models on the half-line:
\begin{align}
\label{eq:BasicNormal}
H^{N,\xi}=D_x^2 + (x+\xi)^2\;,
\end{align}
on $L^2({\mathbb R}_{+})$ and
with Neumann boundary conditions at $x=0$. 
Important results about the operators $H^{N,\xi}$ will be recalled in
Appendix~\ref{AppA}; here we only define the notation that will be
used throughout the text.
Let $\hat{\mu}^{(1)}(\xi)$ be
the lowest eigenvalue of $H^{N,\xi}$. Then $\xi \mapsto
\hat{\mu}^{(1)}(\xi)$ has a unique minimum $\Theta_0$ attained at a
point that we will denote by
$\xi_0$. The corresponding unique positive, normalized
eigenfunction of $H^{N,\xi_0}$ will be denoted by $u_0$. We also introduce~:
\begin{align}\label{eq:C1}
C_1 = \frac{u_0^2(0)}{3}\;.
\end{align}

The main result of the paper gives the asymptotic expansion of the lowest eigenvalues of ${\mathcal H}$. We define $\mu^{(n)}(h)$ to be  the $n$-th eigenvalue of ${\mathcal H}$, in particular,
$$
\mu^{(1)}(h) = \inf \Spec {\mathcal H}_{h, \Omega}\;,
$$
and prove the following result.
\begin{theorem}
\label{thm:main}~\\
Suppose that $\Omega$ is a smooth bounded domain, that its curvature $\pa \Omega\ni s\mapsto \kappa(s)$ at the boundary has a unique maximum,
\begin{align}
\label{eq:uniqueMax}
\kappa(s) < \kappa(s_0) =: k_{\rm max}\;, \text{ for all } s \neq s_0\;,
\end{align}
and that the maximum is non-degenerate, i.e.
\begin{align}
\label{eq:non-deg}
k_2 := - \kappa''(s_0) \neq 0\;.
\end{align}
Then, for all $n \in {\mathbb N}$, there exists 
a sequence $\{ \zeta_j^{(n)}\}_{j=1}^{\infty} \subset {\mathbb R}$ (which can be calculated recursively to any order) such that $\mu^{(n)}(h)$ admits the following asymptotic expansion (for $h \searrow 0$)~:
\begin{align}
\mu^{(n)}(h)\sim \Theta_0 h - k_{\rm max} C_1 h^{3/2} + C_1
\Theta_0^{1/4} \sqrt{\tfrac{3 k_2}{2}} (2n-1)h^{7/4}  + h^{15/8}\sum_{j=0}^\infty  h^{j/8} \zeta_j^{(n)}\;.
\end{align}
\end{theorem}

\begin{remark}~\\
The semiclassical limit $h \searrow 0$ is clearly equivalent to a large magnetic field limit, since
\begin{align*}
\int_{\Omega} | (-ih\nabla_{z} - B A(z)) u(z) |^2\,dz = B^2 \int_{\Omega} | (-i\tfrac{1}{B}\nabla_{z} -  A(z)) u(z) |^2\,dz\;.
\end{align*}
\end{remark}
~
\begin{remark}~\\
Previous results on the bottom of the spectrum of ${\mathcal H}_{h, \Omega}$ were obtained  in \cite{HeMo3}, who gave the  two first terms in the expansion of $\mu^{(1)}(h)$ (see \cite[Theorems~10.3 and 11.1]{HeMo3}):
\begin{align}
\label{eq:previousEst}
\mu^{(1)}(h) = \Theta_0 h - k_{\rm max} C_1 h^{3/2} +{\mathcal O}(h^{5/3})\;.
\end{align}
\end{remark}

\begin{remark}~\\
It is rather reasonable to believe that  the proof of Theorem~\ref{thm:main} 
 can be adapted for getting  a similar result 
 under the weaker assumption that there exists $J\in {\mathbb N}$, such that
\begin{equation}
\left\{
\begin{array}{lll}
\kappa^{(2j)}(s_0) =0\;, & \text{ for }& j=1,2, \ldots, J-1\;, \\
\kappa^{(2J)}(s_0) \neq 0\;,&&
\end{array}
\right. 
\end{equation}
i.e.~the maximum is non-degenerate of order $2J$. However we will not pursue this further.

If the uniqueness condition in \eqref{eq:uniqueMax} is replaced by the assumption that there is a finite number
of maxima (for which \eqref{eq:non-deg} is assumed to hold), we expect the existence of  sequences of eigenvalues $z^{(n)}(h)$ corresponding to each maximum. This also follows from the techniques applied in the present paper with a little extra work.
\end{remark}

For applications to bifurcations from the normal state in superconductivity it seems important to calculate the splitting between the ground and first excited states of ${\mathcal H}(h)$. Let us define
\begin{equation}\label{defDelta}
\Delta(h) = \mu^{(2)}(h) - \mu^{(1)}(h)\;.
\end{equation}

\begin{cor}~\\
Under the hypothesis from Theorem~\ref{thm:main},   $\Delta(h)$ admits the following asymptotics~:
\begin{align}\label{expDelta}
\Delta(h) \sim  C_1 \Theta_0^{1/4} \sqrt{6 k_2} h^{7/4} + h^{15/8}\sum_{j=0}^\infty  h^{j/8} \xi_j\;.
\end{align}
where $\xi_j = \zeta_j^{(2)}-\zeta_j^{(1)}$\;.
\end{cor}

The case where $\Omega$ is a disc has been analyzed in great detail in
\cite{BaPhTa}, using the radial symmetry to reduce the problem to
ordinary differential equations. In this case the splitting
$\Delta(h)$ turns out to become zero for a sequence of values of $h$ 
tending to $0$. This is a complication in the analysis of
bifurcation. Thus, in some sense, the more `generic' situation
considered in this paper has a nicer property.
We recall that for the disc it is reasonable to conjecture from \cite{BaPhTa}
 that~:
$$
0=\liminf_{h \rightarrow 0} \frac{\Delta(h)}{h^2} <  \limsup_{h \rightarrow 0} \frac{\Delta(h)}{h^2} \;<+\infty\;.
$$
We recall also that in the case of a domain with a  unique corner, with a sufficiently small angle, one has (\cite{Bo2}, \cite{BoDa})~:
$$
\liminf_{h\rightarrow 0} \frac{\Delta(h)}{h} >0\;.
$$
In our case, \eqref{expDelta} implies~:
$$
\lim_{h\rightarrow 0} \frac{\Delta(h)}{h^\frac  74} >0\;.
$$
Of course (see Bonnaillie \cite{Bo1} for a discussion inspired by
Helffer-Sj\"ostrand \cite{HeSj1, HeSjII}), if there are multiple
minima and symmetries, one expects an exponentially small gap
between the two lowest eigenvalues.

The plan of the paper is as follows. In Section~\ref{simple} we prove a simple non-optimal upper bound to the ground state energy. This calculation motivates the more systematic treatment in Section~\ref{GrushinUpper}, where we introduce a `Grushin problem' in order to reduce the analysis to an effective model on the boundary. The effective model allows us to construct quasimodes whose energy corresponds to the lowest eigenvalues of ${\mathcal H}$ to any order in $h$. Thus we get the upper bound inherent in Theorem~\ref{thm:main}. In order to prove that the Grushin approach also gives a lower bound, we need to prove suitable localization results in phase space. That is carried through in sections~\ref{locPhase} and \ref{phasespace}. Finally, in Section~\ref{lower2} we finish the proof of Theorem~\ref{thm:main}. Appendix~\ref{AppA} recalls a number of results from the analysis of the half-plane model that are needed in the calculations. Appendix~\ref{AppB} contains definitions concerning the coordinate system near the boundary in which all the calculations will take place.

\section{A simple upper  bound to the ground state energy } 
\label{simple}
This section contains a simple variational estimate of the ground
state energy $\mu^{(1)}(h)$. The motivation for giving this result is
a number of remarks and calculations appearing in the literature. It turns out
that the `obvious' choice of trial functions does not give as good energy estimates as one might expect. This motivates the more systematic approach in later sections.

Recall that we have defined the constants $\Theta_0$ and $C_1$ in the introduction.
\begin{theorem}
\label{thm:variational}~\\
Suppose $\Omega$ is a smooth bounded domain.
Let 
$$
k_{\rm max} = \sup_{s} \kappa(s) = \max_{s} \kappa(s)\;,
$$
be the maximal curvature of the boundary and let
$$
k_2 = \inf_{s\in \kappa^{-1}(k_{\rm max} )} (-\kappa''(s))\;.
$$
Then the ground state energy $\mu^{(1)}(h)$ of the operator ${\mathcal H}$ (defined in \eqref{defHNeumann}) satisfies
$$
\limsup_{h \ar 0_{+}}
h^{-7/4} \Big\{ \mu^{(1)}(h) - 
\Big(\Theta_0 h - k_{\rm max} C_1 h^{3/2} + \sqrt{\frac{k_2 C_{1}}{2}} h^{7/4}\Big) \Big\}
\leq 0\;.
$$
\end{theorem}
\begin{remark}~\\
Theorem~\ref{thm:variational} does not give the correct coefficient to
the $h^{7/4}$-term (compare with Theorem~\ref{thm:main}). The trial
function used in the proof below is too simple since it only uses the
ground state $u_0$ in the normal variable. This seems to disprove a belief  
stated in a remark in del~Pino, Felmer and Sternberg \cite{PiFeSt}. See also Remark~\ref{rem:74coeff} below.
\end{remark}

\begin{proof}~\\
The proof consists of an explicit calculation with a suitably chosen test function.
(This is the same test function used in the remark in \cite{PiFeSt}).

Let us consider a point $x_0$ on the boundary $\partial \Omega$ such that the curvature of $\partial \Omega$ at $x_0$ is $k_{\rm max}$,  the maximal curvature of the boundary.
We choose our boundary coordinates $(s,t)$ (see Appendix~\ref{AppB}) such that $x_0$ has coordinates $(0,0)$. 
Let $\chi \in C_0^{\infty}({\mathbb R})$ be a standard cut-off function~:\\
 $$\chi(t) = 1\mbox{  for } |t|\leq 1/2\;, \; \mbox{ and } \supp \chi \subset (-1,1)\;.
$$
Consider now the  test function
\begin{equation}\label{formesimplea}
\phi_\chi(s,t;h) = \phi_0(t,s;h) \,\chi(2 s/|\partial \Omega|)\, \chi(t/t_0)\;,
\end{equation}
where,  for $\alpha >0$ to be chosen below, 
\begin{equation}\label{formesimpleb}
\phi_0(t,s;h) :=
(2\alpha)^{1/4} h^{-5/16} e^{-\alpha s^2/h^{1/4}} e^{i\xi_0 s/h^{1/2}} u_{0}(h^{-1/2}t) \;.
\end{equation}
and  $t_0$ is the constant from Appendix~\ref{AppB} defining the tubular neighborhood of the boundary on which one may use boundary coordinates.

We will get an upper bound to the ground state energy of the Neumann problem by calculating the Rayleigh quotient $ \frac{\langle \phi\;|\;{\mathcal H} \phi \rangle}{\| \phi \|^2}$ for a suitable $\phi$ in the domain of $\mathcal H$
 $D(\mathcal H)$. Actually, one could also work with $\phi$ in the form domain
 of the corresponding quadratic form $q_\mathcal H$.\\
 After a gauge transformation, we can assume 
that in 
our  boundary coordinates (see Appendix~\ref{AppB})~:
$$
\tilde{A} = (\tilde A_1,\tilde A_2)= \left( -t(1-\frac{t}{2}\kappa(s)), 0 \right).
$$
From now on, we fix the gauge such that this property is satisfied.\\
Then 
\begin{align*}
\langle \phi\;|\; {\mathcal H} \phi \rangle &=
\int_{-|\partial \Omega|/2}^{|\partial \Omega|/2} \int_0^{\infty}
\left\{ |(hD_t - \tilde{A}_2)\phi|^2 + (1-t\kappa(s))^{-2} |(hD_s - \tilde{A}_1)\phi |^2 \right\}\\
& \quad \quad \quad \quad\quad \quad\quad \quad\quad \quad\quad \quad
\times
(1-t\kappa(s))\,ds\,dt \;.
\end{align*}
Now, using the decay properties of $u_0$ and the exponential decay of the Gaussian, we first get~:
\begin{align*}
\langle \phi\;|\; {\mathcal H} \phi \rangle
& = 
\int_{-|\partial \Omega|/2}^{|\partial \Omega|/2} \int_0^{\infty}
\left\{ |(hD_t)\phi_0|^2 + (1-t\kappa(s))^{-2} |(hD_s - \tilde{A}_1)\phi_0 |^2 \right\}\\
& \quad \quad \quad \quad\quad \quad\quad \quad\quad \quad\quad \quad
\times
(1-t\kappa(s))\chi(2 s/|\partial \Omega|)^2\, \chi(t/t_0)^2\,ds\,dt \\
& \quad +\Og(h^\infty)\;.
\end{align*}
Again using the properties of $u_0$ (see (\ref{propu0})) and of the Gaussian, we get 
\begin{align}
\langle \phi\;|\; {\mathcal H} \phi \rangle &= 
 h^{-5/8}\sqrt{2\alpha}  \int_{-|\partial \Omega|/2}^{|\partial \Omega|/2} \int_0^{\infty}
e^{-2\alpha s^2/h^{1/4}} h | u_0'(h^{-1/2}t)|^2 (1-t\kappa(s))\,ds\,dt \nonumber \\
&\quad+
h^{-5/8}\sqrt{2\alpha}  \int_{-|\partial \Omega|/2}^{|\partial \Omega|/2} \int_0^{\infty}
e^{-2\alpha s^2/h^{1/4}} | u_0(h^{-1/2}t)|^2 \nonumber \\
& \quad \quad \quad \quad\quad \quad \quad \quad \times
\Big| h^{1/2} \xi_0 + i 2\alpha s h^{3/4} + t(1-\frac{t}{2}\kappa(s)) 
\Big|^2 \nonumber \\
& \quad \quad \quad \quad\quad \quad \quad \quad \times\;(1-t\kappa(s))^{-1}\,\chi(2 s/|\partial \Omega|)^2\, \chi(t/t_0)^2\,ds\,dt \nonumber\\
&\quad+ {\mathcal O}(h^{\infty})\;.
\end{align}
It is then clear that by interpreting $(1-t\kappa(s))^{-1}$ as 
$\sum_{n\geq 0} t^n \kappa(s)^n$ and computing term by term, the cut-off function in $t$ does not
 affect the computation modulo $\Og(h^\infty)$. So we get
\begin{align}
&\langle \phi\;|\; {\mathcal H} \phi \rangle \sim \\
&\quad h^{-5/8}\sqrt{2\alpha}  \int_{-|\partial \Omega|/2}^{|\partial \Omega|/2} \int_0^{\infty}
e^{-2\alpha s^2/h^{1/4}} h | u_0'(h^{-1/2}t)|^2 (1-t\kappa(s))\,ds\,dt \nonumber\\
&\quad+
h^{-5/8}\sqrt{2\alpha}  \int_{-|\partial \Omega|/2}^{|\partial \Omega|/2} \int_0^{\infty}
e^{-2\alpha s^2/h^{1/4}}\; | u_0(h^{-1/2}t)|^2 \times \nonumber\\ 
&\quad\quad \times
\Big| h^{1/2} \xi_0 + i 2\alpha s h^{3/4} + t(1-\frac{t}{2}\kappa(s)) \Big|^2
 \Big(\sum_n t^n \kappa(s)^n\Big) \,\chi(2 s/|\partial \Omega|)^2\,\,ds\,dt \;. \nonumber
\end{align}
The next step is to replace $\kappa (s)$ by its Taylor expansion $\kappa^{\rm Tay}(s)$ at $0$,
 which leads to the equality (modulo $\Og(h^\infty)$)~:
\begin{align}
&\langle \phi\;|\; {\mathcal H} \phi \rangle \nonumber\\
& \sim 
h^{-5/8}\sqrt{2\alpha}  \int_{-\infty}^{+\infty} \int_0^{\infty}
e^{-2\alpha s^2/h^{1/4}} h | u_0'(h^{-1/2}t)|^2 (1-t\kappa^{\rm Tay}(s))\,ds\,dt \nonumber \\
&\quad +
h^{-5/8}\sqrt{2\alpha}  \int_{-\infty}^{+\infty} \int_0^{\infty}
e^{-2\alpha s^2/h^{1/4}} | u_0(h^{-1/2}t)|^2  \nonumber\\ 
&\quad \quad \quad \times
\Big| h^{1/2} \xi_0 + i 2\alpha s h^{3/4} + t(1-\frac{t}{2}\kappa^{\rm Tay}(s)) \Big|^2 \Big(\sum_n t^n \kappa^{\rm Tay}(s)^n\Big) \,ds\,dt\;.
\end{align}
Here the cut-off functions have completely disappeared and the integration in the $s$ variable is now over $(-\infty,+\infty)$.\\
We omit in what follows the reference to Taylor expansions written in superscript ``${\rm Tay}$'' for $\kappa$ and we use for
 shortness $(1-t\kappa(s))^{-1}$ instead of   $\sum_n t^n \kappa(s)^n$ in the next computations.\\
With the change of variables $\sigma = \sqrt{2\alpha} h^{-1/8} s$, $\tau = h^{-1/2} t$, we can continue the calculation as
\begin{align*}
&\langle \phi\;|\; {\mathcal H} \phi \rangle \sim 
h \int_{-\infty}^{+\infty}  \int_0^{\infty} e^{-\sigma^2} |u_0'(\tau)|^2 
\left(1-h^{1/2} \tau \kappa(\tfrac{h^{1/8}\sigma}{\sqrt{2\alpha}})\right) \,d\sigma\, d\tau\\
&+
h \int_{-\infty}^{+\infty}  \int_0^{\infty} e^{-\sigma^2} |u_0(\tau)|^2
\Big| \xi_0 + \tau(1- \frac{h^{1/2}\tau}{2} \kappa^{\rm Tay}(\tfrac{h^{1/8}\sigma}{\sqrt{2\alpha}})) + i\sqrt{2\alpha} h^{3/8} \sigma \Big|^2 \\
& \quad \quad \quad \quad\quad \quad\quad \quad\quad \quad\quad \quad
\times
(1-h^{1/2} \tau \kappa(\tfrac{h^{1/8}\sigma}{\sqrt{2\alpha}}))^{-1}
\,d\sigma d\tau \\
&= h\left\{ T_1 + T_2 + T_3 + T_4 +T_5 +\Og(h^\frac 78)\right\},
\end{align*}
with (using that $\kappa'(0)=0$, since $\kappa(0)= k_{\rm max}$)
\begin{align*}
T_1 &= \int_{-\infty}^{+\infty} e^{-\sigma^2} \int_0^{\infty} |u_0'(\tau)|^2 + (\xi_0 + \tau)^2 |u_0(\tau)|^2 \, d\tau \,d\sigma\;, \\
\displaybreak[0]
T_2 &= - h^{1/2} \int_{-\infty}^{+\infty}  \int_0^{\infty} e^{-\sigma^2} \tau |u_0'(\tau)|^2 \Big(
\kappa(0) + \frac{1}{2} \kappa''(0) \frac{h^{1/4}\sigma^2}{2\alpha} \Big)\, d\tau \,d\sigma\;, \\
\displaybreak[0]
T_3 &= 2 \alpha h^{3/4} \int_{-\infty}^{+\infty}  \int_0^{\infty} e^{-\sigma^2} |u_0(\tau)|^2 \sigma^2 \, d\tau \,d\sigma\;, \\
\displaybreak[0]
T_4 &= h^{1/2} \int_{-\infty}^{+\infty}  \int_0^{\infty} e^{-\sigma^2} |u_0(\tau)|^2
(\xi_0 + \tau)^2 \tau \Big(
\kappa(0) + \frac{1}{2} \kappa''(0) \frac{h^{1/4}\sigma^2}{2\alpha} \Big)\, d\tau \,d\sigma\;, \\
\displaybreak[0]
T_5 &= - h^{1/2} \int_{-\infty}^{+\infty}  \int_0^{\infty} e^{-\sigma^2} |u_0(\tau)|^2 
(\xi_0 +\tau) \tau^2 \Big(
\kappa(0) + \frac{1}{2} \kappa''(0) \frac{h^{1/4}\sigma^2}{2\alpha} \Big)\, d\tau \,d\sigma\;.
\end{align*}
Therefore, up to $\Og(h^{\frac {15}{8}})$, 
we get the equivalence
\begin{align*}
&\langle \phi\;|\;{\mathcal H} \phi \rangle 
\sim 
h \big\{ S_0 + h^{1/2} S_{1/2} + h^{3/4} S_{3/4}\big\}\;,
\end{align*}
with
\begin{align*}
S_0 & = \Theta_0 \int e^{-\sigma^2}\,d\sigma\;, \\
S_{1/2} &= \kappa(0)  \int e^{-\sigma^2}\,d\sigma \Big[
-\int_0^{\infty} \tau |u_0'(\tau)|^2\, d\tau 
+ \int_0^{\infty} \tau (\xi_0 + \tau)^2 |u_0(\tau)|^2\, d\tau \\
& \quad \quad \quad \quad\quad \quad\quad \quad\quad \quad\quad \quad
- \int_0^{\infty} \tau^2 (\xi_0 + \tau) |u_0(\tau)|^2\, d\tau \Big]\;,\\
S_{3/4} &= 2\alpha \int \sigma^2 e^{-\sigma^2}\,d\sigma \\
&\quad
+ \kappa''(0) \int \frac{\sigma^2}{4\alpha} e^{-\sigma^2}\,d\sigma
\Big[
-\int_0^{\infty} \tau |u_0'(\tau)|^2\, d\tau 
+ \int_0^{\infty} \tau (\xi_0 + \tau)^2 |u_0(\tau)|^2\, d\tau \\
& \quad \quad \quad \quad\quad \quad\quad \quad\quad \quad\quad \quad
- \int_0^{\infty} \tau^2 (\xi_0 + \tau) |u_0(\tau)|^2\, d\tau \Big]\;.
\end{align*}
From the known moments of $u_0$ (see Lemma~\ref{Mom} below or Fournais-Helffer \cite[(6.15), (6.16) and (6.17)]{FoHe}) we have
\begin{align*}
\int_0^{\infty} \tau  |u_0(\tau)|^2\, d\tau
&= \sqrt{\Theta_0}\;, &
\int_0^{\infty} \tau (\xi_0 + \tau)^2 |u_0(\tau)|^2\, d\tau
&=
\frac{1}{2}(C_1 + \Theta_0^{3/2})\;,\\
\int_0^{\infty} \tau |u_0'(\tau)|^2\, d\tau &= 
C_1 + \frac{\Theta_0^{3/2}}{2}\;, &
\int_0^{\infty} \tau^2 (\xi_0 + \tau) |u_0(\tau)|^2\, d\tau
&=
\frac{C_1}{2} + \Theta_0^{3/2}\;.
\end{align*}
So with $I_0 = \int e^{-\sigma^2}\,d\sigma$, $I_2 = \int \sigma^2 e^{-\sigma^2}\,d\sigma$, we get
\begin{align*}
S_0 & = \Theta_0 I_0\;, \\
S_{1/2} &= \kappa(0)  I_0
 \Big[ -(C_1 + \frac{\Theta_0^{3/2}}{2}) + \frac{1}{2}(C_1 + \Theta_0^{3/2})
 - (\frac{C_1}{2} + \Theta_0^{3/2})
 \Big] \\
 &= - \kappa(0)  I_0(C_1 + \Theta_0^{3/2}) \;,\\
S_{3/4} 
&=I_2 \Big[ 2\alpha - \frac{\kappa''(0)}{4\alpha}( C_1+\Theta_0^{3/2}) \Big] \;.
\end{align*}
Therefore, we finally find
\begin{align*}
\langle \phi\;|\; {\mathcal H} \phi \rangle 
&=
h \Theta_0 I_0
-  h^{3/2} \kappa(0)  I_0(C_1 + \Theta_0^{3/2}) \\
& \quad+ h^{7/4} I_2 \big[ 2\alpha - \frac{\kappa''(0)}{4\alpha}( C_1+\Theta_0^{3/2}) \big] + \Og(h^{\frac{15}{8}})\;.
\end{align*}
We now compute the asymptotics of $\| \phi \|_2^2$. Along the same lines
 as the previous computations and with the same conventions,  we obtain
\begin{align*}
\| \phi \|_2^2 &\sim 
\int_{-\infty}^{+\infty}  \int_0^{\infty} e^{-\sigma^2} |u_0(\tau)|^2 
(1-h^{1/2} \tau \kappa(\tfrac{h^{1/8}\sigma}{\sqrt{2\alpha}})) \,d\sigma d\tau + {\mathcal O}_{\rm unif}(h^{\infty})\\
&=
\int_{-\infty}^{+\infty} e^{-\sigma^2} \Big(
1-h^{1/2}  \kappa(\tfrac{h^{1/8}\sigma}{\sqrt{2\alpha}}) \int_0^{\infty}\tau |u_0(\tau)|^2\,d\tau \Big) \,d\sigma  + {\mathcal O}_{\rm unif}(h^{\infty})\\
&= 
\int_{-\infty}^{+\infty} e^{-\sigma^2} \Big(
1-h^{1/2}  \kappa(\tfrac{h^{1/8}\sigma}{\sqrt{2\alpha}}) \sqrt{\Theta_0} \Big) \,d\sigma  + {\mathcal O}_{\rm unif}(h^{\infty})\\
&=I_0 - h^{1/2} \sqrt{\Theta_0} \kappa(0) I_0 - h^{3/4} \sqrt{\Theta_0} \frac{\kappa''(0)}{4\alpha} I_2 + \Og(h^{\frac 78})\;. 
\end{align*}
So the Rayleigh quotient becomes
\begin{align*}
\frac{\langle \phi\;|\; {\mathcal H} \phi \rangle}{\| \phi \|_2^2}
=
\Theta_0 h - \kappa(0) C_1 h^{3/2} + (2\alpha -\frac{\kappa''(0)C_1}{4\alpha})\frac{I_2}{I_0} h^{7/4} + \Og(h^{\frac{15}{8}})\;.
\end{align*}

Since the curvature $\kappa$ has a maximum at $s=0$,  we see that $\kappa''(0) \leq 0$. We recall that $\phi$ depends on $\alpha$ and that
 we can now optimize over $\alpha$. We recover first the fact that the term in $\Og(h^\frac 32)$ is obtained without  having  to specify
 $\alpha$. In the case when $k_2=-\kappa''(0)\neq 0$, which is our main interest,  the optimal choice of $\alpha$ is
$$
\alpha = \sqrt{\frac{k_2 C_1}{8}}
$$ and we get
\begin{align*}
\frac{\langle \phi\;|\; {\mathcal H} \phi \rangle}{\| \phi \|_2^2}
=
\Theta_0 h - \kappa(0) C_1 h^{3/2} + \sqrt{\frac{k_2 C_1}{2}}\frac{I_2}{I_0} h^{7/4} + \Og(h^{\frac{15}{8}})\;.
\end{align*}

In the case where $\kappa''(0)=0$, we can choose $\alpha$ as small as we wish and therefore get
\begin{align*}
\frac{\langle \phi\;|\; {\mathcal H} \phi \rangle}{\| \phi \|_2^2}
=
\Theta_0 h - \kappa(0) C_1 h^{3/2} + o(h^{7/4})\;.
\end{align*}
Using
\begin{align*}
I_0 &= \int e^{-\sigma^2}\,d\sigma = \sqrt{\pi}\;, &I_2 &= \int \sigma^2 e^{-\sigma^2}\,d\sigma = \frac{\sqrt{\pi}}{2}\;,
\end{align*}
we therefore get the result of the theorem.
\end{proof}

\begin{remark}~\\
In the case where $k_2 = 0$,  one would expect that the error term $o(h^{7/4})$ could be replaced by (the stronger) ${\mathcal O}(h^s)$ for some $s \in (7/4, 2]$ depending on the order to which the Taylor expansion of $\kappa(s) - \kappa(0)$ vanishes at $0$. We will not pursue this further. See however also Remark~\ref{rem:Degenerate}.
\end{remark}

\section{Grushin type approach for upper bounds}
\label{GrushinUpper}
\subsection{Main statements}~\\
In this section we will prove the following accurate result.

\begin{theorem}
\label{thm:grushin}~\\
Let $\Omega$ satisfy the assumptions of Theorem \ref{thm:main}, and let $n \in {\mathbb N}$. There exist a sequence
 $\{\zeta_j^{(n)}\}_{j=0}^{\infty} \subset {\mathbb R}$
 and a sequence of functions $\{\phi_j^{(n)}\}_{j=0}^{\infty}$ 
in $L^2(\Omega)$ such that, for all $N > 0$, 
there exists  $M>0$ such that, if
\begin{align}\label{eq:formz}
z_M^{(n)}(h) & = \Theta_0 h - k_{\rm max} C_1 h^{3/2} 
+ C_1 \sqrt{\frac{3}{2}} \Theta_0^{1/4} \sqrt{k_2} (2n-1) h^{7/4} + h^{15/8}\sum_{j=0}^M h^{j/8} \zeta_j^{(n)}\;, 
\end{align}
and 
\begin{align}\label{eq:formquasimode}
\phi_M^{(n)}( x,h ) &=  \sum_{j=0}^{M} h^{j/8} \phi_j^{(n)}(x)\;,
\end{align}
then (for $h \searrow 0$)
\begin{align}\label{eq:quasimode}
\| ({\mathcal H}-z_M^{(n)}) \phi_M^{(n)} \|_{L^2} = {\mathcal O}(h^N)\| \phi_M^{(n)} \|_{L^2} \;.
\end{align}
\end{theorem}
With the notations of the theorem, we define $z^{(n)}_{\infty}(h)$ as the asymptotic sum
\begin{align}
\label{eq:defzn}
z^{(n)}_{\infty}(h) &:= \Theta_0 h - k_{\rm max} C_1 h^{3/2} + C_1 \sqrt{\frac{3}{2}} \Theta_0^{1/4} \sqrt{k_2} (2n-1) h^{7/4}
+ h^{15/8}\sum_{j=0}^{\infty} h^{j/8} \zeta_j^{(n)}\;.
\end{align}
Consequently, $z^{(n)}_M (h)$ is the truncated sum of  $z^{(n)}_\infty (h)$
 at rank $M$.

\begin{remark}
\label{rem:74coeff} ~\\
The lowest approximate eigenvalue $z^{(1)}(h)$ agrees with the calculation from Bernoff-Sternberg \cite{BeSt} (see also \cite{St}) up to the order that they calculate (term of order $h^{7/4}$). However, it is different from the result stated in 
del~Pino, Felmer and Sternberg \cite[Remark 4.2]{PiFeSt} that they claim to be sharp.
\end{remark}

Since the operator ${\mathcal H}$ is self-adjoint, we can deduce the existence of eigenvalues near the points with asymptotics $z_{\infty}^{(n)}$.

\begin{cor}~\\
\label{cor:Upper}
Let $n \in  {\mathbb N}$, $M \in {\mathbb N}$ and let $z_M^{(n)}(h)$ be as above. Then there exist $C>0$ and $h_0>0$  such that
$$
\dist( z_M^{(n)}(h), \Spec({\mathcal H})) \leq C h^{\frac{15 +M}{8}}\;,\; \forall h\in (0,h_0]\;.
$$
\end{cor}

\begin{proof}~\\
This is clear by the Spectral Theorem.
\end{proof}

\begin{remark}~\\
In particular, the upper bound announced  in Theorem~\ref{thm:main} is a direct consequence of Theorem~\ref{thm:grushin}.
\end{remark}
~
\begin{proof}[Proof of Theorem~\ref{thm:grushin}]~\\
The proof is fairly long; so we will split it in different steps described in the next subsections. From now on we will assume that the maximum of $\kappa$, $k_{\rm max}$\,, is attained at $s=0\,$.
\subsection{Expanding operators in fractional powers of $h$}~\\
From \cite[(B.8)]{HeMo3} we get that the operator ${\mathcal H}$ in boundary coordinates becomes
\begin{align}
\label{eq:bdrycoor}
{\mathcal H} = a^{-1}\Big[ (hD_s - \tilde{A}_1) a^{-1} (hD_s - \tilde{A}_1) + (hD_t - \tilde{A}_2)a (hD_t - \tilde{A}_2)\Big]\;,
\end{align}
with 
\begin{align}
a(s,t) & = 1 -t \kappa(s)\;.
\end{align}
\begin{remark}~\\
The representation of ${\mathcal H}$ given in \eqref{eq:bdrycoor} is only defined on functions with support in $[0,t_0)\times [- |\partial \Omega|/2, +|\partial \Omega|/2]$.
We will only apply our operator on functions which are a product
of cut-off functions with functions in the form of linear combination of
terms like 
$h^\nu w( h^{-\frac 14} s, h^{-\frac 12}t)$,
with $w$ in $ \mathcal S(\rz\times \overline{\rz^+})$. These functions are consequently  $O(h^\infty)$ outside a fixed neighborhood of $(0,0)$. This is similar to the calculations in the previous section.
We will do the computations formally in the sense that~:
\begin{itemize}
\item Everything is determined modulo $\Og(h^\infty)$;
\item $a^{-1}(s,t)$ will be replaced by $\sum_{n\geq 0} (t \kappa (s))^n $;
\item $\kappa (s)$ will be replaced by its Taylor's expansion.
\end{itemize}

\end{remark}
For any $n$ and $N$, 
we will find $M$ and construct trial functions $\tilde{\phi}_M^{(n)}$ (expressed in boundary coordinates $(s,t)$ and in the form \eqref{eq:formquasimode}), localized near $(s,t)=(0,0)$ and satisfying
\begin{align}
\label{eq:quasimode_bdry}
\| ({\mathcal H} - z_M^{(n)})\tilde{\phi}_M^{(n)} \|_{L^2} &= {\mathcal O}(h^N)\| \tilde{\phi}_M^{(n)} \|_{L^2}\;, &
(hD_t - \tilde{A}_2) \tilde{\phi}_M^{(n)}  \big|_{(s,t)=(s,0)} &= 0\;.
\end{align}
By changing back to the original coordinates, this clearly implies \eqref{eq:quasimode}. We will omit the tilda's in the following and thus denote by $\phi$ the trial function in boundary coordinates.

As in the preceding section (see also Appendix \ref{AppB}),  we choose a gauge where
\begin{align*}
\tilde{A}_1 & = -t a_2(s,t)\;, & \tilde{A}_2& = 0\;; & a_2(s,t) & = 1 -t \kappa(s)/2\;.
\end{align*}
We make the scaling $\tau = h^{-1/2} t$, $\sigma = h^{-1/8} s$. Then ${\mathcal H}$ becomes
\begin{align}\label{avectilda}
\tilde{P} = \tilde{a}^{-1} (h^{7/8} D_{\sigma} +
 h^{1/2}\tau \tilde{a}_2) \tilde{a}^{-1} (h^{7/8} D_{\sigma}
 + h^{1/2}\tau \tilde{a}_2) + h \tilde{a}^{-1}D_{\tau} \tilde{a} D_{\tau}\;,
\end{align}
with
\begin{align}
\label{eq:as}
\tilde{a}(\sigma,\tau) & = 1 -h^{1/2}\tau \kappa(h^{1/8}\sigma)\;,&
\tilde{a}_2(\sigma,\tau) & = 1 -h^{1/2}\tau \kappa(h^{1/8}\sigma)/2\;.
\end{align}
Thus
\begin{align*}
h^{-1} \tilde{P} = \tilde{a}^{-1}(h^{3/8} D_{\sigma} + \tau \tilde{a}_2) \tilde{a}^{-1} (h^{3/8} D_{\sigma} + \tau \tilde{a}_2) + \tilde{a}^{-1} D_{\tau} \tilde{a}D_{\tau}\;.
\end{align*}
We now define 
$$
P = e^{-i \sigma \xi_0/h^{3/8}} h^{-1} \tilde{P} e^{i \sigma
  \xi_0/h^{3/8}} - \Theta_0\;,
$$ 
and get, after removing the tilda's from the $a$'s,
\begin{align}
\label{eq:Def_of_P_scaled}
P & = a^{-1}\big((\tau + \xi_0) + h^{3/8} D_{\sigma} - \tau (1- a_2)\big) a^{-1} \big((\tau + \xi_0) + h^{3/8} D_{\sigma} - \tau (1- a_2)\big) \nonumber\\
&\quad\quad+ a^{-1}D_{\tau} a D_{\tau} - \Theta_0 \;.
\end{align}
We assume that $\kappa$ is $C^{\infty}$ and has a non-degenerate maximum at $s=0$. Then, in the sense of asymptotic series in powers of $h^\frac 18$, we obtain
\begin{align}\label{eq:SomeTaylora}
a(\sigma,\tau) & = 1 -h^{1/2}\tau \kappa(h^{1/8}\sigma) = 1 - h^{1/2}\tau \kappa(0) - 
\tau \sum_{j=2}^{\infty} h^{1/2 + j/8} \frac{\sigma^j \kappa^{(j)}(0)}{j!}
\;,\end{align}
and
\begin{align}\label{eq:SomeTaylora2}
a_2(\sigma,\tau) & = 1 -h^{1/2}\tau \frac{\kappa(h^{1/8}\sigma)}{2} = 1 - h^{1/2}\tau \frac{\kappa(0)}{2} - 
\tau \sum_{j=2}^{\infty} h^{1/2 + j/8} \sigma^j \frac{\kappa^{(j)}(0)}{2 (j!)}\;.
\end{align}
From the asymptotics of $a$ and $a_2$, we get (remember the definition of $k_2$ from \eqref{eq:non-deg})
\begin{align}\label{eq:SomeTaylor2}
a(\sigma,\tau)^{-1} & = 1 + h^{1/2}\tau \kappa(0) -
\tau h^{3/4} \frac{\sigma^2 k_2 }{2} + \Og(h^{7/8})\;, \nonumber \\
a(\sigma,\tau)^{-2} & = 1 + 2h^{1/2}\tau \kappa(0) -
\tau h^{3/4} \sigma^2 k_2 + \Og(h^{7/8})\;, \nonumber\\
-\tau(1-a_2(\sigma,\tau)) &= - h^{1/2}\tau^2 \frac{\kappa(0)}{2} +
\tau^2  h^{3/4} \sigma^2 \frac{k_2}{4} + \Og(h^{7/8})\;.
\end{align}
Thus, we can write
\begin{align}
\label{eq:PExp}
P &= P_0 + h^{3/8}P_1 + h^{1/2}P_2 + h^{3/4} P_3 + h^{7/8} Q(h)\;,
\end{align}
where
\begin{align}
P_0 & = D^2_{\tau} + (\tau + \xi_0)^2-\Theta_0\;, 
\label{eq:P0s}\\
P_1 &= 2 D_{\sigma}(\tau + \xi_0)\;, \label{eq:P1s}\\
P_2 &= - 2 \tau^2 \frac{\kappa(0)}{2} (\tau + \xi_0) + 2 \tau \kappa(0) (\tau + \xi_0)^2 +\kappa(0)(\tau D_{\tau}^2 - D_{\tau} \tau D_{\tau})  \nonumber\\
&=  \kappa(0)\big( 2 \tau (\tau + \xi_0)^2 - \tau^2 (\tau + \xi_0)\big) +i \kappa(0) D_{\tau}\;,\label{eq:P2s}\\
P_3 &= D_{\sigma}^2 - \tau  \sigma^2 k_2 (\tau + \xi_0)^2 + 2 \tau^2 \sigma^2 \frac{k_2}{4}(\tau + \xi_0) -\frac{k_2 \sigma^2}{2}(\tau D_{\tau}^2 - D_{\tau} \tau D_{\tau})\nonumber \\
&= D_{\sigma}^2 -  \big(2 \tau (\tau + \xi_0)^2 
- \tau^2 (\tau + \xi_0)\big) \frac{k_2 \sigma^2}{2}
- \frac{k_2 \sigma^2}{2} i D_{\tau}\;, \label{eq:P3s}
\end{align}
and where $Q(h)$ admits a complete expansion~:
\begin{align*}
Q(h) \sim \sum_{j=0}^{\infty} h^{j/8} Q_j\;.
\end{align*}
We define $\delta P$ by
\begin{align}
\label{eq:DefDeltaP}
\delta P = P - P_0\;,
\end{align}
We search for functions $\phi^{(n)}(h)$ having an asymptotic expansion in $h^{1/8}$ and such that
\begin{align}
\label{eq:sloppy}
(P - \frac{z^{(n)}(h) + \Theta_0 h }{h} ) \phi^{(n)}(h) &\sim 0\;,&D_{\tau} \phi^{(n)}(h; \sigma,0)=0\;.
\end{align}
The constructed functions will have sufficient decay properties to allow interpreting \eqref{eq:sloppy} in the $L^2$ sense and therefore, after multiplying by the cutoff appearing in \eqref{formesimplea}, we
 get \eqref{eq:quasimode_bdry} (which implies \eqref{eq:quasimode}).

\subsection{Reduction to the boundary}~\\
\label{RedBdry}
We will now explain the strategy initiated by  Grushin \cite{Gru} (and reference therein)  and Sj\"ostrand in the context of hypoellipticity \cite{Sj}. Here we use this strategy for 
producing good trial functions and thereby results for the magnetic Neumann Laplacian.

Let us define the operators $R_0^{+}$, $R_0^{-}$ and $E_0$ by~:
\begin{align}
\label{eq:DefR0+}
R_0^{+}:~& {\mathcal S}({\mathbb R}_{\sigma}) \rightarrow {\mathcal S}({\mathbb R}_{\sigma} \times (\overline{{\mathbb R}_{+}})_{\tau}) \\
&\phi(\sigma)  \mapsto \phi(\sigma) u_0(\tau) = \phi \otimes u_0\;, \nonumber
\end{align}
\begin{align}
\label{eq:DefR0-}
R_0^{-}:~&  {\mathcal S}({\mathbb R}_{\sigma} \times (\overline{{\mathbb R}_{+}})_{\tau})\rightarrow {\mathcal S}({\mathbb R}_{\sigma}) \\
&f  \mapsto \int_0^{\infty} f(\sigma,\tau) u_0(\tau) \,d\tau\;,\nonumber
\end{align}
\begin{align}
\label{eq:DefE0}
E_0 :~& {\mathcal S}({\mathbb R}_{\sigma} \times (\overline{{\mathbb R}_{+}})_{\tau})\rightarrow {\mathcal S}({\mathbb R}_{\sigma} \times (\overline{{\mathbb R}_{+}})_{\tau}) \\
&f \otimes \phi  \mapsto \begin{cases} f\otimes (P_0^{-1} \phi)\;, & \text{ if } \phi \perp u_0\;, \\
0\;, & \text{ if } \phi \parallel u_0\;. \end{cases}\nonumber
\end{align}
Here we abused notation and considered $P_0$ as an operator on $ L^2(({\mathbb R}_{+})_{\tau})$ in order to define $E_0$. That $E_0$ respects the Schwartz space ${\mathcal S}({\mathbb R}_{\sigma} \times (\overline{{\mathbb R}_{+}})_{\tau})$ follows from Lemma~\ref{lem:regresolvent}.

Notice that $R_0^{+}$ is the Hilbertian adjoint of $R_0^{-}$ (seen
 as an operator from $L^2 ( \rz_\sigma \times \rz^+_\tau; d\sigma d\tau)$
 into $L^2(\rz_\sigma)$). On the other hand $(P-z)$ is for $z\in \mathbb R$ formally selfadjoint  for the original ($h$-dependent) 
 $L^2$ scalar product  inherited from the change of variable $z\mapsto (s,t)\mapsto (\sigma,\tau)$  (that is
 associated to the measure $(1 - h^{-\frac 12} \tau \kappa(h^{-\frac 14} \sigma) d\sigma d \tau)$)
 but not for the usual $L^2$ associated to the standard Lebesgue measure $d\sigma d \tau$.

With the above notations  we define matrices of operators
\begin{align}
\label{eq:def_matrices}
{\mathcal P}(z) &= \left(\begin{matrix} P-z & R_0^{+} \\ R_0^{-} & 0 \end{matrix}\right)\;,&
{\mathcal E}_0 &= \left(\begin{matrix} E_0 & R_0^{+} \\ R_0^{-} & 0 \end{matrix}\right)\;.
\end{align}
These operators act on ${\mathcal S} (\rz_\sigma\times(\overline {\rz_+})_\tau)
 \times  {\mathcal S}({\mathbb R}_\sigma)$. Actually
 we should better think of operators applied to formal expansions
 in suitable fractional powers of $h$ with coefficients
 in these $\mathcal S$ spaces. These infinite formal expansions will then be 
 truncated at a suitable rank for defining our quasimodes. So we prefer
 to write formal expansions to infinite order, having in mind that we could actually go back to truncated  expansions if we want a given, arbitrarily small, remainder estimate.\\
We note first that~:
\begin{equation}\label{inverse}
\left(\begin{matrix} P_0 & R_0^{+} \\ R_0^{-} & 0 \end{matrix}\right) \circ {\mathcal E}_0= I\;.
\end{equation}
An easy calculation gives then~:
\begin{align*}
{\mathcal P}(z) {\mathcal E}_0 = I + {\mathcal K}\;,
\end{align*}
where
\begin{align*}
{\mathcal K} = \left(\begin{matrix} (\delta P-z) E_0 & (\delta P-z)R_0^{+} \\ 0& 0 \end{matrix}\right)\;.
\end{align*}
If we look for $z=z(h)$ satisfying
\begin{equation}\label{assz}
z(h) \sim  \sum_{\ell\geq 3} \hat  z_\ell h^\frac \ell 8\;,
\end{equation}
and having in mind the expansion \eqref{eq:PExp}, 
we observe that 
$
(\delta P -z) = \Og(h^{3/8})\;,
$
when acting on a fixed function in 
$\mathcal S(\rz_\sigma \times (\overline {\rz^+})_\tau)$ and can actually be expanded in powers of $h^\frac 18$, starting from $h^\frac 38 ( P_1-\hat z_3)$. So, if we define
\begin{align*}
{\mathcal Q}_\infty \sim \sum_{j=0}^{+\infty} (-1)^j {\mathcal K}^j\;,
\end{align*}
then the operator is well defined (after reordering) 
as a formal expansion in powers of $h^\frac 18$ and 
\begin{align}
\label{eq:almostInverse}
{\mathcal P}(z) {\mathcal E}_0 {\mathcal Q}_\infty \sim  I\;.
\end{align}
Now,
\begin{align*}
{\mathcal K}^j \sim \left(\begin{matrix} [(\delta P-z) E_0]^j & [(\delta P-z) E_0]^{j-1} (\delta P-z)R_0^{+} \\ 0& 0 \end{matrix}\right)\;,
\end{align*}
and therefore, if we write
\begin{align*}
{\mathcal E}(z):= {\mathcal E}_0 {\mathcal Q}_\infty
 = \left( \begin{matrix} E_\infty(z) & E_\infty^+(z) \\ E_\infty^-(z) & E_\infty^{\pm}(z) \end{matrix} \right)\;,
\end{align*}
we get, in the sense of formal expansions in powers of $h^\frac 18$,
\begin{subequations}
\begin{align}
(P-z) E_\infty(z) + R_0^+ E_\infty^-(z) & \sim 1\;, \\
\label{eq:328b}
(P-z) E_\infty^+(z) + R_0^+ E_\infty^{\pm}(z) & \sim 0\;,\\
R_0^{-} E_\infty(z) & \sim 0\;, \\
R_0^{-} E_\infty^{+}(z) & \sim 1\;.
\end{align}
\end{subequations}
So, in particular, if $\sigma \mapsto \phi_{\infty} (\sigma;h)$ is a function\footnote{More precisely $\phi_{\infty} (\,\cdot\,;h)\sim \sum_{j\in  \mathbb N} h^{\frac j8} \phi_j(\,\cdot\,)\,$, so all the computations have to be expanded in powers of $h^\frac 18$.} such that 
\begin{equation}\label{phiinfinia}
E_\infty^{\pm}(z) \phi_{\infty}\sim 0\;,
\end{equation}
then inserting $\phi_{\infty}$ in \eqref{eq:328b} (i.e. inserting $\left(\begin{matrix} 0 \\ \phi_{\infty}\end{matrix}\right)$ in \eqref{eq:almostInverse}), we find
\begin{align}
\label{eq:AlmostEigenfunction}
(P-z) E_\infty^+(z) \phi_{\infty}  \sim 0\;,
\end{align}
where everything is well defined modulo $\Og (h^\infty)\,$.
\subsection{Construction of trial functions}~\\
\label{ConstTrial}
From the above, we see that $E_\infty^{\pm}(z)$ is the following asymptotic series, 
\begin{align}
\label{eq:Epm}
E_\infty^{\pm}(z) = \sum_{j=1}^{\infty} (-1)^j R_0^{-} [(\delta P-z)E_0]^{j-1} (\delta P-z)R_0^{+}\;.
\end{align}
We look as before for
\begin{align*}
\phi_{\infty}(\sigma;h) &\sim  \sum_{j=0}^{\infty} \phi_j(\sigma) h^{j/8}\;, \\
z_{\infty}(h) &\sim h^{3/8} z_1 + h^{1/2} z_2 + h^{3/4} z_3 + h^{7/8}\sum_{j=0}^{\infty} \zeta_j h^{j/8}\;,
\end{align*}
such that
\begin{equation}\label{eq:formalsolution}
E_\infty^{\pm}(z_{\infty}(h))\phi_{\infty}(\sigma;h) \sim 0\;,
\end{equation}
in the sense of asymptotic series in powers of $h^\frac 18$.
Here the functions $\phi_j$ are supposed to be in $\mathcal S(\rz_\sigma)$.

\begin{lemma}
\label{lem:asymptoticSolutions}~\\
For each $n \in {\mathbb N}$, there exists a unique solution $(z^{(n)}(h),\phi^{(n)}(h))$ to  equation \eqref{eq:formalsolution}, 
in the sense of asymptotic series, and such that
$$
z^{(n)}(h) \sim C_1 \sqrt{\frac 32 \sqrt{\Theta_0} k_2} \; (2n-1)h^{3/4} + h^{7/8}\sum_{j=0}^{\infty} \zeta_j^{(n)} h^{j/8}\;. 
$$
Conversely, for any pair $(z(h),\varphi(h))$ such that \eqref{eq:formalsolution} is satisfied, with \break $z(h) \sim C h^\frac 34
+ h^\frac 78\sum_{j\geq 0}  \zeta_j h^{\frac j8}$ and $\varphi(h) \sim \sum_{j\geq 0}h ^\frac j8 \varphi_j$,  there exist $n$ and \break
$c(h)\sim \sum_j c_j h^\frac j8$ such that $z(h)= z^{(n)}(h)$ and 
$\varphi(h)= c(h) \phi^{(n)} (h)\,$.
\end{lemma}
~\\
\begin{proof}[Proof of Lemma~\ref{lem:asymptoticSolutions}]~\\
Let us write
\begin{align}
\label{eq:Epmdef}
E_\infty^{\pm}(z_{\infty}(h)) \sim h^{3/8} E_1 + h^{1/2} E_2 + h^{3/4} E_3 + h^{7/8} \sum_{j=0}^{\infty} h^{j/8} F_j\;.
\end{align}
The terms in this sum will be given in \eqref{eq:E1}, \eqref{eq:E2}, \eqref{eq:E3}, and \eqref{eq:Fj} below. Using the definitions \eqref{eq:PExp} and \eqref{eq:DefDeltaP}, and the fact that $R_0^{-} E_0 = 0$ and $E_0 R_0^{+} = 0$, we get modulo terms of order $\Og(h^\frac 78)\,$,
\begin{align*}
E_\infty^{\pm}(z_{\infty}(h)) &\sim  - R_0^{-}\Big( h^{3/8}(P_1-z_1)  + h^{1/2} (P_2-z_2)  + h^{3/4} (P_3-z_3)\Big) R_0^{+} \\
&\quad + h^\frac 34 R_0^{-}P_1 E_0 P_1 R_0^{+} + {\mathcal O}(h^{7/8})\;.
\end{align*}
Since also $ R_0^{-} (\tau+\xi_0) R_0^{+} = 0\,$, we find, using \eqref{eq:P1s},
\begin{align}
\label{eq:E1}
E_1 &= - R_0^{-}(P_1 - z_1) R_0^{+} = z_1\;.
\end{align}
Furthermore, using again \eqref{eq:P2s},
\begin{align*}
E_2 &= - R_0^{-}(P_2 - z_2) R_0^{+} 
=
z_2 - \kappa(0) (I_{1,1}+I_{1,2})\;,
\end{align*}
with
$$
I_{1,1} = \int_0^{\infty} [2\tau (\tau+\xi_0)^2 -\tau^2(\tau+\xi_0)] u_0^2(\tau)\,d\tau\;,
$$
and
$$
I_{1,2} = i \int_0^{\infty} u_0(\tau) D_{\tau} u_0(\tau)\,d\tau\;.
$$
Using Proposition~\ref{prop:Iij}, we get
$$
I_{1,1}+I_{1,2} = -C_1\;,
$$
and therefore
\begin{align}
\label{eq:E2}
E_2 
=
z_2 + \kappa(0) C_1\;.
\end{align}
The term $E_3$ becomes, inserting $P_1$ and $P_3$ from \eqref{eq:P1s} and \eqref{eq:P3s},
\begin{align}
\label{eq:whyImp}
E_3 &= - R_0^{-}(P_3 - z_3) R_0^{+} + R_0^{-} P_1 E_0 P_1 R_0^{+} \nonumber \\
&=
z_3 - D_{\sigma}^2 + \frac{k_2\sigma^2}{2} (I_{1,1}+I_{1,2}) + 4 D_{\sigma}^2 I_2\;,
\end{align}
where we have introduced
\begin{align}
\label{eq:I2}
I_2 = \int_0^{\infty} (\tau+\xi_0) u_0(\tau) P_0^{-1} (\tau+\xi_0) u_0(\tau)\, d\tau\;.
\end{align}
Using Proposition~\ref{prop:Iij} and Proposition~\ref{prop:secondPerturb},
 we have
\begin{align*}
1-4I_2 & = 3 C_1 \sqrt{\Theta_0}\;,& I_{1,1}+I_{1,2} &= -C_1\;,
\end{align*}
and we therefore get
\begin{align}
\label{eq:E3}
E_3 = z_3 - 3 C_1 \sqrt{\Theta_0}D_{\sigma}^2 -  C_1 \frac{k_2 \sigma^2}{2}\;.
\end{align}
Remember that $\kappa(s)$ has a non-degenerate maximum at $s=0$,
 so $k_2=-\kappa''(0) > 0$.\\
\noindent
{\bf The first terms}.\\
In order to get the equation \eqref{eq:formalsolution}
to be satisfied, we choose
\begin{align}
\label{eq:firstcoeffs}
z_1  = 0\;,\; z_2 = - \kappa(0) C_1\;,
\end{align}
which implies
\begin{align}
E_1 =0\;,\; E_2 = 0\;.
\end{align}
With this choice, \eqref{eq:formalsolution}  becomes
\begin{align*}
0 \sim & h^{3/4} E_3 \phi_0 + {\mathcal O}(h^{7/8})\;.
\end{align*}
So we determine $z_3$ and $\phi_0$ by
\begin{equation}\label{3.26a}
E_3 \phi_0 = 0\;.
\end{equation}
Let us solve the equation $E_3 \phi = 0\,$. It reads,  with $k_2=-\kappa''(0)\,$, 
\begin{align*}
z_3 \phi & = C_1(3  \sqrt{\Theta_0}D_{\sigma}^2
 + \frac{k_2 \sigma^2}{2})\phi\;.
\end{align*}
So, after  the scaling $\tilde{s} = \sqrt[4]{\frac{k_2}{6\sqrt{\Theta_0}}} \sigma\,$, we find  that $z_3$ should be an eigenvalue of the harmonic oscillator
$$
C_1 \sqrt{\frac 32 \sqrt{\Theta_0} k_2}  \; \left(D_{\tilde{s}}^2 + \tilde{s}^2\right)\;.
$$
Thus the possible values of $z_3$ are~:
\begin{align}
\label{eq:HarmOsc}
z_3^{(n)} = C_1 \sqrt{\frac 32 \sqrt{\Theta_0} k_2} \; (2n-1)\;, 
\text{ where } n\in \mathbb N\setminus \{0\}\;.
\end{align}
In particular, using the inequality 
$3C_1 \sqrt{\Theta_0} = 1 - 4 I_2 < 1$ (see Proposition~\ref{prop:secondPerturb}), we get that $z_3^{(1)}$ is smaller than
 the value in   Theorem~\ref{thm:variational}.
\begin{remark}~\\
A second look at the calculations above (comparing with Section~\ref{simple})
 shows why Theorem~\ref{thm:variational} does not give 
the correct ground state energy to order $h^{7/4}$. By using 
a trial state which has the simple form
 \eqref{formesimplea} and  \eqref{formesimpleb},  we would 
not see the term $R_0^{-} P_1 E_0 P_1 R_0^{+}$ in the 
first line of \eqref{eq:whyImp} and therefore the last term, 
$4 D_{\sigma}^2 I_2$, would be missing in the second line 
of \eqref{eq:whyImp}. Thus the harmonic oscillator 
discussed above  would become
$$
D_{\tilde{s}}^2+ \frac{k_2 C_1}{2} \tilde{s}^2 \;,
$$
instead. This harmonic oscillator has ground state 
energy $\sqrt{\frac{k_2 C_1}{2}}$ in agreement 
with the result of Theorem~\ref{thm:variational}.
\end{remark}
\noindent
{\bf The iteration procedure.}~\\
Let us define $\Pi$ to be the orthogonal projection on $\{ \phi_0 \}^{\perp}$. 
Notice that $\phi_0$ depends on the $n$ chosen in \eqref{eq:HarmOsc} even though we do not explicitly recall this dependence in the notation.
We will choose $\phi_j$ such that 
\begin{equation}\label{condorth}
\phi_j \perp \phi_0 \mbox{ for all } j>0\;.
\end{equation}
The term of order $h^{\frac 34  + \frac j8 }$ in \eqref{eq:formalsolution} becomes
\begin{align}
\label{eq:transport}
E_3 \phi_j + \sum_{k=0}^{j-1} F_k \phi_{j-1-k} = 0\;.
\end{align}
Notice that
\begin{align}
\label{eq:Fj}
F_j = \zeta_j - R_0^{-} Q_j R_0^+ + \tilde{F}_j\;,
\end{align}
where $\tilde{F}_j$ only depends on $z_1, z_2, z_3$ and 
$\{ \zeta_k\}_{k=0}^{j-1}$. By taking the scalar product 
with $\phi_0$ in the equation \eqref{eq:transport}, 
we therefore get, by using \eqref{3.26a}
and the property that $E_3$ is self adjoint,
\begin{align*}
\zeta_{j-1} \| \phi_0 \|^2 = \langle \phi_0\;, 
R_0^{-} Q_{j-1} R_0^+ \phi_0 \rangle - 
\langle \phi_0\;|\; \tilde{F}_{j-1} \phi_0 \rangle
 - \langle \phi_0\;|\;\sum_{k=0}^{j-2} F_k \phi_{j-1-k} \rangle\;.
\end{align*}
Since $\phi_0 \neq 0$, this equation determines $\zeta_{j-1}\in \mathbb C$
 as a function of $z_1, z_2, z_3$, $\{ \zeta_k\}_{k=0}^{j-2}$
 and $\{ \phi_k\}_{k=0}^{j-1}$\;. The property that $\zeta_{j-1}\in \mathbb R$
 will be only proved later.

Upon projecting the equation \eqref{eq:transport} 
on $\{ \phi_0 \}^{\perp}$, and 
remembering the choice \eqref{condorth}, we get
\begin{align}
\label{eq:function}
\Pi E_3 \Pi \phi_j = - \Pi \Big( \sum_{k=0}^{j-1} F_k \phi_{j-1-k} \Big)\;.
\end{align}
Since $\Pi E_3 \Pi $ is invertible on $\{ \phi_0 \}^{\perp}$, 
\eqref{eq:function} together with \eqref{condorth} determines $\phi_j$.
\noindent
{\bf Uniqueness.}~\\
Suppose that $z_1, z_2$ are not given by the choice 
in \eqref{eq:firstcoeffs}. For concreteness,
 let us suppose that $z_1 \neq 0$. Then the 
equation \eqref{eq:formalsolution} implies that $\phi \sim 0$.
 Thus \eqref{eq:firstcoeffs} is the only nontrivial choice.

Furthermore, in the construction above we imposed that
 $\phi_j \perp \phi_0$ for all $j>0\,$. Suppose we do
 not impose that condition. Let $\overline{\phi}_j$
 be the solution constructed above and let $\phi_j$ be the new solution.
Then we can write each $\phi_j$ as 
\begin{align}
\phi_j &= \phi_j' + c_j \phi_0\;, &\text{ with } \phi'_j & \perp \phi_0, \quad c_j \in {\mathbb C}\;.
\end{align}
We now write
\begin{align*}
&&\phi(h) &\sim  \phi_0 + \sum_{j\geq 1} h^{j/8} \phi_j 
\sim  c(h) \phi_0 + \sum_{j \geq 1} h^{j/8} \phi'_j\;, \\
\text{ with } &&c(h) &\sim 1 + \sum_{j\geq 1} h^{j/8} c_j\;, \\
\text{ and } && \phi_j' & \perp \phi_0\;.
\end{align*}
By linearity, we therefore find that $\frac{\phi(h)}{c(h)}$ 
is the solution $\phi_0 + \sum_{j\geq 1} h^{j/8} \overline{\phi}_j$ 
constructed above. 

This finishes the proof of Lemma~\ref{lem:asymptoticSolutions}
\end{proof}

Using Lemmas~\ref{lem:asymptoticSolutions}, \ref{lem:Schwarzbdry-Schwarzdomain}, and \ref{lem:regresolvent}, 
we can finish the proof of Theorem~\ref{thm:grushin}.
 Let $(z^{(n)}(h), \phi^{(n)}(h))$ be one of the formal
 solutions from Lemma~\ref{lem:asymptoticSolutions}.
 By stopping the formal sum at a finite number of
 terms we obtain partial sums $(z_M^{(n)}(h), \phi_M^{(n)}(h))$,
 solutions to
\begin{align}
\label{eq:finitesum}
E^{\pm}_M(z_M^{(n)}(h))  \phi_M^{(n)}(h) = h^{M} R_M(h)\;,
\end{align}
where $E^{\pm}_M$ is also defined by stopping the expansion
 of $E^{\pm}_\infty$.
Since the $\phi_j$'s are Schwartz functions and all involved
 operators respect the space ${\mathcal S}$ 
(they are differential operators whose coefficients 
are smooth with polynomially bounded derivatives), 
the remainder  $R_M(h)$ in \eqref{eq:finitesum} 
is bounded in ${\mathcal S}$. 
Using Lemma~\ref{lem:regresolvent}, 
Lem\-ma~\ref{lem:Schwarzbdry-Schwarzdomain} 
and the fact that all terms in $E_M^{+}$ preserve 
the Schwartz space (differential operators 
with polynomially bounded, smooth derivatives),
 we see that $E_M^{+} \phi_M^{(n)}(h)$ defines 
a finite sum,
whose coefficients are in the space 
${\mathcal S}({\mathbb R}\times
 \overline{{\mathbb R}_{+}}\,)$.
Thus, the procedure described in Subsection~\ref{RedBdry} above (reduction to the boundary) 
gives a solution $\psi_M(h)=E_M^{+} \phi_M^{(n)}(h)$ 
to the equation
$$
(P-z_M(h))\psi_M(h) = {\mathcal O}(h^{M})\;.
$$
Here the right hand side is in  $\mathcal S$ and controlled
 in ${\mathcal O}(h^{M})$ for any semi-norm on 
$\mathcal S$, thus in particular in the $L^2$ norm.

Moreover, ${\mathcal H}$ being selfadjoint, we can now prove that $\zeta_j\in \mathbb R$ and this finishes the proof of Theorem~\ref{thm:grushin}.
\end{proof}

\section{Space localization}~\\
\label{locPhase}
In this section we will prove that the ground state is well localized both in $s$ and $t$. In the following Section~\ref{phasespace} we will prove a similar (slightly weaker) localization result in the frequency variable $\xi$ corresponding to $s$.

\subsection{Agmon estimates in the normal direction}
\label{Agmont}~\\
If $\phi$ is a function with compact support in $\Omega$, i.e. $\phi \in C^{\infty}_0(\Omega)$, then 
\begin{equation}\label{ineqfond}
\int_{\Omega} | (-ih\nabla-A)\phi|^2\,dx = \int_{{\mathbb R}^2} | (-ih\nabla-A)\phi|^2\,dx \geq h \| \phi \|^2_{L^2({\mathbb R}^2)} = h \| \phi \|^2_{L^2(\Omega)}\;.
\end{equation}
Since $1 > \Theta_0\,$, this implies (compare $1\cdot h$ with $\Theta_0 \cdot h$) that functions with energy below our upper bounds from Theorem~\ref{thm:grushin} cannot be localized in the interior of $\Omega$ (i.e. away from the boundary), as $h\rightarrow 0\,$. The powerful method of Agmon estimates can be applied to strengthen this property into an exponential localization of the eigenfunctions (corresponding to the bottom of the spectrum)
in a neighborhood of the boundary. This is the content of the following proposition.

\begin{theorem}[Normal Agmon estimates]
\label{prop:Agmon-t}~\\
Let $h_0>0, M\in (\Theta_0, 1)$. Then there exists $C, \alpha>0$ and $h_1 \in (0, h_0]$ such that if 
$(u_h)_{h\in (0,h_0]}$ is a family of  normalized eigenfunctions of ${\mathcal H}_{h, \Omega}$ with corresponding eigenvalue $\mu(h)$ satisfying
$\mu(h) \leq M h $,
then, for all $h \in (0,h_1]$, 
\begin{align}
\int_{\Omega} e^{2\alpha \dist(x, \partial \Omega)/h^{1/2}} 
\big( |u_h(x)|^2 + h^{-1} |(-ih\nabla -A) u_h(x)|^2 \big) \,dx \leq C\;.
\end{align}
\end{theorem}

\begin{proof}~\\ 
The proof is similar to (but easier than) the proof of Theorem~\ref{thm:Agmons} below. We omit the details and refer to Helffer-Morame \cite[Section~6.4, p.~621-623]{HeMo3} or Helffer-Pan \cite{HePa}. In \cite{HeMo3} only the ground state is considered, but it is immediate to see that the analysis goes through for the eigenfunctions corresponding to higher eigenvalues.
\end{proof}
As a corollary, we get the weaker but useful estimate for $u_h$ near the boundary.
\begin{corollary}[Weak normal Agmon estimates]~\\
Let the assumptions be as in Theorem~\ref{prop:Agmon-t}.
For any integer $k$, there exist $C >0$ and $h_0$, such that
\begin{equation}\label{coragmont}
|| \dist(x, \partial \Omega)^k u_h ||_{L^2(\Omega)} \leq C\; h^\frac k2\;,\;\forall h\in (0,h_0]\;.
\end{equation}
\end{corollary}

\begin{remark}~\\
The $L^2$ statement in Theorem~\ref{prop:Agmon-t} can be converted to an $L^{\infty}$ result using the Sobolev imbedding theorem. See \cite[Theorem~6.3]{HeMo3} for details.
\end{remark}

\subsection{First lower bound}
\label{FirstLower}~\\
In order to get good localization properties of the eigenfunctions in the variable parallel to the boundary (the $s$ variable), we need to improve the lower bound on the ground state energy from \eqref{eq:previousEst}. We will prove the following improvement of \cite[Theorem~10.3]{HeMo3}.

\begin{theorem}
\label{thm:HMimproved}
Let $\Omega$ be a bounded region with smooth boundary satisfying the assumptions of Theorem~\ref{thm:main}. Then
\begin{align}
\label{eq:GoodLower}
\mu^{(1)}(h) \geq \Theta_0 h - C_1 k_{{\rm max}} h^{3/2} + {\mathcal O}(h^{7/4})\;.
\end{align}
\end{theorem}

\begin{proof}
Since $\Omega$ is bounded, we have $k_{{\rm max}} >0$. Using the results from \cite{HeMo2}, we may localize to the region near boundary points with $\kappa(s) >0$, so we may assume without loss of generality in the proof that $\kappa(s) >0$ for all $s$.

The proof of Theorem~\ref{thm:HMimproved} is similar to the proof of  \cite[Theorem~10.3]{HeMo2}. We just need to do one of the estimates slightly more carefully. In particular, the proof goes by comparison with the case of a disc. For a disc, one can calculate the ground state energy with great precision by using the rotational symmetry. This was carried through in \cite{BaPhTa}. We state one of their results in the following form.

\begin{theorem}
\label{lem:BPT}
Let $\mu^{(1)}(h,b,D(0,R))$ be the ground state energy of the operator in \eqref{defHNeumann} in the case where $\curl A = b$ (independent of $x \in \Omega$) and $\Omega = D(0,R)$; the disc of radius $R$. Then there exists $C>0$ such that if
$$
bR^2/h \geq C\;,
$$
then
\begin{align}
\mu^{(1)}(h,b,D(0,R)) \geq \Theta_0 b h - C_1 b^{1/2} h^{3/2}/R - C h^2 R^{-2}\;.
\end{align}
\end{theorem}
\noindent Notice that in the case of a disc, the curvature $\kappa$ is constant, $\kappa = R^{-1}$.

Let $\rho > 0$. We can find a sequence $\{ s_{j,h} \}_{j=0}^{N(h)}$ in ${\mathbb R}/|\partial \Omega|$ and a partition of unity $\{ \tilde{\chi}_{j,h} \}_{j=0}^{N(h)}$ on ${\mathbb R}/|\partial \Omega|$ such that
$\supp \tilde{\chi}_{j,h} \cap \supp \tilde{\chi}_{k,h} = \emptyset$ if $j \notin \{k-1, k , k+ 1\}$ (with the convention that $N(h) + 1 = 0\,$, $0-1 = N(h)$). Furthermore, we may impose the conditions
\begin{align}
\supp \tilde{\chi}_{j,h} &\subset s_{j,h}  + [-h^{\rho}, h^{\rho}]\;,
&
\sum_{j} \tilde{\chi}_{j,h}^2 &= 1 \;,
&
\sum_{j} | \nabla \tilde{\chi}_{j,h}|^2 &\leq  C h^{-2\rho} \;.
\end{align}
We will always choose the $s_{j,h}$ such that $|s_{j,h}| \leq |\partial \Omega|/2\,$.

Let $\chi_1, \chi_2$ be a standard partition of unity on ${\mathbb R}$~:
\begin{align}\label{eq:chi12}
\chi_1^2 + \chi_2^2 &= 1\;, &
\supp \chi_1 &\subset (-2,2)\;, &
\chi_1 &= 1 \text{ on a nbhd of } [-1,1]\;.
\end{align}
Let us define
$$
\chi_{j,h}(s,t) = \tilde{\chi}_{j,h}(s) \chi_1(t/h^{\rho})\;.
$$
We will also consider $\chi_{j,h}$ as a function on $\Omega$ (by passing to boundary coordinates) without changing the notation.
For $\psi \in D(\mathcal H)$, we can write
\begin{equation}\label{decpsi}
\psi = \sum_j \chi_{j,h}^2 \psi + \theta_{2,h}^2 \psi\;,
\end{equation}
with
\begin{equation}\label{defthetajh}
\theta_{j,h} (x)= \chi_j(t(x)/h^{\rho})\;,\; \mbox{for } j=1,\,2\;.
\end{equation}
We get by the `IMS'-formula~:
\begin{align}
\langle \psi \,\big |\, {\mathcal H} \psi \rangle & =
\big\langle \psi \,\big |\, {\mathcal H}  \big(\sum_j \chi_{j,h}^2 \psi + 
\theta_{2,h}^2\psi\big) \big\rangle \\
&= \sum_j \langle \chi_{j,h} \psi\,\big |\, 
{\mathcal H} \chi_{j,h} \psi \rangle - h^2 \int |\nabla \chi_{j,h}|^2 | \psi|^2 \,dx \nonumber\\
&\quad + \langle \theta_{2,h} \psi\, \big |\, {\mathcal H} \theta_{2,h} \psi \rangle - h^2 \int |\nabla \theta_{2,h}|^2 | \psi|^2 \,dx\;.
\end{align}
In particular, for $\psi = u_h^{(1)}$ being a  normalized ground state wave function, we get using the weak normal Agmon estimates
 (in the $t$ variable)  and the condition $\rho < 1/2$ (we will choose $\rho = 1/8$ in the end).
\begin{align}\label{eq:IMS}
\mu^{(1)}(h) = \sum_j \langle \chi_{j,h} u_h^{(1)} \, \big |\, {\mathcal H} \chi_{j,h} u_h^{(1)} \rangle + {\mathcal O}(h^{2-2\rho})\;.
\end{align}
We will write each of the terms in $\langle \cdot \,|\, \cdot \rangle$ in boundary coordinates and compare with the similar term with fixed curvature. 
Define 
\begin{align*}
\tilde{A}_{1}(s,t) &= -t(1-t\kappa(s)/2)\;, &\;a(s,t)&=1-t\kappa(s)\;,
\end{align*}
and
\begin{equation}
{\mathcal B}_{j,h} := \int e[\chi_{j,h} u_h^{(1)}](s,t) \,ds\,dt \;, \label{defBjh}
\end{equation}
with
\begin{equation}
e[f] := a^{-1} |(hD_s - \tilde{A}_{1}) f |^2 + a |h D_t f|^2\;.
\end{equation}
Similarly, we define 
\begin{align*}
\kappa_{j,h} &= \kappa(s_{j,h})\;, &
\;\tilde{A}_{1,j,h}(s,t)& = -t(1-t\kappa_{j,h}/2)\;, &
\;a_{j,h}&=1-t\kappa_{j,h}\;,
\end{align*}
and
\begin{equation}
{\mathcal A}_{j,h} := \int e_{j,h}[\chi_{j,h} u_h^{(1)}](s,t) \,ds\,dt \;,\label{defAjh}\end{equation}
with
\begin{equation}
e_{j,h}[f] := a_{j,h}^{-1} |(hD_s - \tilde{A}_{1,j,h}) f |^2 + a_{j,h} |h D_t f|^2\;.
\end{equation}
We observe that  ${\mathcal B}_{j,h} = \langle \chi_{j,h} u_h^{(1)}\, |\, {\mathcal H} \chi_{j,h} u_h^{(1)} \rangle$ and will compare ${\mathcal B}_{j,h}$
 and ${\mathcal A}_{j,h}$.
We clearly have
\begin{align}
\label{eq:ef}
e[\chi_{j,h} u_h^{(1)}](s,t) &= e_{j,h}[\chi_{j,h} u_h^{(1)}](s,t) + f_1(s,t) + f_2(s,t) + f_3(s,t)\;,\end{align}
and
\begin{align*}
f_1 &= (a^{-1}-a_{j,h}^{-1}) |(hD_s - \tilde{A}_{1}) \chi_{j,h} u_h^{(1)} |^2 
+ (a - a_{j,h}) |h D_t (\chi_{j,h} u_h^{(1)})|^2\;,\\
f_2 &= a_{j,h}^{-1} | (\tilde{A}_{1} - \tilde{A}_{1,j,h})  \chi_{j,h} u_h^{(1)}|^2\;,\\
f_3 &= 2 a_{j,h}^{-1} \Re \Big\{(\tilde{A}_{1} - \tilde{A}_{1,j,h})  \chi_{j,h} \overline{u_h^{(1)}}
 (hD_s - \tilde{A}_{1,j,h}) \chi_{j,h} u_h^{(1)} \Big\}\;.
\end{align*}
Notice that for $s \in s_{j,h} + [-h^{\rho}, h^{\rho}]\,$, we have, since $\kappa'(0) = 0\,$,
\begin{align}\label{eq:kappa}
|\kappa(s) - \kappa_{j,h}| = |s - s_{j,h}| \cdot \Big| \int_0^1 \kappa'((1-\ell) s_{j,h} + \ell s)\,d\ell \Big| 
\leq C h^{\rho} \big(| s_{j,h}| + h^{\rho}\big)\; .
\end{align}
Thus,
\begin{align}\label{eq:comp_a}
&|a - a_{j,h}| = t |\kappa(s) - \kappa_{j,h}| \leq C h^{\rho} \big(| s_{j,h}| + h^{\rho}\big) t\;, \nonumber\\
&|a^{-1} - a_{j,h}^{-1} | \leq C h^{\rho} \big(| s_{j,h}| + h^{\rho}\big) t\;, \quad \text{ for } \quad t<2h^{\rho}\;.
\end{align}
We estimate, using \eqref{eq:comp_a}, for any $\epsilon > 0\,$,
\begin{align}\label{eq:f1}
|f_1&(s,t)| \leq C h^{\rho} \big(| s_{j,h}| + h^{\rho}\big) t \,e[\chi_{j,h}u_h^{(1)}](s,t)\nonumber \\&\leq
C' \epsilon h^{1/4+ 2\rho} \big(| s_{j,h}| + h^{\rho}\big)^2 \, e[\chi_{j,h}u_h^{(1)}](s,t) +
C' \epsilon^{-1} h^{-1/4} t^2 \, e[\chi_{j,h}u_h^{(1)}](s,t) \nonumber\\
&\leq C'' \epsilon h^{1/4+ 2\rho} \big(| s_{j,h}| + h^{\rho}\big)^2 \, e_{j,h} [\chi_{j,h}u_h^{(1)}](s,t) +
C' \epsilon^{-1}  h^{-1/4} t^2 \, e[\chi_{j,h}u_h^{(1)}](s,t)\;. 
\end{align}
We also  estimate $f_2$ and $f_3$ by
\begin{align}\label{eq:f2}
f_2(s,t) &\leq C h^{2\rho} t^4 |\chi_{j,h} u_h^{(1)}|^2\;, 
\end{align}
and
\begin{align}\label{eq:f3}
|f_3(s,t)| &\leq 2 C t^2 h^{\rho} \big(| s_{j,h}| + h^{\rho}\big) a_{j,h}^{-1} \Big|\chi_{j,h} \overline{u_h^{(1)}}
 (hD_s - \tilde{A}_{1,j,h}) \chi_{j,h} u_h^{(1)} \Big| \nonumber\\
 &\leq
 C' \epsilon^{-1} h^{-1/4} t^4 |\chi_{j,h} u_h^{(1)}|^2+
 C' \epsilon h^{1/4+2\rho} \big(| s_{j,h}| + h^{\rho}\big)^2 e_{j,h} [\chi_{j,h}u_h^{(1)}](s,t)\;. 
\end{align}
Thus, we get by combining \eqref{eq:ef} with \eqref{eq:f1}, \eqref{eq:f2}, and \eqref{eq:f3} and integrating,
\begin{align}\label{eq:aB}
{\mathcal B}_{j,h} &\geq \{1 - C \epsilon h^{1/4 + 2\rho}  \big(| s_{j,h}| + h^{\rho}\big)^2 \}
{\mathcal A}_{j,h} \nonumber\\
&\quad- C \epsilon^{-1} h^{-1/4} \int t^2 e[\chi_{j,h} u_h^{(1)}](s,t)\,ds\,dt \nonumber\\
&\quad- C ( h^{2\rho} +  \epsilon^{-1} h^{-1/4}) \int t^4 |\chi_{j,h} u_h^{(1)}|\,ds\,dt\;.
\end{align}
From Theorem~\ref{lem:BPT} we get the estimate
\begin{align}\label{eq:BPT}
{\mathcal A}_{j,h} &\geq \Big(\Theta_0 h - C_1 \kappa_{j,h} h^{3/2} - C h^2\Big) \| \chi_{j,h} u_h^{(1)}\|^2 \nonumber \\
& \geq \Big( \Theta_0 h - C_1 \big(k_{{\rm max}}- c_0 \min(|s_{j,h}|^2,1)\big) h^{3/2} - C h^2\Big) \| \chi_{j,h} u_h^{(1)}\|^2\;,
\end{align}
for some $c_0>0\,$, using the non-degeneracy of the maximum. Therefore, using that $|s_{j,h}| \leq |\partial \Omega|/2\,$, we get that, for $\epsilon$ sufficiently small and $\rho = 1/8\,$,
\begin{align}\label{eq:pos2ndDer}
\big\{1 - C \epsilon h^{1/4 + 2\rho}  \big(| s_{j,h}| + h^{\rho}\big)^2 \big\}
{\mathcal A}_{j,h} \geq 
\big(\Theta_0 h - C_1 k_{{\rm max}} h^{3/2} - C h^{7/4}\big) \| \chi_{j,h} u_h^{(1)}\|^2.
\end{align}
Therefore, we get from \eqref{eq:aB}, for the given choice of $\epsilon$ and $\rho = 1/8\,$,
\begin{align}\label{eq:AlmostThere}
\sum_j {\mathcal B}_{j,h} &
\geq \Big(\Theta_0 h - C_1 k_{{\rm max}} h^{3/2} - C h^{7/4}\Big) \| u_h^{(1)}\|^2 \\
&\quad - C h^{-1/4} \int \sum_j  t^2 e[\chi_{j,h} u_h^{(1)}](s,t)\,ds\,dt 
- C h^{-1/4} \int t^4 | u_h^{(1)}|^2\,ds\,dt\;.\nonumber
\end{align}
The weak normal Agmon estimates and easy manipulations (as in \cite{HeMo3})  give that the last two terms in \eqref{eq:AlmostThere} are bounded by $C h^{7/4}$. Therefore, the theorem follows from \eqref{eq:AlmostThere} and \eqref{eq:IMS}, remembering that $\rho =1/8\,$.
\end{proof}

\begin{remark}
\label{rem:Degenerate}
The above proof actually extends to the case where $k_{{\rm max}}$ is a non-degenerate maximum of higher order, i.e.
$$
\kappa(s) = k_{{\rm max}} + a s^{2N} + {\mathcal O}(|s|^{2N+1})\;,
$$
with $N\geq 2$ and $a \neq 0$. In that case \eqref{eq:kappa} becomes
\begin{align}\label{eq:kappaDeg}
|\kappa(s) - \kappa_{j,h}| 
\leq C h^{\rho} \big(| s_{j,h}|^{2N-1} + h^{\rho}\big)\;.
\end{align}
This implies that \eqref{eq:aB} becomes
\begin{align}\label{eq:aBDeg}
{\mathcal B}_{j,h} &\geq \big\{1 - C \epsilon h^{1/4 + 2\rho}  \big(| s_{j,h}|^{2N-1} + h^{\rho}\big)^2 \big\}
{\mathcal A}_{j,h} \nonumber\\
&\quad- C \epsilon^{-1} h^{-1/4} \int t^2 e[\chi_{j,h} u_h^{(1)}](s,t)\,ds\,dt \nonumber\\
&\quad- C ( h^{2\rho} +  \epsilon^{-1} h^{-1/4}) \int t^4 |\chi_{j,h} u_h^{(1)}|^2\,ds\,dt\;.
\end{align}
But instead of \eqref{eq:BPT}, we  get
\begin{align}\label{eq:BPTDeg}
{\mathcal A}_{j,h} 
& \geq \big\{ \Theta_0 h - C_1 \big(k_{{\rm max}}- c_0 \max(|s_{j,h}|^{2N},1)\big) h^{3/2} - C h^2\big\} \| \chi_{j,h} u_h^{(1)}\|^2\;,
\end{align}
Since $|s|^{2N} \geq |s|^{2(2N-1)}$ for small $s$, this implies the result. Actually, in this case one should be able to optimize the proof above (in particular choose $\rho < 1/8$) and get a better error bound than $\mathcal O(h^{7/4})$ in \eqref{eq:GoodLower}.
\end{remark}

It is convenient to have a lower bound of the operator ${\mathcal H}$ in terms of a potential $U_h$. That is our next statement.

\begin{theorem}\label{thm:PotBelow}
There exist $\epsilon_0\,,\, C_0 >0$ such that, if
\begin{align}\label{eq:DefWtilde}
\tilde{\kappa}(s) :=  k_{{\rm max}} - \epsilon_0 s^2\;,
\end{align}
and
\begin{align}
\label{eq:defW}
U_h(x) = \begin{cases} h\;, & \text{ if } \quad t(x) > 2h^{1/8}\;, \\ \Theta_0h - C_1 \tilde{\kappa}(s) h^{3/2}- C_0 h^{7/4}\;, & \text{ if } \quad t(x) \leq 2h^{1/8}\;,
\end{cases}
\end{align}
then
\begin{align}
\langle u \, \big| \,{\mathcal H} u \rangle \geq \int_{\Omega} U_h(x) |u(x)|^2\,dx\;,
\end{align}
for all $u \in D(\mathcal H)$ 
and all $h\in (0,1]\,$.
\end{theorem}

\begin{remark} It is very likely  that one could replace $\tilde \kappa$ by $\kappa$ by $\kappa$ in \eqref{eq:defW} (see also \cite[Proposition~10.2]{HeMo2}). However, we do not need this improvement.
\end{remark}

\begin{proof}
With $\theta_{1,h}\,,\, \theta_{2,h}$ as in \eqref{defthetajh} and $\rho=1/8$, we have
\begin{align*}
\langle u \, \big | \, {\mathcal H} u \rangle &= 
\langle \theta_{1,h} u \, \big | \, {\mathcal H}  \theta_{1,h} u \rangle +
\langle  \theta_{2,h}u \, \big | \, {\mathcal H}  \theta_{2,h} u \rangle \\
& \quad
- C h^{7/4} \int_{\{h^{1/8} \leq t(x) \leq 2 h^{1/8}\}} | u |^2 \, dx\;.
\end{align*}
Since $\langle  \theta_{2,h}  u \, |\,  {\mathcal H}  \theta_{2,h}  u \rangle \geq h \| \theta_{2,h}  u\|^2$, it therefore suffices to prove that
\begin{align}
\langle u \, | \,{\mathcal H} u \rangle \geq \int_{\Omega} \widetilde{U}_h(x) |u(x)|^2\,dx\;,
\end{align}
for all $u \in H^1(\Omega)$ and all $h\in (0,1]$, where
\begin{align}
\widetilde{U}_h(x) = \begin{cases} \gamma h\;, & \text{ if } \quad t(x) > 2h^{1/8}\;, \\ \Theta_0h - C_1 \tilde{\kappa}(s) h^{3/2}- C_0' h^{7/4}\;, & \text{ if } \quad t(x) \leq 2h^{1/8}\;,
\end{cases}
\end{align}
and $\gamma = (1+ \Theta_0)/2$ and $C_0'$ is some positive constant.

Let $\tilde{u}_h^{(1)}$ be a ground state for ${\mathcal H} - \widetilde{U}_h$ with ground state energy $\tilde{\mu}^{(1)} (h)$. 
We will prove that $\tilde{\mu}^{(1)} (h)\geq 0\,$. 

Since $\Theta_0 < \gamma < 1$, the normal Agmon estimates, Theorem~\ref{prop:Agmon-t}, are also valid for $\tilde{u}_h^{(1)}\,$. 

Using the `IMS'-formula, and notations as in the proof of Theorem~\ref{thm:HMimproved}, we get
\begin{align*}
\tilde{\mu}^{(1)} (h) &\geq
\sum_{j} \langle \chi_{j,h} \tilde{u}_h^{(1)}  \, | \,\big( {\mathcal H} - \widetilde{U}_h\big) \chi_{j,h} \tilde{u}_h^{(1)} \rangle
+ \int (h-\widetilde{U}_h) |\theta_{2,h} \tilde{u}_h^{(1)}|^2\,dx \\
&\quad\quad -
C\,h^{7/4} \int_{\{h^{1/8} \leq t(x) \leq 2h^{1/8}\}} |\tilde{u}_h^{(1)}|^2\,dx\;.
\end{align*}
Modulo choosing $C_0'$ sufficiently big, it therefore suffices to prove that
\begin{align}
\label{eq:suffices}
 \sum_{j} \Big\langle \chi_{j,h} \tilde{u}_h^{(1)}  \, | \, \big( {\mathcal H} - \Theta_0h + 
C_1 \tilde{\kappa}(s) h^{3/2} \big)&\chi_{j,h} \tilde{u}_h^{(1)} \Big \rangle \nonumber\\
&\geq - C h^{7/4} \int \sum_j |\chi_{j,h} \tilde{u}_h^{(1)}|^2\,dx\;.
\end{align}
Since the normal Agmon estimates hold for $\tilde{u}_h^{(1)}$, we can now go through the proof of Theorem~\ref{thm:HMimproved} with $u_h^{(1)}$ replaced everywhere by $\tilde{u}_h^{(1)}$. We replace  $u_h^{(1)}$ everywhere by $\tilde{u}_h^{(1)}$, in particular in the definition of ${\mathcal A}_{j,h}$ and ${\mathcal B}_{j,h}$, which are then denoted by  $\tilde {\mathcal A}_{j,h}$ and $\tilde {\mathcal B}_{j,h}$\,. In particular, we get
 as in \eqref{eq:BPT}
\begin{align}\label{eq:BPTCopy}
\tilde {\mathcal A}_{j,h} 
& \geq \big\{ \Theta_0 h - C_1 \big(k_{{\rm max}}- c_0 \min(|s_{j,h}|^2,1)\big) h^{3/2} - C h^2\big\} \| \chi_{j,h} \tilde{u}_h^{(1)}\|^2\;.
\end{align}
So
\begin{align}\label{eq:pos2ndDerCopy}
\big\{1 - C \epsilon h^{1/4 + 2\rho}  \big(| s_{j,h}| + h^{\rho}\big)^2 \big\}
\tilde {\mathcal A}_{j,h} &
- \big(\Theta_0h - C_1 \tilde{\kappa}(s) h^{3/2}\big) 
\| \chi_{j,h} \tilde{u}_h^{(1)}\|^2 \nonumber\\
&\geq 
- C h^{7/4} \| \chi_{j,h} \tilde{u}_h^{(1)}\|^2\;.
\end{align}
Therefore,
\begin{align*}
\sum_j & \Big\langle \chi_{j,h} \tilde{u}_h^{(1)}  \, | \, 
\big( {\mathcal H} - \Theta_0h + 
C_1 \tilde{\kappa}(s) h^{3/2} \big) \chi_{j,h} \tilde{u}_h^{(1)} \Big \rangle \\
&=\sum_j \Big( \tilde {\mathcal B}_{j,h} - \big(\Theta_0h - C_1 \tilde{\kappa}(s) h^{3/2}\big) 
\| \chi_{j,h} \tilde u_h^{(1)}\|^2 \Big)\\
&\geq
- C h^{7/4} \sum_j  \| \chi_{j,h} \tilde{u}_h^{(1)}\|^2 \\
&\quad - C h^{-1/4} \int \sum_j  t^2 \, e[\chi_{j,h} \tilde{u}_h^{(1)}](s,t)\,ds\,dt 
- C h^{-1/4} \int t^4 \,| \tilde{u}_h^{(1)}|^2\,ds\,dt\;.
\end{align*}
Using the weak normal Agmon estimates to bound the last terms by ${\mathcal O}(h^{7/4})$, this implies \eqref{eq:suffices} and therefore finishes the proof of Theorem~\ref{thm:PotBelow}.
\end{proof}

\subsection{Agmon estimates in the tangential direction}~\\
Theorem~\ref{thm:PotBelow} can be used to obtain exponential localization estimates in the tangential ($s$-)variable.

\begin{theorem}[Tangential Agmon estimates]\label{thm:Agmons}~\\
Let $h_0 >0$, $M >0$. Then there exist $C$, $\alpha>0$  and $h_1\in  (0,h_0]$, such that if
$(u_h)_{h\in]0,h_0]}$ is a family of  normalized eigenfunctions of ${\mathcal H}$ with corresponding eigenvalue $\mu(h)$ satisfying the bound
\begin{align}\label{eq:EVbound}
\mu(h) \leq \Theta_0 h - C_1 k_{{\rm max}} h^{3/2} + M h^{7/4}\;\;,\; \forall h\in (0,h_0]\;,
\end{align} 
and if $\chi_1 \in C_0^{\infty}$ is the function from \eqref{eq:chi12}, then, for all $ h\in (0,h_1]$, 
\begin{align}
\label{eq:Agmons}
\int_{\Omega} e^{2\alpha |s(x)|^2/h^{1/4}} \chi_1^2(t(x)/h^{1/8}) \Big\{
|u_h(x)|^2 + h^{-1}\big|(-ih\nabla - A(x)) u_h(x)\big|^2 \Big\}\,dx \leq C\;.
\end{align}
\end{theorem}

\begin{proof} First we observe that there exists $\beta >0$ such that, 
for all $S>0$, we have (with $\chi_2$ from \eqref{eq:chi12})
\begin{align}
\label{eq:AgmonStart}
\mu(h) & \Big\| \chi_2(\tfrac{s}{Sh^{1/8}})  \chi_1(\tfrac{t}{h^{1/8}}) e^{\alpha |s|^2/h^{1/4}} u_h \Big\|^2 \nonumber\\
&=
\Big \langle \chi_2^2(\tfrac{s}{Sh^{1/8}}) \chi_1^2(\tfrac{t}{h^{1/8}}) e^{2\alpha |s|^2/h^{1/4}} u_h \, | \, {\mathcal H} u_h \Big \rangle \nonumber \\
&\geq 
\Big \langle \chi_2(\tfrac{s}{Sh^{1/8}}) \chi_1(\tfrac{t}{h^{1/8}}) e^{\alpha |s|^2/h^{1/4}} u_h \, | \, {\mathcal H} \chi_2(\tfrac{s}{Sh^{1/8}}) \chi_1(\tfrac{t}{h^{1/8}}) e^{\alpha |s|^2/h^{1/4}} u_h \Big \rangle  \nonumber\\
&\quad- C h^2 \int  \Big| \nabla (\chi_2(\tfrac{s}{Sh^{1/8}}) \chi_1(\tfrac{t}{h^{1/8}}) ) \Big|^2 e^{2\alpha |s|^2/h^{1/4}} |u_h|^2 \,dx \nonumber\\
&\quad -\alpha^2 \beta h^{3/2} \int \chi_2^2(\tfrac{s}{Sh^{1/8}}) \chi_1^2(\tfrac{t}{h^{1/8}})
s^2 e^{2\alpha |s|^2/h^{1/4}} |u_h|^2 \,dx\;.
\end{align}
But it follows from Theorem~\ref{thm:PotBelow} and \eqref{eq:EVbound} that if $\alpha$ is chosen such that $\beta \alpha^2 \leq \epsilon_0 C_1/2\,$, then
\begin{align}\label{eq:compEnergies}
\Big \langle \chi_2(\tfrac{s}{Sh^{1/8}}) \chi_1(\tfrac{t}{h^{1/8}}) e^{\alpha |s|^2/h^{1/4}} u_h \, \Big | \, \big({\mathcal H} - \mu(h) &-\beta \alpha^2 s^2 h^{3/2}\big) \nonumber\\
& \times
\chi_2(\tfrac{s}{Sh^{1/8}}) \chi_1(\tfrac{t}{h^{1/8}}) e^{\alpha |s|^2/h^{1/4}} u_h \Big \rangle\nonumber\\
\geq
(\frac{\epsilon_0 C_1 S^2}{2} - C_0-M) h^{7/4} \big\| \chi_2(\tfrac{s}{Sh^{1/8}})  &\chi_1(\tfrac{t}{h^{1/8}}) e^{\alpha |s|^2/h^{1/4}} u_h \big\|^2\; .
\end{align}
Therefore, it follows from \eqref{eq:AgmonStart} and \eqref{eq:compEnergies} that, for $\alpha$ sufficiently small and $S$ sufficiently big,
\begin{multline}\label{eq:SplitAgmon}
\big\| \chi_2(\tfrac{s}{Sh^{1/8}})  \chi_1(\tfrac{t}{h^{1/8}}) e^{\alpha |s|^2/h^{1/4}} u_h \big\|^2 \\
\leq
C h^{1/4} \int  \big| \nabla (\chi_2(\tfrac{s}{Sh^{1/8}}) \chi_1(\tfrac{t}{h^{1/8}}) )\big|^2 e^{2\alpha |s|^2/h^{1/4}} |u_h|^2 \,dx\;.
\end{multline}
Now,
\begin{align}
h^{1/4} \int  \big| &\nabla (\chi_2(\tfrac{s}{Sh^{1/8}}) \chi_1(\tfrac{t}{h^{1/8}})) \big|^2 e^{2\alpha |s|^2/h^{1/4}} |u_h|^2 \,dx
\leq I + II\;, \\
I &:= C \int  |\chi_2'(\tfrac{s}{Sh^{1/8}})|^2 \chi_1^2(\tfrac{t}{h^{1/8}})  e^{2\alpha |s|^2/h^{1/4}} |u_h|^2 \,dx \;, \nonumber\\
II &:= C \int_{\{h^{1/8} \leq t \leq 2h^{1/8}\}} e^{2\alpha |s|^2/h^{1/4}} |u_h|^2 \,dx\;.\nonumber
\end{align}
On $\{h^{1/8} \leq t \leq 2h^{1/8}\}$ we have $|s| \leq |\partial \Omega| t/h^{1/8}$, and clearly $|s| \leq |\partial \Omega|/2$, so we have
$$
II \leq C \int_{\{h^{1/8} \leq t \leq 2h^{1/8}\}} e^{\alpha |\partial \Omega|^2 t/h^{3/8}} |u_h|^2 \,dx\;.
$$
By the normal Agmon estimates  (Theorem~\ref{prop:Agmon-t}), this implies that
$$
II \leq C\;.
$$
To estimate $I$, we use that $|\chi_2'(\tfrac{s}{Sh^{1/8}})|^2 e^{2\alpha |s|^2h^{1/4}}$ is bounded uniformly in $h$ and get
$$
I \leq C \int |u_h|^2 \,dx = C\;.
$$
Since also $\chi_1^2(\tfrac{s}{Sh^{1/8}}) e^{2\alpha |s|^2/h^{1/4}}$ is bounded uniformly in $h$, \eqref{eq:SplitAgmon} implies that
\begin{align}
\label{eq:AgmonL2}
\big\| \chi_1(\tfrac{t}{h^{1/8}}) e^{\alpha |s|^2/h^{1/4}} u_h \big\|^2 \leq C\;.
\end{align}
The bound on 
$$
\int_{\Omega} e^{2\alpha |s|^2/h^{1/4}} \chi_1^2(t/h^{1/8}) \big|(-ih\nabla - A(x)) u_h(x)\big|^2\,dx
$$
in \eqref{eq:Agmons} now follows in the same way by inserting \eqref{eq:AgmonL2} in \eqref{eq:AgmonStart}.
\end{proof}

\begin{cor}[Weak tangential Agmon estimates]~\\
Let the assumptions be as in Theorem~\ref{thm:Agmons}. Let $\chi \in C_0({\mathbb R})$, $\supp \chi \subset (-t_0,t_0)$, with the constant $t_0$ from the definition of the boundary coordinates in Appendix~\ref{AppB}. Then, for all $k>0$, there exists  $C>0$ such that
$$
\int_{\Omega} |s(x)|^k \chi(t(x)) |u_h(x)|^2\,dx \leq C h^{k/8}\;.
$$
\end{cor}
The proof of the corollary is immediate.

\section{A phase space bound}
\label{phasespace}
For our analysis of the low lying eigenvalues of ${\mathcal H}$, we need, apart from the localizations in $s$ and $t$, to have a precise localization in $D_s$. This is the goal of the present section.
\subsection{Main statement and main step of the proof}
\begin{theorem}[Localization in $D_s$]
\label{prop:LocXi}~\\
Let $M >0$, $h_0 >0$ and  $\chi_1, \chi_2 \in C^{\infty}({\mathbb R})$ be a standard partition of unity as in \eqref{eq:chi12}. 
Let $(s,t)$ be the boundary coordinates introduced in Appendix~\ref{AppB} chosen such that $\kappa(0) = k_{\rm max}$ and let $\epsilon$ in $(0,3/8)$.
Then for all $N>0$ exists $C_N>0$ such that if
$(u_h)_{h\in (0,h_0]}$ is a family of normalized eigenfunctions of ${\mathcal H} = {\mathcal H}(h)$ with eigenvalue $\mu(h)$ satisfying
\begin{align}\label{apub}
\mu(h) \leq \Theta_0 h - C_1 k_{{\rm max}} h^{3/2} + M h^{7/4}\;,
\end{align}
and the operator $W_s$ acting on functions localized near the boundary is defined by
\begin{align}
W_s \chi_1(t/t_0) = \chi_1(4s/|\partial \Omega|) 
\chi_2\big(\frac{|h^{1/2}D_s -\xi_0|}{h^{\epsilon}}\big) \chi_1(4s/|\partial \Omega|) \chi_1(4t/t_0)
\end{align}
with $t_0$ from \eqref{eq:t0},
then
\begin{align}\label{eq:AgmonDs}
&\big\| W_s \chi_1(4t/t_0) u_h \big\|_{L^2} \leq C_N h^N\;. 
\end{align}
and
\begin{align}\label{eq:AgmonDsEnergy}
\big| \big\langle W_s \chi_1(4t/t_0) u_h\;\big |\; {\mathcal H}(h) W_s \chi_1(4t/t_0) u_h 
\big \rangle \big |
\leq C_N h^N\;.
\end{align}
\end{theorem}

Let us be more explicit about  the meaning of the operator $W_s \chi_1(t/t_0)\,$. On the support of $t\mapsto \chi_1(4t/t_0)$,  we can use boundary coordinates $(s,t)$ (see Appendix~\ref{AppB}). Thus, for each $\phi \in L^2(\Omega)$, $f(s,t):=\chi_1(4t/t_0)\phi$ is a $|\partial \Omega|$-periodic function in $s$. 
After multiplication by $\chi_1(4s/|\partial \Omega|)$ we find a function with support in 
$(-\frac{|\partial \Omega|}{2}, \frac{|\partial \Omega|}{2}) \times \overline{{\mathbb R}_{+}}$ which we extend by zero to a function (with compact support) on ${\mathbb R} \times \overline{{\mathbb R}_{+}}$. This function we still denote by $\chi_1(4s/|\partial \Omega|) \chi_1(t/t_0) \phi$. On ${\mathbb R} \times \overline{{\mathbb R}_{+}}$ the meaning of the operator $\chi_2(\frac{|h^{1/2}D_s -\xi_0|}{h^{\epsilon}})$ is obvious (for example using the Fourier transformation). After multiplying a second time by $\chi_1(4s/|\partial \Omega|)$ we get a new function $\chi_1(4s/|\partial \Omega|) \chi_2(\frac{|h^{1/2}D_s -\xi_0|}{h^{\epsilon}}) \chi_1(4s/|\partial \Omega|) \chi_1(t/t_0) \phi$ with support in $(-\frac{|\partial \Omega|}{2}, \frac{|\partial \Omega|}{2}) \times \overline{{\mathbb R}_{+}}$ which we may reinterpret as a function on a neighborhood of the boundary of $\Omega$, expressed in boundary coordinates.

Thus, $W_s \chi_1(t/t_0)\,$ is an $h$-pseudo\-differential operator (or rather $h^{1/2-\epsilon}$-pseudo\-differential operator). We will use elementary commutation properties of such operators. The relevant results (and much more) can be found in introductions to the subject, such as \cite{DiSj} and \cite{Robert}.

\begin{remark}
As a shorter notation, instead of \eqref{eq:AgmonDs} and \eqref{eq:AgmonDsEnergy} we will write
\begin{align*}
\| W_s \chi_1(4t/t_0) u_h \big\|_{L^2} + 
\big| \big\langle W_s \chi_1(4t/t_0) u_h\;\big |\; {\mathcal H}(h) W_s \chi_1(4t/t_0) u_h 
\big \rangle \big |
= {\mathcal O}_{\rm unif}(h^{\infty})\;.
\end{align*}
Here the subscript `unif' is included to remind us that the constants (in \eqref{eq:AgmonDs} and \eqref{eq:AgmonDsEnergy}) are uniform for eigenfunctions in a suitable energy interval (as given in \eqref{apub}).
\end{remark}
~
\begin{proof}[Proof of Theorem~\ref{prop:LocXi}]~\\
Let $0<\delta<1/2$ and define $W$, $\chi_s$, $\chi_{s,0}$, $\chi_t$,  and $\chi_0$ by
\begin{align}
W &:= \chi_2\big(\frac{|h^{1/2}D_s -\xi_0|}{h^{\epsilon}}\big)\;, &
\chi_t &:= \chi_1(t/h^{1/2-\delta})\;,&&
\label{eq:definitioner} \\
\chi_0 &:= \chi_1(4t/t_0)\;, & \chi_{s,0}&:=\chi_1(4s/|\partial \Omega|)\;,& \chi_s &:= \chi_1(s/h^{1/8-\delta})\;.
\label{eq:definitionera}
\end{align}
We will choose $\delta$ small such that~:
\begin{align}
\label{eq:littledelta}
0<\delta < (\tfrac{3}{8} - \epsilon)/4\;.
\end{align}
By using the normal Agmon estimates (Theorem~\ref{prop:Agmon-t}),  it suffices to prove the following localized versions of \eqref{eq:AgmonDs} and \eqref{eq:AgmonDsEnergy}~:
\begin{align}
\label{eq:AgmonDsLocal}
&\| W_s\chi_t u_h \|_{L^2} = {\mathcal O}_{\rm unif}(h^{\infty})\;, \\
\label{eq:AgmonDsEnergyLocal}
&\langle W_s\chi_t u_h\;\big |\; {\mathcal H}(h) W_s \chi_t u_h  \rangle 
 = {\mathcal O}_{\rm unif}(h^{\infty})\;.
\end{align}

We start the proof of \eqref{eq:AgmonDsLocal} and \eqref{eq:AgmonDsEnergyLocal} by the easy identities
\begin{align}
\label{eq:AgmonDsStart}
\mu(h) \| W_s \chi_t u_h \|^2 &= \Re \langle \chi_t  W_s^* W_s  \chi_t u_h \, | \, {\mathcal H} u_h \rangle = T_1(u_h) + T_2(u_h)\;,  
\end{align}
with\footnote{Notice that $T_j$ depends also on a choice of a pair
 $(\chi_t\,,\,W_s)$ and that we will have to consider different
 pairs in the induction argument.}
\begin{align}
T_1(u_h)&:= \langle W_s  \chi_t u_h \, | \, {\mathcal H} W_s  \chi_t u_h \rangle\;, \label{defT1}\\
T_2(u_h) &:= \frac{1}{2} \big\langle u_h \,\big| \, \big( \chi_t W_s^* W_s \chi_t {\mathcal H} - 2 \chi_t W_s^{*} {\mathcal H} W_s \chi_t + {\mathcal H} \chi_t W_s^* W_s \chi_t
\big) u_h\big\rangle\;.  \label{defT2}
\end{align}

\noindent We will also use the following estimates
\begin{align}
\label{eq:obvious}
\| \chi_t u_h \|^2 &\leq 1\;, & \langle \chi_t u_h\;|\; {\mathcal H} \chi_t u_h \rangle \leq C h\;.
\end{align}
Only the second  estimate in \eqref{eq:obvious} deserves comment. 
It is however an easy consequence of the standard identity
\begin{align*}
\chi_t {\mathcal H} \chi_t = \tfrac{1}{2}( \chi_t^2 {\mathcal H} + {\mathcal H} \chi_t^2) + h^2 | \nabla \chi_t|^2\;,
\end{align*}
and of the estimate : 
\begin{equation}\label{Ctildeh}
\mu(h) \leq \tilde C h \;,
\end{equation}
resulting from Assumption \eqref{apub}.

\noindent{\bf Induction argument.}~\\
The proof of \eqref{eq:AgmonDsLocal} and \eqref{eq:AgmonDsEnergyLocal} will be obtained  by proving by induction 
 that $p(N)$ is satisfied for any $N \in {\mathbb N}$, where $p(N)$ is the following statement.\\

\noindent{\bf Statement  $p(N)$:}\\
For any $\chi_t$ and $W$  as in \eqref{eq:definitioner}, then
\begin{align}
\label{eq:AgmonDsLocalM}
&\| W_s\chi_t u_h \|_{L^2} = {\mathcal O}_{\rm unif}(h^{3N(\frac{3}{8} - \epsilon-\delta)})\;, \\
\label{eq:AgmonDsEnergyLocalM}
&\langle W_s\chi_t u_h\;\big |\; {\mathcal H}(h) W_s \chi_t u_h  \rangle 
 = {\mathcal O}_{\rm unif}(h^{3N(\frac{3}{8} - \epsilon-\delta)+1})\;.
\end{align}
{\bf Initialization $N=0$}.\\
The estimate \eqref{eq:AgmonDsLocalM} is trivially satisfied for $N=0$ and
\eqref{eq:AgmonDsEnergyLocalM} is a consequence of \eqref{defT1}, \eqref{eq:AgmonDsStart},
\eqref{Ctildeh}, \eqref{eq:littledelta} and Proposition~\ref{lem:ResultT2} (Proposition~\ref{lem:ResultT2} is somewhat stronger than needed at this step).\\
{\bf From $N$ to $N+1$.}~\\
Suppose now that we have proved $p(N)$ for some $N\geq 0$. Given $\chi_t$ and $W$, choose $\tilde{\chi}_t$ and $\widetilde{W}$ satisfying the same assumptions, but being slightly `larger', i.e. 
\begin{align*}
\tilde{\chi}_t \chi_t &= \chi_t\;, & \widetilde{W} W &= W\;.
\end{align*}
We introduce $\widetilde{W}_s := \chi_{s,0} \widetilde{W} \chi_{s,0}$.
Then we consider $\phi_h:=\tilde \chi_t \widetilde W_s u_h$ instead of $u_h$
 and assume $p(N)$, with the pair $(\widetilde W_s, \tilde \chi_t)$. We come back to \eqref{eq:AgmonDsStart} and observe, using the rough $h$-pseudodifferential calculus,  that $T_j(u_h)=T_j(\phi_h) + {\mathcal O}_{\rm unif}(h^{\infty})$ ($j=1,2)$.
We notice also that Proposition~\ref{lem:AgmonAfterLoc} implies that one can take $\varphi_h=\phi_h$ in Propositions~\ref{lem:ResultT1} and \ref{lem:ResultT2}. Therefore, \eqref{eq:AgmonDsStart} and Proposition~\ref{lem:ResultT2} applied with $\varphi_h=\phi_h$ together with $p(N)$ leads to \eqref{eq:AgmonDsEnergyLocalM}$_{N+1}$. Finally, 
Proposition~\ref{lem:ResultT1} and \eqref{eq:AgmonDsEnergyLocalM}$_{N+1}$
 give (using \eqref{apub}) \eqref{eq:AgmonDsLocalM}$_{N+1}$
and this finishes the induction.

Thus, \eqref{eq:AgmonDs} and \eqref{eq:AgmonDsEnergy} are proved and 
we have reduced the proof of Theorem~\ref{prop:LocXi} to the proof of the three Propositions~\ref{lem:ResultT1}, \ref{lem:AgmonAfterLoc}, and \ref{lem:ResultT2}, which  will be given  in the next subsections.
\end{proof}
\subsection{Step 2 : Lower bound for the local energy $T_1(\varphi_h)$. }
\begin{proposition}
\label{lem:ResultT1}
Let $\Xi \in C_0^{\infty}({\mathbb R})$, $\Xi \equiv 1$ on $[-\frac{1}{2}, \frac{1}{2}]$, $\Xi \equiv 0$ on ${\mathbb R} \setminus [-1,1]$.
Suppose that $\epsilon \in (0,3/8)$, that $\delta$ satisfies \eqref{eq:littledelta} 
and let $C >0\,$.
Then there exists $c_0 >0$ (depending also on the constants implicit  in ${\mathcal O}(h^{\infty})$ in \eqref{eq:AgmonsT1} and \eqref{eq:FourierFar} below) 
and for all $N \in {\mathbb N}$ exists $C_N>0$
such that if $\varphi_h \in D(\mathcal H)$ satisfies
\begin{align}
\label{eq:SuppAgmont}
&\supp \varphi_h \subset \{ t(x) \leq 2 h^{1/2 - \delta} \} \;, \\
& \int_{\Omega}  \big( |\varphi_h|^2 + h^{-1} |(-ih\nabla - A(x))\varphi_h|^2\big)\,dx \leq C\;,\\
\label{eq:AgmonsT1}
& \int_{\Omega} \chi_2(s/h^{1/8-\delta}) \big\{ |\varphi_h|^2 + h^{-1} |(-ih\nabla - A(x))\varphi_h|^2\big\}\,dx = {\mathcal O}(h^{\infty})\;, 
\end{align}
and
\begin{align}
\label{eq:FourierFar}
\Big\| \Xi(\frac{4s}{|\partial \Omega|}) \Xi(\frac{h^{1/2}D_s - \xi_0}{h^{\epsilon}})\Xi(\frac{4s}{|\partial \Omega|}) \varphi_h \Big\|_{L^2(\Omega)}= {\mathcal O}(h^{\infty})\;.
\end{align}
Then,
\begin{align}
\label{eq:ResT1}
T_1(\varphi_h) := \langle \varphi_h \, | \, {\mathcal H} \varphi_h \rangle
\geq \Big( \Theta_0 h - C_1 k_{\rm max} h^{3/2}  +  c_0 h^{1+2\epsilon}\Big) \| \varphi_h \|^2
-C_N h^N\;.
\end{align}
\end{proposition}
~\\
\begin{proof}[Proof of Proposition~\ref{lem:ResultT1}]
The term $T_1(\varphi_h)$ is an integral
\begin{align}
T_1(\varphi_h) &= \int  \Big(a^{-1} | (hD_s - \tilde{A}_1)\varphi_h |^2 + a |(hD_t) \varphi_h |^2\Big) \,ds\,dt \;.
\end{align}
We introduce a localized version of $T_1(\varphi_h)$,
\begin{align}
\tilde{T}_1(\varphi_h) &= \int \chi_s^2 \Big(a^{-1} | (hD_s - \tilde{A}_1)\varphi_h |^2 + a|(hD_t) \varphi_h |^2\Big) \,ds\,dt \;.
\end{align}
Using the localization estimates in $s$ (see \eqref{eq:AgmonsT1}), we obtain~:
\begin{align}
\label{eq:locT1}
T_1(\varphi_h) = \tilde{T}_1(\varphi_h) + {\mathcal O}(h^{\infty})\;.
\end{align}
On the set $$ \{|s| \leq h^{1/8-\delta}\}\cap \{ t \leq 2h^{1/2-\delta}\}\;,$$ we have
\begin{align}
a (s,t)& = 1 - t k_{\rm max} + {\mathcal O}(h^{3/4-3\delta})\;, &
a^{-1}(s,t) &= 1 + t k_{\rm max} + {\mathcal O}(h^{3/4-3\delta})\;, \nonumber\\
\tilde{A}_1(s,t) &= -t(1-t k_{\rm max}) + {\mathcal O}(h^{5/4-4\delta})\;.&&
\end{align}
Therefore, with 
\begin{align}
a_1(t)&:= 1 - t k_{\rm max}\;, &
a_2(t)&:= 1 + 2 t k_{\rm max}\;, &
A(t) &:= -t(1-t k_{\rm max})\;,
\end{align}
we get, using \eqref{eq:locT1}, 
\begin{align}
T_1(\varphi_h) &\geq (1-h^{3/4- 3\delta}) \tilde{Q}[ \varphi_h ] + {\mathcal O}(h^{7/4-5\delta})\| \varphi_h\|^2 + {\mathcal O}(h^{\infty})\;,
\end{align}
where
\begin{align}
\tilde{Q}[f] &:= \int \chi_s^2 \Big(  a_2(t) \big| (hD_s - A(t))f \big|^2 
+  \big|(hD_t) f \big|^2 \Big) a_1(t)\,ds\,dt \;.
\end{align}
It is clear, using again \eqref{eq:AgmonsT1}, that we can remove the localization $\chi_s$ and get
\begin{align}
\label{eq:IntroQ}
T_1(\varphi_h) &\geq (1-h^{3/4 - 3 \delta}) Q[ \Xi(\frac{4s}{|\partial \Omega|}) \varphi_h ] + {\mathcal O}(h^{7/4-5\delta})\| \Xi(\frac{4s}{|\partial \Omega|}) \varphi_h\|^2 + {\mathcal O}(h^{\infty})\;,
\end{align}
where
\begin{align}\label{eq:QuadForm}
Q[f] &:= \int_{{\mathbb R}^2_{+}} \Big(  a_2(t) \big| (hD_s - A(t))f \big|^2 
+  \big|(hD_t) f \big|^2 \Big)a_1(t) \,ds\,dt \;.
\end{align}
Now the coefficients in $Q$ do not depend on $s$, so we can make a Fourier decomposition of the quadratic form. Let us define
\begin{align}
\label{eq:FourierSeries}
\tilde{f}(s,\tau) &:= h^{1/4} \chi_1(h^{\delta}\tau) f(s,h^{1/2}\tau)\;, \nonumber \\
\ell &:= k_{\rm max} h^{\delta}\tau \chi_1(h^{\delta}\tau/2) \;, \nonumber \\
g_{\zeta}(\tau) &:= (2\pi)^{-1/2} \int_{{\mathbb R}}
e^{-i \zeta s} \tilde{f}(s,\tau)\,ds\;.
\end{align}
Notice that the function $\ell$  is uniformly bounded on $\mathbb R^+$. 
Then
\begin{align}
Q[ f ] &= h \int_{{\mathbb R}} q_{\zeta}[g_{\zeta}] \,d\zeta \;,\nonumber \\
q_{\zeta}[g] &:= \int_0^{\infty} \Big\{(1+ 2h^{1/2-\delta} \ell(\tau) ) \big[h^{1/2} \zeta +\tau(1 - h^{1/2-\delta} \ell(\tau) /2) \big] ^2 \, | g(\tau)|^2 \nonumber\\
&\quad\quad\quad\quad\quad\quad+
|g'(\tau)|^2\Big\} (1- h^{1/2-\delta} \ell(\tau) ) \,d\tau\;.
\end{align}
The form $q_{\zeta}$  is the quadratic form on 
$H^1\big({\mathbb R}_+, (1- h^{1/2-\delta} \ell ) \,d\tau\big)
 \cap L^2\big({\mathbb R}_+, \tau^2 (1- h^{1/2-\delta} \ell ) \,d\tau\big)$ defining a selfadjoint  unbounded operator ${\mathfrak h}(\zeta)$ on the space $L^2\big({\mathbb R}_+, (1- h^{1/2-\delta} \ell ) \,d\tau\big)$~:
$$
\mathfrak h (\zeta) = - \frac{1}{(1- h^{1/2-\delta} \ell ) } \frac{d}{d \tau} \{ (1- h^{1/2-\delta} \ell )  \} \frac{d}{d \tau}
 +   \big[h^{1/2} \zeta +\tau(1 - h^{1/2-\delta} \ell(\tau) /2) \big] ^2\;.
$$
Similarly, we can introduce
 the quadratic form on $H^1(\mathbb R^+)\cap L^2(\mathbb R^+, \tau^2 d\tau)$~:
$$ 
q_{\zeta}^0[g] := \int_0^{\infty} \Big\{ (h^{1/2} \zeta +\tau )^2 \, | g(\tau)|^2 +
|g'(\tau)|^2\Big\}  \,d\tau\;.
$$
with associated operator $\mathfrak h_0(\zeta)$  on $L^2({\mathbb R}_+,d\tau)$
which is the Neumann selfadjoint realization of~:
$$
{\mathfrak h}_0(\zeta):= - \frac{d^2}{d\tau^2} + (h^{1/2} \zeta + \tau)^2 \;.
$$
In the two cases, the form domain is the same space and the operator domain
 involves the Neumann condition at $\tau=0$.
\begin{lemma}
\label{lem:PertTheory} 
There exists $c_0\,, \;C\,,\; M>0\,$, such that if $|h^{1/2}\zeta - \xi_0| \geq M h^{1/4 - 3\delta/2}\,$, then
\begin{align}
\inf \Spec {\mathfrak h}(\zeta) \geq \Theta_0 + c_0 \min\big(1, |h^{1/2} \zeta - \xi_0|^2\big),
\end{align}
and if $|h^{1/2}\zeta - \xi_0| \leq M h^{1/4 - 3\delta/2}$, then
\begin{align}
\inf \Spec {\mathfrak h}(\zeta) &\geq \Big\{\Theta_0 +3 C_1 |\xi_0| (h^{1/2}\zeta -\xi_0)^2 -C_1 k_{\rm max} h^{1/2}\Big\}  \nonumber\\
&\quad\quad - C \big( |h^{1/2}\zeta -\xi_0|^3 + h^{1/2} |h^{1/2}\zeta -\xi_0 | \big)\;.
\end{align}
\end{lemma}
~\\
\begin{proof}[Proof of Lemma~\ref{lem:PertTheory}]
The proof is similar to a calculation given in \cite[Section~11]{HeMo3}, so we will be rather brief.
Since $0\leq \ell \leq C$, $0\leq \ell \tau \leq h^{-\delta}$ and $\delta \in (0, 1/2)$, we get for all $f$ in the form domain of ${\mathfrak h}(\zeta)$,
\begin{align}
q_\zeta [f] \leq (1+ C h^{1/2 - \delta}) q_\zeta^0 [f] +  h^{1/2 -3\delta} \| f \|^2\;,
\end{align}
and the same inequality is true by exchanging $q_\zeta$ and $q_\zeta^0$.
Thus, by the variational characterization of the eigenvalues,
\begin{align}
\label{eq:RayleighRitz}
\big| \mu_j( {\mathfrak h}(\zeta)) -  \mu_j( {\mathfrak h}_0(\zeta))\big| \leq C h^{1/2 - 3 \delta} \big\{ 1 +  \mu_j( {\mathfrak h}_0(\zeta))\big\}\;.
\end{align}
Here $\mu_j(\mathfrak h)$ denotes the $j$th eigenvalue of the self-adjoint operator $\mathfrak h$ (with $\mathfrak h = \mathfrak h(\zeta)$ or $\mathfrak h_0(\zeta)$) .
Now it follows from \eqref{lw21} that, for some $c_0 >0\,$, 
$$
\mu_1({\mathfrak h}_0(\zeta)) \geq \Theta_0 + c_0 \min\big(1, |h^{1/2} \zeta - \xi_0|^2\big)\;,
$$
and therefore we get from \eqref{eq:RayleighRitz}, if $M$ is chosen sufficiently big, that
\begin{align}
\mu_1({\mathfrak h}(\zeta)) \geq \Theta_0 
+ \frac{c_0}{2} \min\big(1, |h^{1/2} \zeta - \xi_0|^2\big)\quad \text{ for } |h^{1/2} \zeta - \xi_0| \geq M h^{1/4-3\delta/2}\;.
\end{align}
Note that from \eqref{eq:littledelta}, $1/4-3\delta/2\geq 0$.
For $|h^{1/2} \zeta - \xi_0| < M h^{1/4-3\delta/2}\,,$ we will construct an explicit trial function for ${\mathfrak h}(\zeta)$.
With $P_0^{-1}$ being the regularized resolvent from \eqref{eq:A14} we write
\begin{align}
f_{\zeta}(\tau) &= u_0(\tau) - 2(h^{1/2}\zeta - \xi_0) P_0^{-1}[ (t+\xi_0) u_0(t) ] (\tau) \\
&\quad + 4(h^{1/2}\zeta - \xi_0)^2 P_0^{-1}\Big\{ (t+\xi_0) P_0^{-1}\big[ (t'+\xi_0) u_0(t') \big](t) - I_2 u_0(t) \Big\} (\tau) \;.\nonumber
\end{align}
We note that $f_\zeta(\tau)$ belongs to the domain of $\mathfrak h(\zeta)$ and 
a  straightforward calculation gives that
\begin{align}
\Big\| \{{\mathfrak h}_0(\zeta) - [\Theta_0 + (h^{1/2}\zeta - \xi_0)^2 (1-4I_2)]\} f_{\zeta} \Big\| \leq C | h^{1/2}\zeta - \xi_0 |^3\;.
\end{align}
where $I_2$ is the constant from \eqref{eq:I2}, i.e.
\begin{align}
I_2 := \int_0^{\infty} (\tau+\xi_0) u_0(\tau) P_0^{-1}[ (t+\xi_0) u_0(t) ](\tau)\,d\tau\;.
\end{align}
Define now
\begin{align}
\tilde{f}_{\zeta} := f_{\zeta} - h^{1/2} k_{\rm max} P_0^{-1}\Big[ \big\{\frac{d}{d\tau} +2 \tau (\xi_0+\tau)^2 - \tau^2 (\xi_0+\tau)\big\} u_0 \Big].
\end{align}
Again a straightforward calculation, using the decomposition
$$
\begin{array}{ll}
{\mathfrak h}(\zeta)-{\mathfrak h}_0(\zeta) &:= h^{\frac 12 - \delta}\left( \frac{\ell'}{1-h^{1/2-\delta}\ell} \frac{d}{d\tau} 
+2\ell (h^{1/2} \zeta + \tau - h^{1/2-\delta}\ell \tau/2)^2\right.\\
&\quad\quad \left.- (h^{1/2} \zeta + \tau)\ell \tau
+ \tau^2 h^{1/2-\delta}\frac{\ell^2}{4} \right)\;,
\end{array}
$$
 gives that
\begin{align}
\label{eq:Straight}
&\Big\| \{{\mathfrak h}(\zeta) - [\Theta_0 + (h^{1/2}\zeta - \xi_0)^2 (1-4I_2) + h^{1/2} k_{\rm max} N ]\} \tilde{f}_{\zeta} \Big\| \nonumber\\
&\quad \leq C\big( | h^{1/2}\zeta - \xi_0 |^3 + h^{1/2} | h^{1/2}\zeta - \xi_0 | \big)\;,
\end{align}
with
\begin{align}
N:= \int_0^{\infty} u_0(\tau) \Big\{\frac{d}{d\tau} +2 \tau (\xi_0+\tau)^2 - \tau^2 (\xi_0+\tau)\Big\} u_0(\tau) \,d\tau\;.
\end{align}
Using Propositions~\ref{prop:Iij} and \ref{prop:secondPerturb}, we get
\begin{align}
N &= -C_1\;,&
1-4I_2 &=-3C_1 \xi_0= 3C_1 |\xi_0|\;.
\end{align}
This, together with \eqref{eq:RayleighRitz} which permits to have
 a lower bound of $\mu_2(h(\zeta))$, finishes the proof of Lemma~\ref{lem:PertTheory} (see \cite{HeMo3} for a similar argument). We have actually obtained the
 better
\begin{align}\mu_1(\mathfrak h(\zeta))  & \sim \Big\{\Theta_0 +3 C_1 |\xi_0| (h^{1/2}\zeta -\xi_0)^2 -C_1 k_{\rm max} h^{1/2}\Big\}  \nonumber\\
&\quad\quad - C \big( |h^{1/2}\zeta -\xi_0|^3 + h^{1/2} |h^{1/2}\zeta -\xi_0 | \big)\;.
\end{align}
\end{proof}

Lemma~\ref{lem:PertTheory} has the following consequence 
\begin{lemma}\label{cor:Summarize}
Let $\varphi_h, \epsilon$ and $\delta$ satisfying the assumptions
 of  Proposition~\ref{lem:ResultT1}.
Then there exists $c_0>0$ such that
\begin{align*}
Q[ \Xi(\frac{4s}{|\partial \Omega|}) \varphi_h] \geq \big( \Theta_0 h &- C_1 k_{\rm max} h^{3/2} + c_0 h^{1+2\epsilon}\big)
\int | \Xi(\frac{4s}{|\partial \Omega|}) \varphi_h |^2 \, (1-tk_{\rm max})\,ds\,dt \\
&+ {\mathcal O}_{\rm unif}(h^{\infty})\; .
\end{align*}

\end{lemma}
The proof of Lemma~\ref{cor:Summarize} is immediate.

\noindent{\bf End of the proof of Proposition~\ref{lem:ResultT1}.}~\\
Using \eqref{eq:SuppAgmont} and \eqref{eq:AgmonsT1}, we get that
$$
\int_{{\mathbb R}^2_{+}} | \Xi(\frac{4s}{|\partial \Omega|}) \varphi_h|^2 (1- tk_{\rm max}) \,ds\,dt = (1 + {\mathcal O}(h^{3/4-3\delta}) )
\| \varphi_h\|^2_{L^2(\Omega)}\;.
$$
Therefore, Lemma~\ref{cor:Summarize} implies that
\begin{align}
\label{eq:TowardsT1}
Q[ \Xi(\frac{4s}{|\partial \Omega|})  \varphi_h ] \geq \big( \Theta_0 h - C_1 k_{\rm max} h^{3/2} + c_0 h^{1+2\epsilon} - C h^{7/4-3\delta}\big)
\| \varphi_h \|_{L^2(\Omega)}^2\;.
\end{align}
Combining \eqref{eq:TowardsT1} with \eqref{eq:IntroQ}, and using the choice of $\delta$ from \eqref{eq:littledelta}, yields \eqref{eq:ResT1}.
\end{proof}
\subsection{Step 3~: Preservation of localization.}
\begin{proposition}
\label{lem:AgmonAfterLoc}
Let $\epsilon \in (0,3/8)$ and let $\delta \in (0, \frac 12)$. Then there exist $\alpha, C>0$ such that if
$(u_h)_{h\in (0,h_0)}$ is the family of functions from Theorem~\ref{prop:LocXi} and $\chi_t, W$ are as in \eqref{eq:definitioner}, then $\phi_h := \chi_t W_s u_h$ satisfies 
\begin{align*}
&\int e^{\alpha t(x)/h^{1/2}}  \{ |\phi_h|^2 + h^{-1} |(-ih\nabla - A(x) \phi_h|^2\}\,dx \leq C\;, \\
&\int \chi_2^2(s/h^{1/8-\delta}) \big\{ |\phi_h|^2 + h^{-1} |(-ih\nabla - A(x))\phi_h|^2\big\}\,dx = {\mathcal O}_{\rm unif}(h^{\infty})\;.
\end{align*}
\end{proposition}
~\\
\begin{proof}[Proof of Proposition~\ref{lem:AgmonAfterLoc}]~\\
We only consider the localization in $s$, since the localization in $t$ is much simpler.
Let us define 
\begin{align*}
T:= \int \chi_2^2(s/h^{1/8-\delta})  |(-ih\nabla - A(x))\phi_h|^2\,dx \;.
\end{align*}
We will only prove that $T = {\mathcal O}_{\rm unif}(h^{\infty})$, the remaining estimate in Proposition~\ref{lem:AgmonAfterLoc} being easier.
We write, with $\overline{\chi}_s:= \chi_2(s/h^{1/8-\delta})$,
\begin{align}
T & = R_1 + R_2 \;,\\
R_1 &:= \int \overline{\chi}_s^2 a^{-1} | (hD_s - \tilde{A}_1)\phi_h |^2 \,ds\,dt \;, \quad\quad
R_2 := \int \overline{\chi}_s^2  a |(hD_t) \phi_h |^2 \,ds\,dt \;. \nonumber
\end{align}
Since $a$ is a bounded function on $\supp \chi_t$ and $W_s$ commutes with $D_t$, we have, with $\langle \cdot \;|\; \cdot \rangle$ being the inner product on $L^2( [-|\partial \Omega|/2, |\partial \Omega|/2] \times {\mathbb R}; ds\,dt)$,
\begin{align}
|R_2| \leq C \langle (hD_t) (\chi_t u_h) \,| \, W_s \overline{\chi}_s^2 W_s (hD_t) (\chi_t u_h)\rangle\;.
\end{align}
Now, with $\tilde{\chi}_j$ defined by 
$$ \tilde{\chi}_j (s)= \chi_j( 2s/h^{1/8-\delta})\;,
$$
\begin{align}
W_s \overline{\chi}_s^2 W_s &= (\tilde{\chi}_1^2 + \tilde{\chi}_2^2) W_s \overline{\chi}_s^2 W_s (\tilde{\chi}_1^2 + \tilde{\chi}_2^2)\;.
\end{align}
Since $\epsilon< 3/8\,$, 
we see by repeated commutations that 
$\tilde{\chi}_1^2 W_s \overline{\chi}_s^2 = {\mathcal O}_{\rm unif}(h^{\infty})$, so
\begin{align}
|R_2| &\leq C \big\langle \tilde{\chi}_2^2 (hD_t) (\chi_t u_h) \,| \, W_s \overline{\chi}_s^2 W_s \, \tilde{\chi}_2^2 (hD_t) (\chi_t u_h)\big\rangle \nonumber\\
&\quad+ {\mathcal O}_{\rm unif}(h^{\infty}) \| (hD_t) (\chi_t u_h) \|^2\;.
\end{align}
Now the Agmon estimates in $s$ and $t$ easily imply that $R_2 = {\mathcal O}_{\rm unif}(h^{\infty})$.

The estimate of $R_1$ is similar but slightly more complicated, since $W_s$ does not commute with $\tilde{A}_1$. We estimate
\begin{align}
| R_1| & \leq R_1^{(1)} + R_1^{(2)} \;,\\
R_1^{(1)} &:= C \int \overline{\chi}_s^2 | (hD_s +t )(\chi_t W_s u_h) |^2 \,ds\,dt \;, \nonumber\\
R_1^{(2)} &:= C \int \overline{\chi}_s^2 |  \tilde{A}_1- t|^2 |\chi_t W_s u_h |^2 \,ds\,dt \;, \nonumber
\end{align}
On $\supp \chi_t$, $|  \tilde{A}_1- t| \leq C$, so $R_1^{(2)}$ can be controlled like $R_2$ by ${\mathcal O}_{\rm unif}(h^\infty)$.\\

For $R_1^{(1)}$ we use that $W$ commutes with $(hD_s +t )$ and the Agmon estimates in $s$, to write
\begin{align}
R_1^{(1)} = C \langle (hD_s +t ) \chi_t  u_h\, | \,  W_s  \overline{\chi}_s^2 W_s (hD_s +t ) \chi_t  u_h \rangle
+ {\mathcal O}_{\rm unif}(h^{\infty})\;.
\end{align}
Notice that
\begin{align}
\| (hD_s +t ) \chi_t  u_h \|^2 \leq C \| (hD_s +\tilde{A}_1 ) \chi_t  u_h \|^2 + C \leq Const\;.
\end{align}
So we can, like for the control of $R_2$, localize modulo ${\mathcal O}_{\rm unif}(h^\infty)$
 on the support of  $\tilde{\chi}_2$ and  get
\begin{align}
|R_1^{(1)}| &\leq C \langle \tilde{\chi}_2^2 (hD_s +t ) \chi_t  u_h\, | \,  W_s  \overline{\chi}_s^2 W_s \tilde{\chi}_2^2 (hD_s +t ) \chi_t  u_h \rangle + {\mathcal O}_{\rm unif}(h^{\infty}) \nonumber\\
&\leq C' \| \tilde{\chi}_2^2 (hD_s +t ) \chi_t  u_h \|^2 + {\mathcal O}_{\rm unif}(h^{\infty}) \nonumber\\
&\leq C'' \big( \| \tilde{\chi}_2^2 (hD_s -\tilde{A}_1 ) \chi_t  u_h \|^2 +
\| \tilde{\chi}_2^2 \chi_t  u_h \|^2\Big)+ {\mathcal O}_{\rm unif}(h^{\infty}) \nonumber\\
&= {\mathcal O}_{\rm unif}(h^{\infty})\;.
\end{align}
Here we used the tangential Agmon estimates to get the last inequality. This finishes the proof of Proposition~\ref{lem:AgmonAfterLoc}.
\end{proof}
\subsection{Step 4 : Control of the commutator $T_2 (\varphi_h)$\,.}
\subsubsection{Main statement}
\begin{proposition}
\label{lem:ResultT2}
Suppose that $\epsilon \in (0,3/8)\,$, that $\delta$ satisfies \eqref{eq:littledelta} and let $\alpha\,,\; C >0\,$.
Then there exists $c_0 >0$ (depending also on the constants implicit in \eqref{eq:AgmonlikeT2s} below) and for all $N \in {\mathbb N}$ there exists $C_N>0$ such that if $\varphi_h \in D(\mathcal H)$ is such that
\begin{align}
\label{eq:Agmonliket}
& \int e^{\alpha t(x)/h^{1/2}}  \{ |\varphi_h|^2 + h^{-1} |(-ih\nabla - A(x))\varphi_h|^2\}\,dx \leq C\;, \\
\label{eq:AgmonlikeT2s}
& \int \chi_2(s/h^{1/8-\delta}) \{ |\varphi_h|^2 + h^{-1} |(-ih\nabla - A(x))\varphi_h|^2\}\,dx = {\mathcal O}(h^{\infty})\;, 
\end{align}
Then, with $W_s$ and $\chi_t$ from \eqref{eq:definitioner} and
\begin{align}
\label{eq:defT2}
T_2(\varphi_h) := \frac{1}{2} \langle \varphi_h \, | \,\big( \chi_t W_s^{*} W_s \chi_t {\mathcal H} - 2 \chi_t W_s^{*} {\mathcal H} W_s \chi_t + {\mathcal H} \chi_t W_s^{*} W_s \chi_t  \big) \varphi_h \rangle\;,
\end{align}
we have
\begin{align}
\label{eq:ResT2}
|T_2(\varphi_h)| \leq c_0 h^{9/8-\epsilon-\delta} \Big( \langle \varphi_h \, | \, {\mathcal H} \varphi_h \rangle + h \| \varphi_h \|^2\Big) + C_N h^N\;.
\end{align}
\end{proposition}
The proof of Proposition~\ref{lem:ResultT2} is based on successive decompositions
 of the `commutator'.
\subsubsection{First decomposition for $T_2(\varphi_h)$.}~\\
Since $\chi_t$ localizes near the boundary, we can use boundary coordinates $(s,t)$. Thus, we get, with $a=1-t\kappa(s)$,
\begin{align}
W_s^* &= a^{-1} W_s a = W_ s+ \widehat{W}_s\;; & \widehat{W}_s:=\chi_{s,0}\widehat{W}\chi_{s,0}\;,&
& \widehat{W}:=\frac{-t}{a} [W, \kappa(s)]\;.
\end{align}
Remember that $W_s = \chi_{s,0} W \chi_{s,0}$.
Let $\Xi \in C_0^{\infty}({\mathbb R})$ satisfy, $\Xi(s) \chi_{s,0}(s) = \chi_{s,0}(s)$, $\supp \Xi \subset (-\frac{|\partial \Omega|}{2}, \frac{|\partial \Omega|}{2})$. Clearly, $T_2(\varphi_h) = T_2(\Xi \varphi_h)$. Now we can calculate the `commutator' in $T_2(\Xi \varphi_h)$;
$$
\chi_t W_s^{*} W_s \chi_t {\mathcal H} - 2 \chi_t W_s^{*} {\mathcal H} W_s \chi_t + {\mathcal H} \chi_t W_s^{*} W_s \chi_t\;;
$$
as an (pseudodifferential) operator on $L^2({\mathbb R}^2_{+})$, where we extend the curvature function $\kappa(s)$ (appearing, for instance, in the expression for ${\mathcal H}$ in boundary coordinates) as a periodic function of $s \in {\mathbb R}$.

Using the localization estimates in $s$ from \eqref{eq:AgmonlikeT2s} 
combined with the fact that $\chi_t$ commutes with $W$ and $\widehat{W}\,$, we therefore get
\begin{align}
T_2(\varphi_h) &= \frac{1}{2} \langle \chi^2_{s,0} \varphi_h \, | \, ({\mathcal C}_1 + {\mathcal C}_2) \chi^2_{s,0} \varphi_h \rangle + {\mathcal O}_{\rm unif}(h^{\infty})\;,\label{t2phi}\\
{\mathcal C}_1 &:= [ \chi_t W, [ \chi_t W, {\mathcal H}]] \;, \quad\quad
{\mathcal C}_2 := {\mathcal C}_{2,1} + {\mathcal C}_{2,2} + {\mathcal C}_{2,3} \;,\label{decc1c2a}\\
{\mathcal C}_{2,1} &:=
\widehat{W}[ \chi_t, [ \chi_t, {\mathcal H}]] W \;,\quad 
{\mathcal C}_{2,2}:=\widehat{W} \chi_t^2 [W,  {\mathcal H}] \;, \quad
{\mathcal C}_{2,3} := [{\mathcal H}, \widehat{W}] W \chi_t^2 \;. \label{decc1c2b}
\end{align}
We will calculate and estimate these commutators. We write
\begin{align}
{\mathcal H}(h) &= a^{-1}(hD_s - \tilde{A}_1) a^{-1}(hD_s - \tilde{A}_1) + a^{-1}(hD_t) a (hD_t) \nonumber\\
&= {\mathcal H}_1 + {\mathcal H}_2 + {\mathcal H}_3 + {\mathcal H}_4\;,
\end{align}
with
\begin{align}\label{defHj}
 \mathcal H_1&:= (hD_s - \tilde{A}_1) a^{-2}(hD_s - \tilde{A}_1)\;, &
\mathcal H_2&:= -ih \frac{\partial_s a}{a^3}(hD_s - \tilde{A}_1)\;,\nonumber \\
\mathcal H_3&:=(hD_t)^2\;, &
\mathcal H_4&:=  - ih\frac{\partial_t a}{a} (hD_t)\;.
\end{align}

\begin{remark}
The commutator ${\mathcal C}_{2,3}$ gives the leading order term. This can be understood by a `back-of-the-envelope' calculation replacing ${\mathcal H}$ by the leading terms from \eqref{eq:PExp}, i.e.
${\mathcal H} \approx  P_0 + h^{3/8}P_1 + h^{1/2}P_2$. We do not give this formal calculation here, since we do not justify this approximation.
\end{remark}
\subsubsection{Control of  $\frac 12 \langle {\mathcal C}_1 \chi^2_{s,0} \varphi_h\,|\, \chi^2_{s,0} \varphi_h\rangle$.}
The terms with derivatives in $D_s$ are the most involved.

\noindent{\bf Commutation with $\mathcal H_1$.}~\\
Since $\chi_t$ commutes with $W$ and ${\mathcal H}_1$, we find
\begin{align}
[\chi_t W, [\chi_t W, {\mathcal H}_1]] = \chi_t^2[W,[W, {\mathcal H}_1]]\;.
\end{align}
The inner commutator becomes
\begin{equation}
[W, {\mathcal H}_1] =: Q_1 + Q_2\;,
\end{equation}
with
\begin{align}
Q_1&:= 
(hD_s - \tilde{A}_1)  [W, a^{-2}] (hD_s - \tilde{A}_1)\;,\\
Q_2&:= -\left\{ [W, \tilde{A}_1] a^{-2}(hD_s - \tilde{A}_1)+(hD_s - \tilde{A}_1)a^{-2}[W, \tilde{A}_1] \right\}\;.
\end{align}
We calculate the double commutators separately
\begin{align}
\label{eq:doubleComm}
[W, Q_1] & = (hD_s - \tilde{A}_1)  [W, [W, a^{-2}] ]
(hD_s - \tilde{A}_1) \nonumber\\
&\quad -\left\{ [W, \tilde{A}_1] [W, a^{-2}] (hD_s - \tilde{A}_1)
+ (hD_s - \tilde{A}_1)[W, a^{-2}] [W, \tilde{A}_1] \right\}, \nonumber\\
[W, Q_2] & = - [W, [W, \tilde{A}_1]] a^{-2}(hD_s - \tilde{A}_1)-
(hD_s - \tilde{A}_1)a^{-2}[W,[W, \tilde{A}_1]]\nonumber\\
&\quad-
[W, \tilde{A}_1] [W,a^{-2}](hD_s - \tilde{A}_1)-
(hD_s - \tilde{A}_1)[W,a^{-2}] [W, \tilde{A}_1] \nonumber\\
&\quad + 2[W, \tilde{A}_1] a^{-2} [W, \tilde{A}_1]\;.
\end{align}
Remember that $\tilde{A}_1(s,t) = -t(1-t\kappa(s)/2)$, $a(s,t)=1-t\kappa(s)$. Therefore, 
\begin{align}
\label{eq:commutators1}
&[W, \tilde{A}_1]  = t^2[W, \kappa(s)]/2 = t^2 h^{1/2-\epsilon} \mathcal O_1\,,&
&[W, [W, \tilde{A}_1]]  = t^2 h^{1-2\epsilon} \mathcal O_2\;, \\
\label{eq:commutators2}
&[W, a^{-2}]  = t h^{1/2-\epsilon} \mathcal O_3\;,& 
&[W,[W,a^{-2}]] = t  h^{1-2\epsilon} \mathcal O_4\;,
\end{align}
where, after a right multiplication by a cutoff function
localizing in $[0,t_0)$, the $\mathcal O_j$'s are
bounded (pseudodifferential) operators commuting with the multiplication by functions $\psi(t)$.

Using \eqref{eq:commutators1}, \eqref{eq:commutators2}, the Cauchy-Schwarz inequality and that $|t| \leq 2 h^{1/2-\delta}$ on $\supp \chi_t$, we find from \eqref{eq:doubleComm},
\begin{align}
\label{eq:finalComm7}
| \langle & \chi^2_{s,0} \varphi_h \;|\; 
[\chi_t W, [\chi_t W, {\mathcal H}_1]] \chi^2_{s,0} \varphi_h \rangle | \nonumber\\
&\leq
C \Big(h^{\frac 32 -2\epsilon-\delta} \| (hD_s - \tilde{A}_1)\chi_t \chi^2_{s,0} \varphi_h \|^2
\nonumber\\
&\quad\quad\quad
+ h^{2-2\epsilon-2\delta} \| (hD_s - \tilde{A}_1)\chi_t \chi^2_{s,0} \varphi_h \| \; \| \chi_t \chi^2_{s,0} \varphi_h \| 
+ h^{3-2\epsilon-4\delta} \| \chi_t \chi^2_{s,0} \varphi_h \|^2\Big) \nonumber\\
&\leq \tilde C h^{3/2-2\epsilon-2\delta}
 \big(  \langle \chi_t  \varphi_h\;|\; {\mathcal H}(h) 
\chi_t \varphi_h \rangle + (h+h^{3/2-2\delta}) \| \chi_t  \varphi_h \|^2\big)\;. 
\end{align}
Using the condition satisfied by  $\delta$ from \eqref{eq:littledelta}, and the localization estimate in $s$ \eqref{eq:AgmonlikeT2s}, this implies \eqref{eq:ResT2}  for the expectation value $\langle \chi^2_{s,0} \varphi_h\;|\; [\chi_t W, [\chi_t W, {\mathcal H}_1]] \chi^2_{s,0} \varphi_h \rangle$.

\noindent{\bf Commutation with $\mathcal H_2$.}~\\
The commutation with ${\mathcal H}_2$ is similar, but easier.
\begin{align}
\big[ \chi_t W, [\chi_t W, {\mathcal H}_2]\big] &= -ih \chi_t^2 \big[W, [W, \frac{\partial_s a}{a^3}](hD_s - \tilde{A}_1) - \frac{\partial_s a}{a^3}[W, \tilde{A}_1)]\big] \nonumber\\
&=
-ih \chi_t^2[W, [W, \frac{\partial_s a}{a^3}]](hD_s - \tilde{A}_1)
+2ih \chi_t^2[W, \frac{\partial_s a}{a^3}][W, \tilde{A}_1)] \nonumber\\
&\quad
+ih \chi_t^2 \frac{\partial_s a}{a^3}[W,[W, \tilde{A}_1)]]\;.
\end{align}
Now, $\partial_s a = -t \kappa'(s)$, so the new terms to control are
\begin{align}
[W, \frac{\partial_s a}{a^3}] &=  h^{1/2-\epsilon} t\,\mathcal O_5, &
[W,[W, \frac{\partial_s a}{a^3}]] &= h^{1-2\epsilon} t\, \mathcal O_6\;,
\end{align}
with
bounded\footnote{after multiplication by a cutoff function
localizing in $[0,t_0)$,}  $\mathcal O_j$'s. Thus, for $[\chi_t W, [\chi_t W, {\mathcal H}_2]]$, we get
\begin{align}
\label{eq:h2Res}
| \langle \chi^2_{s,0}& \varphi_h  \;|\; [\chi_t W, [\chi_t W, {\mathcal H}_2]]  \chi^2_{s,0} \varphi_h \rangle |  \nonumber \\
&\leq
C h^{2-2\epsilon} (\| t \chi_t \chi^2_{s,0} \varphi_h \| \,\|(hD_s - \tilde{A}_1) \chi_t \chi^2_{s,0} \varphi_h \|
+ \| t^2 \chi_t \chi^2_{s,0} \varphi_h \|\, \| t \chi_t \chi^2_{s,0} \varphi_h \|) \nonumber\\
&\leq
\tilde C \left(h^{2-2\epsilon -\delta} + h^{\frac 52-2\epsilon -3\delta}\right)\big( \langle \chi_t  \varphi_h\;|\; {\mathcal H}(h) \chi_t  \varphi_h \rangle + h \| \chi_t  \varphi_h \|^2 \big)\;.
\end{align}
This implies \eqref{eq:ResT2} for $ \langle \chi^2_{s,0} \varphi_h\;|\; [\chi_t W, [\chi_t W, {\mathcal H}_2]]  \chi^2_{s,0} \varphi_h \rangle$.

\noindent{\bf Commutation with $\mathcal H_3$.}~\\
Since $W$ commutes with $\chi_t$ and $D_t$, we can calculate
\begin{align}
[\chi_t W,[\chi_t W, {\mathcal H}_3]] & = h^2 W^2 [\chi_t, [\chi_t, D_t^2]] = -2h^2 W^2 |\partial_t \chi_t|^2 \nonumber\\
&= 
-2 h^{1+2\delta} W^2 |\chi_1'(\frac{t}{h^{1/2-\delta}})|^2\;.
\end{align}
The expectation value of this term in the state $\chi^2_{s,0} \varphi_h$
 will be exponentially small, due to the normal Agmon estimates, \eqref{eq:Agmonliket}. Explicitly, \begin{align}
 \label{eq:h3Res}
|\langle \chi^2_{s,0} &\varphi_h\;|\; [\chi_t W,  [\chi_t W, {\mathcal H}_3]] 
\chi^2_{s,0} \varphi_h \rangle | \nonumber\\
&=
2 h^{1+2\delta} \Big |\langle e^{\alpha t/h^{1/2}}
 \chi^2_{s,0} \varphi_h\;\big|\; W^2 e^{-2\alpha t/h^{1/2}} 
|\chi_1'(\frac{t}{h^{1/2}-\delta})|^2 
e^{\alpha t/h^{1/2}} \chi^2_{s,0} \varphi_h \rangle \Big|\nonumber \\
&\leq 2 h^{1+2\delta} \| W \|^2 \|e^{-2\alpha t/h^{1/2}} |
\chi_1'(\frac{t}{h^{1/2-\delta}})|^2\|_{\infty} \|
e^{\alpha t/h^{1/2}} \varphi_h\|^2 \nonumber\\
&\leq C h^{1+2\delta} e^{-2\alpha h^{-\delta}} = {\mathcal O}_{\rm unif}(h^{\infty})\;.
\end{align}
In particular, \eqref{eq:ResT2} is satisfied for the term $\langle \chi^2_{s,0} \varphi_h\;|\; [\chi_t W, [\chi_t W, {\mathcal H}_3]] 
\chi^2_{s,0} \varphi_h \rangle$.

\noindent{\bf Commutation with $\mathcal H_4$.}~\\
When we calculate $[\chi_t W,[\chi_t W, {\mathcal H}_4]] $,  we will use the discussion of the previous paragraph 
 to conclude that if a derivative falls on $\chi_t$, then the resulting expectation value  becomes exponentially small. Thus,
\begin{align}
\langle \chi^2_{s,0} \varphi_h\;|&\; [\chi_tW,[\chi_t W, {\mathcal H}_4]] \chi^2_{s,0} \varphi_h \rangle \nonumber\\
&= 
\langle \chi^2_{s,0} \varphi_h\;|\; -ih [W,[W,\frac{\partial_t a}{a} ] ]\chi_t^2 (hD_t) \chi^2_{s,0} \varphi_h \rangle + {\mathcal O}_{\rm unif}(h^{\infty}) \nonumber\\
&= -ih \langle \chi_t \chi^2_{s,0} \varphi_h\;|\; -ih [W,[W,\frac{\partial_t a}{a} ] ] (hD_t) \chi_t \chi^2_{s,0} \varphi_h \rangle + {\mathcal O}_{\rm unif}(h^{\infty})\;.
\end{align}
From the formula $\frac{\partial_t a}{a} = \frac{-\kappa(s)}{1-t\kappa(s)}\,$, we see that all derivatives (in $s$) of $\frac{\partial_t a}{a}$ are uniformly bounded on the support of $\chi_t$. 
We therefore find
\begin{align}
[W,[W,\frac{\partial_t a}{a} ] ]= h^{1-2\epsilon} \mathcal O_7\;,
\end{align}
where $\mathcal O_7$ is a bounded (pseudodifferential) operator. Thus
\begin{align}
\label{eq:h4res}
|\langle \chi^2_{s,0}\varphi_h\;|&\; [\chi_t W, [\chi_t W, {\mathcal H}_4]] \chi^2_{s,0}\varphi_h \rangle|
\\
&\leq 
C h^{2-2\epsilon}\big( \| (hD_t) \chi_t \chi^2_{s,0}\varphi_h \| \| \chi_t \chi^2_{s,0}\varphi_h \|\big) + {\mathcal O}_{\rm unif}(h^{\infty}) \nonumber\\
&\leq 
\tilde C h^{\frac 32 -2\epsilon}\big( \langle \chi_t\varphi_h\;|\; {\mathcal H}(h) \chi_t \varphi_h \rangle + h \| \chi_t \varphi_h \|^2 \big) + {\mathcal O}_{\rm unif}(h^{\infty})\;.
\nonumber
\end{align}
This implies \eqref{eq:ResT2} for $\langle \chi^2_{s,0}\varphi_h\;|\; [\chi_t W, [\chi_t W, {\mathcal H}_4]] \chi^2_{s,0} \varphi_h \rangle$, and therefore \eqref{eq:ResT2}  is established for the expectation of ${\mathcal C}_1$.\\
\subsubsection{Control of  $\frac 12 \langle {\mathcal C}_2 \chi^2_{s,0} \varphi_h\,|\, \chi^2_{s,0} \varphi_h\rangle$.}~\\
To estimate this term,  we use similar calculations and the decomposition given
 in \eqref{decc1c2a} and \eqref{decc1c2b}.

\noindent{\bf Estimate of $\langle {\mathcal C}_{2,1} \chi^2_{s,0} \varphi_h\,,\, \chi^2_{s,0} \varphi_h\rangle$.}~\\
For the first term $\widehat{W} [ \chi_t, [\chi_t, {\mathcal H}]]W$, we clearly get, using \eqref{eq:Agmonliket} and as for \eqref{eq:h3Res},
\begin{align}
\label{eq:C21res}
\langle \chi^2_{s,0} \varphi_h \, | \, \widehat{W} [ \chi_t, [\chi_t, {\mathcal H}]]W \chi^2_{s,0}\varphi_h \rangle &= 
- h^2 \langle \chi^2_{s,0}\varphi_h \, | \, \widehat{W} | \partial_t \chi_t|^2 W \chi^2_{s,0}\varphi_h \rangle 
\nonumber\\
&= {\mathcal O}_{\rm unif}(h^{\infty})\;.
\end{align}
Thus, \eqref{eq:ResT2} holds for the expectation of ${\mathcal C}_{2,1}$.

\noindent{\bf Estimate of 
$\langle {\mathcal C}_{2,2}\chi^2_{s,0}\varphi_h\,|\, \chi^2_{s,0}\varphi_h\rangle$.}~\\
For this term, we notice that $[W, {\mathcal H}_3] = 0$, and calculate, using the ${\mathcal O}_j$'s introduced previously.
\begin{align}
{\mathcal C}_{2,2} & = \widehat{W} \chi_t^2 [ W, {\mathcal H}_1+ {\mathcal H}_2 + {\mathcal H}_4] \nonumber\\
&= \widehat{W} \chi_t^2 \Big(
(hD_s - \tilde{A}_1) [W, a^{-2}] (hD_s - \tilde{A}_1) - [W, \tilde{A}_1] a^{-2} (hD_s - \tilde{A}_1) 
\nonumber\\
&\quad\quad\quad\quad - (hD_s - \tilde{A}_1) a^{-2} [W, \tilde{A}_1]
-ih [W, \frac{\partial_s a}{a^3}] (hD_s - \tilde{A}_1)\nonumber\\
&\quad\quad\quad\quad\quad\quad+ ih \frac{\partial_s a}{a^3} [W, \tilde{A}_1] -
ih [W, \frac{\partial_t a}{a}] (hD_t) \Big) \nonumber\\
&= \widehat{W} \chi_t^2 \Big( (hD_s - \tilde{A}_1) t h^{1/2-\epsilon} 
{\mathcal O}_3 (hD_s - \tilde{A}_1) -t^2 h^{1/2-\epsilon} {\mathcal O}_1  a^{-2} (hD_s - \tilde{A}_1)
\nonumber\\
&\quad\quad\quad\quad
-(hD_s - \tilde{A}_1)a^{-2} t^2 h^{1/2-\epsilon} {\mathcal O}_1
-i t h^{3/2-\epsilon} {\mathcal O}_5 (hD_s - \tilde{A}_1)\nonumber\\
&\quad\quad\quad\quad
- ih^{3/2-\epsilon} t \frac{\kappa'(s)}{a^3} t^2  {\mathcal O}_1
-ih^{3/2-\epsilon} {\mathcal O}_8 (hD_t)\Big)\;.
\end{align}
Here ${\mathcal O}_8$ (which we used in the last line) and ${\mathcal O}_9$ (that will be used below), are
\begin{align}
[ W, \frac{-\kappa(s)}{a}] &= h^{1/2-\epsilon} {\mathcal O}_8\;,&
\widehat{W} &= t h^{1/2-\epsilon} {\mathcal O}_9\;.
\end{align}
Therefore, we get (using the pseudo-differential calculus for showing that
 $[D_s,\mathcal O_9]=\mathcal O_{10}$)
\begin{align}
\label{eq:C22res}
|\langle \chi^2_{s,0} \varphi_h \, |&\, {\mathcal C}_{2,2} \chi^2_{s,0}\varphi_h \rangle | \nonumber\\
&\leq C\Big( 
h^{1-2\epsilon} h^{1-2\delta}\| (hD_s - \tilde{A}_1) {\mathcal O}_9 \chi^2_{s,0}\varphi_h \|\, \| (hD_s - \tilde{A}_1) \chi^2_{s,0}\varphi_h \| \nonumber\\
&\quad\quad+ h^{1-2\epsilon} h^{3/2-3\delta} \| \chi^2_{s,0}\varphi_h\|\, \| (hD_s - \tilde{A}_1) \chi^2_{s,0}\varphi_h \| \nonumber\\
&\quad\quad+ h^{1-2\epsilon} h^{3/2-3\delta} 
\| (hD_s - \tilde{A}_1) {\mathcal O}_9 \chi^2_{s,0}\varphi_h \|\, \| \chi^2_{s,0} \varphi_h \| \nonumber\\
&\quad\quad+ h^{2-2\epsilon} h^{1-2\delta}  \|  \chi^2_{s,0}\varphi_h \| \, \| (hD_s - \tilde{A}_1) \chi^2_{s,0}\varphi_h \|
+ h^{2-2\epsilon} h^{2-4\delta}  \|  \chi^2_{s,0} \varphi_h \|^2\nonumber\\
&\quad\quad
+ h^{2-2\epsilon} h^{1/2-\delta} \|  \chi^2_{s,0}\varphi_h \|\, \| hD_t \chi^2_{s,0}\varphi_h\| \Big) \nonumber\\
&\leq \tilde C h^{3/2-2\epsilon}  \big(\langle \varphi_h \, | \, {\mathcal H}
 \varphi_h \rangle + h \|  \varphi_h \|^2\big)\;.
\end{align}
So, by \eqref{eq:littledelta},  \eqref{eq:ResT2} also holds for the expectation of ${\mathcal C}_{2,2}$.

\noindent{\bf Estimate of $\langle {\mathcal C}_{2,3}\chi^2_{s,0} \varphi_h\,|\,\chi^2_{s,0} \varphi_h\rangle$.}~\\
For this last term the approach is equally direct. 
We calculate~:
\begin{align}
[\widehat{W},{\mathcal H} ] &=
[\widehat{W},  (hD_s - \tilde{A}_1)] a^{-2} (hD_s - \tilde{A}_1)
+ (hD_s - \tilde{A}_1) [\widehat{W}, a^{-2}] (hD_s - \tilde{A}_1) \nonumber\\
&\quad\quad\quad+
(hD_s - \tilde{A}_1) a^{-2} [\widehat{W},  (hD_s - \tilde{A}_1)]  \nonumber\\
&\quad -ih [\widehat{W}, \frac{\partial_s a}{a^3} ] (hD_s - \tilde{A}_1) -ih \frac{\partial_s a}{a^3}  [\widehat{W},  (hD_s - \tilde{A}_1)] \nonumber\\
&\quad
- [\frac{t}{a}, (hD_t)^2] [W,\kappa(s)] \nonumber\\
&\quad +ih\frac{t}{a}  \big[ [W,\kappa(s)] ,\frac{\partial_t a}{a}\big] (hD_t) 
+ih \frac{\partial_t a}{a}  \big[\frac{t}{a}, (hD_t)\big]  [W,\kappa(s)] \;.
\end{align}
The commutators in the above expression can be estimated~:
\begin{align}
[\widehat{W},  (hD_s - \tilde{A}_1)] &= -t \big[ a^{-1}[W, \kappa(s)] ,hD_s \big] + t \big[ a^{-1}[W, \kappa(s)], \tilde{A}_1 \big] \nonumber\\
&= -i ht \frac{\partial_s a}{a^2}[W, \kappa(s)]
 + ht a^{-1} [[W, \kappa(s)] ,D_s ] \nonumber\\
&\quad\quad- t^3 a^{-1} [[W, \kappa(s)] ,\kappa(s)]/2 \nonumber\\
&= -i ht^2 \frac{\kappa'(s)}{a^2}[W, \kappa(s)] 
+ ht a^{-1} [W, [\kappa(s) , D_s ] ]\nonumber\\
&\quad\quad- t^3 a^{-1} [[W, \kappa(s)] ,\kappa(s)]/2 \nonumber\\
&= t^2 h^{3/2-\epsilon} {\mathcal U}_{(1)} + t h^{3/2-\epsilon} {\mathcal U}_{(2)} + t^3 h^{1-2\epsilon} {\mathcal U}_{(3)}\;,
\end{align}
with bounded ${\mathcal U}_{(j)}$'s  when composed with $t$-cut-off functions.\\
Thus, when we localize in  $\{|t|\leq C h^{1/2-\delta}\}$, we get
\begin{align}
[\widehat{W},  (hD_s - \tilde{A}_1)] = h^{2-\delta-\epsilon} {\mathcal U}_1\;,
\end{align}
where ${\mathcal U}_1$ is uniformly bounded.
Similarly,
\begin{align}
[\widehat{W}, a^{-2}] &= [\widehat{W}, \frac{1-a^2}{a^2}] = t^2 a^{-1} \big[ [W, \kappa(s)], \frac{2\kappa(s) -t \kappa(s)}{a^2}\big] \nonumber\\
&= h^{1-2\delta} h^{1-2\epsilon} {\mathcal U}_2\;,
\end{align}
\begin{align}
\big[ \widehat{W}, \frac{\partial_s a}{a^3} \big] = t^2 a^{-1} \big[ [W, \kappa(s)], \frac{\kappa'(s)}{a^3} \big]
= h^{1-2\delta} h^{1-2\epsilon} {\mathcal U}_3\;,
\end{align}
\begin{align}
\big[ \frac{t}{a}, (hD_t) \big] = ih\partial_t(\frac{t}{a})\;.
\end{align}
\begin{align}
\big[ \frac{t}{a}, (hD_t)^2 \big] &=
ih (hD_t)  \partial_t(\frac{t}{a}) + ih \partial_t(\frac{t}{a}) (hD_t)
= 2ih (hD_t)  \partial_t(\frac{t}{a}) - h^2 \partial_t^2 (\frac{t}{a}) \nonumber\\
&= (hD_t) h {\mathcal U}_4 + h^2 {\mathcal U}_5\;,
\end{align}
\begin{align}
\big[ [W, \kappa(s)], \frac{\partial_t a}{a} \big] = h^{1-2\epsilon} {\mathcal U}_6\;,
\end{align}
Thus, we can estimate, after additional controls of commutators,
\begin{align}\label{eq:AlmostT2}
& |\langle \chi^2_{s,0}  \varphi_h \,| \, {\mathcal C}_{2,3} \chi^2_{s,0}\varphi_h \rangle| \nonumber\\
& \leq 
C \Big(
h^{2-\delta-\epsilon} \| \chi^2_{s,0}\varphi_h \| \,\| (hD_s -\tilde{A}_1) 
W \chi_t^2 \chi^2_{s,0}\varphi_h \| \nonumber\\
&\quad\quad+ h^{1-2\delta} h^{1-2\epsilon} 
\| (hD_s -\tilde{A}_1)  \chi^2_{s,0}\varphi_h \|\, \| (hD_s -\tilde{A}_1) 
W \chi_t^2 \chi^2_{s,0}\varphi_h \| \nonumber\\
&\quad\quad
+ h^{2-\delta-\epsilon}  \| (hD_s -\tilde{A}_1)  \chi^2_{s,0} \varphi_h \|\, \| \chi^2_{s,0} \varphi_h \|\nonumber\\
&\quad\quad
+ h^{2-2\delta} h^{1-2\epsilon} \| \chi^2_{s,0}\varphi_h \|\, \| (hD_s -\tilde{A}_1) W \chi_t^2 \chi^2_{s,0}\varphi_h \| \nonumber\\
&\quad\quad
+ h^{3-\delta-\epsilon} \| \chi^2_{s,0}\varphi_h \|^2
+  h | \langle \chi^2_{s,0}\varphi_h \,| \, (hD_t) {\mathcal U}_4 [W, \kappa(s)]
 \chi^2_{s,0}\varphi_h  \rangle |
+ h^{5/2-\epsilon} \| \chi^2_{s,0}\varphi_h \|^2\nonumber\\
&\quad\quad
+ h^{3/2-\delta} h^{1-2\epsilon} \| \chi^2_{s,0}\varphi_h \|\, \| (hD_t) W \chi_t^2 \chi^2_{s,0}\varphi_h \|
+ h^{5/2-\epsilon}  \| \chi^2_{s,0}\varphi_h \|^2\Big) \nonumber\\
&\leq
\tilde C\Big( h^{3/2-\delta-\epsilon} \langle \varphi_h \,|\, {\mathcal H} \varphi_h \rangle 
+ h^{5/2-\delta-\epsilon} \| \varphi_h \|^2 \nonumber\\
&\quad\quad\quad\quad\quad\quad
+ h | \langle \chi^2_{s,0} \varphi_h \,| \, (hD_t) {\mathcal U}_4 [W, \kappa(s)]
 \chi^2_{s,0} \varphi_h  \rangle | \Big)\;. 
\end{align}

\begin{remark}
If, in \eqref{eq:AlmostT2}, we estimate $[W, \kappa(s)]$ by the easy pseudodifferential
result, i.e. use the bound
$$
\big\| [W, \kappa(s)] \big \| = {\mathcal O}(h^{1/2-\epsilon})\;,
$$ 
we can continue \eqref{eq:AlmostT2} to get
$$
|\langle \chi^2_{s,0} \varphi_h \,| \, {\mathcal C}_{2,3} \chi^2_{s,0} \varphi_h \rangle|
\leq 
C\big( h^{1-\epsilon} \langle \varphi_h \,|\, {\mathcal H} \varphi_h \rangle 
+ h^{2-\epsilon} \| \varphi_h \|^2 \big)\;.
$$
This would only allow us to take $\epsilon \in (0,1/3)$ instead of $\epsilon \in (0,3/8)$ as claimed 
in Theorem~\ref{prop:LocXi}. In order to get the optimal range for $\epsilon$, we estimate the commutator $[W, \kappa(s)]$ slightly better using the `Agmon estimates' in $s$, i.e. \eqref{eq:AgmonlikeT2s}. 
\end{remark}

We write $\kappa(s) = k_{\rm max} - s^2\hat{\kappa}(s)$, with $\hat{\kappa}$ a smooth function with bounded derivatives of all orders. 
Now,
\begin{align}
[W, \kappa(s)] &= [\hat{\kappa}(s)s^2, W ]
= \hat{\kappa}(s) [s, [s,W]] + 2 \hat{\kappa}(s) [s, W] s + [\hat{\kappa}(s), W ]s^2\nonumber\\
& = \widetilde{{\mathcal U}}_1 h^{1-2\epsilon} + \widetilde{{\mathcal U}}_2 h^{1/2-\epsilon}s +
\widetilde{{\mathcal U}}_3 h^{1/2-\epsilon}s^2\;.
\end{align}
Therefore, using the estimate \eqref{eq:AgmonlikeT2s}, we get the inequality,
\begin{align}\label{eq:tricky}
\big\| [W, \kappa(s)] \chi^2_{s,0} \varphi_h \big\| \leq C h^{5/8-\epsilon-\delta} \| \varphi_h \|
 + \Og_{\rm unif}(h^\infty)\;,
\end{align}
and finally, \eqref{eq:ResT2} follows (for the final term ${\mathcal C}_{2,3}$) by inserting \eqref{eq:tricky} in \eqref{eq:AlmostT2}. 

This ends the proof of Proposition~\ref{lem:ResultT2}.

\section{Lower bounds in Grushin's approach}~\\
\label{lower2} 
In this section we  finish the proof of Theorem~\ref{thm:main}.

\begin{theorem}
\label{thm:GrushinLower2}
Let $\mu^{(n)}(h)$ be the $n$-th eigenvalue  of ${\mathcal H}(h)$  and let $z_{\infty}^{(n)}(h)$ be the asymptotic sum given in \eqref{eq:defzn}.
Then  $\mu^{(n)}(h)$ has  $z_{\infty}^{(n)}(h)$ as  asymptotic expansion.
\end{theorem}

It is clear that Theorem~\ref{thm:GrushinLower2} implies Theorem~\ref{thm:main}. From the estimates in the previous sections~\ref{locPhase} and \ref{phasespace}, we know the low-lying
eigenfunctions to be localized in phase space, around the following set (where we denote by $\xi$ the variable in phase space dual to $s$)~:
\begin{itemize}
\item $s\in (-2h^{1/8-\epsilon_1}\;,\;2h^{1/8-\epsilon_1})$\;;
\item $t \in (-2h^{1/2-\epsilon_2}\;, \;2h^{1/2-\epsilon_2})$\;;
\item $\xi \in(-2h^{\epsilon_3-1/2}+\xi_0 h^{-1/2}\;,\;2h^{\epsilon_3-1/2}+\xi_0 h^{-1/2})$\;.
\end{itemize}
We will choose $\epsilon_1 = \epsilon_2  =: \eta\,$, $\epsilon_3 = 3/8 - \eta\,$, where $\eta < 1/8\,$. Thus, in the remainder  of the section, $\eta$ will be a fixed (small) constant satisfying,
\begin{align}\label{eq:ChoiceEpsilon2}
\eta \in (0, 1/8)\;.
\end{align}

\begin{remark}~\\
We only need a space localization in $t$. There is no localization in the corresponding frequency. This is pleasant, since it avoids possible complications due to the boundary condition.
\end{remark}

More precisely, the localization of $u_h^{(n)}$ is analyzed in  the following lemma.

\begin{lemma}
\label{lem:SummarizeLoc2}~\\
Let $M>0$, $h_0 >0$ and let $\chi$ be a standard cut-off function:
\begin{align}\label{eq:locFunc2}
\chi & \in C_0^{\infty}({\mathbb R})\;, &
\chi(x) & = 1\;,\; \mbox{ on a nbd.~of~} [-\frac 32, +\frac 32]\;,\;&
\supp \chi &\subset (-2,2)\;.
\end{align}
Then for all $K>0$ there exists $b_K>0$ such that if $(u_h)_{h \in (0,h_0)}$ is a family of normalized eigenfunctions of ${\mathcal H}$ with eigenvalue $\mu(h)$ satisfying
\begin{align}
\label{eq:EVBOUND}
\mu(h) \leq \Theta_0h - C_1 k_{{\rm max}}h^{3/2} + M h^{7/4}\;,
\end{align}
then, with $\eta$ from \eqref{eq:ChoiceEpsilon2},
\begin{align}
\label{eq:microlocal2-2}
\Big\|u_h - \chi(\frac{t}{h^{1/2-\eta}})
 \chi(\frac{s}{h^{1/8-\eta}}) 
\chi\big(\frac{|h^{1/2}D_s-\xi_0|}{h^{3/8-\eta}}\big) 
\chi(\frac{4s}{|\partial \Omega|}) u_h\Big\|_2 \leq b_K h^K\;.
\end{align}
\end{lemma}

\begin{proof}[Proof  of Lemma \ref{lem:SummarizeLoc2}]~\\
Define $\chi_2 = 1 - \chi$. 
From Theorems~\ref{prop:Agmon-t} and~\ref{thm:Agmons} we know that 
$$
(1-\chi(\frac{t}{h^{1/2-\eta}})
\chi(\frac{s}{h^{1/8-\eta}}) ) u_h= {\mathcal O}_{\rm unif}(h^{\infty})\;.
$$ 
So it suffices to prove that
\begin{align}
\label{eq:microlocal3}
\Big\| \chi(\frac{t}{h^{1/2-\eta}})
 \chi(\frac{s}{h^{1/8-\eta}})  u_h& - 
\chi(\frac{t}{h^{1/2-\eta}})
 \chi(\frac{s}{h^{1/8-\eta}}) 
\chi\big(\frac{|h^{1/2}D_s-\xi_0|}{h^{3/8-\eta}}\big) 
\chi(\frac{4s}{|\partial \Omega|}) u_h\Big\|_2 \nonumber\\
&= {\mathcal O}_{\rm unif}(h^{\infty})\;.
\end{align}
By writing 
\begin{align*}
 & \chi(\frac{t}{h^{1/2-\eta}})
 \chi(\frac{s}{h^{1/8-\eta}})  u_h=
 \chi(\frac{t}{h^{1/2-\eta}})
 \chi(\frac{s}{h^{1/8-\eta}}) \chi(\frac{4s}{|\partial \Omega|}) u_h \\
 &\quad=
\chi(\frac{t}{h^{1/2-\eta}})
 \chi(\frac{s}{h^{1/8-\eta}}) 
  \big\{\chi(\frac{|h^{1/2}D_s-\xi_0|}{h^{3/8-\eta}}) + \chi_2(\frac{|h^{1/2}D_s-\xi_0|}{h^{3/8-\eta}})\big\}
  \chi(\frac{4s}{|\partial \Omega|}) u_h\;,
\end{align*}
and appealing to Theorem~\ref{prop:LocXi} we get \eqref{eq:microlocal3} and thereby Lemma~\ref{lem:SummarizeLoc2}.
\end{proof}
~\\
\begin{proof}[Proof of Theorem~\ref{thm:GrushinLower2}]~\\
For a definite choice of $\chi$ (fixed once and for all) as in Lemma~\ref{lem:SummarizeLoc2}, and $u_h$ an eigenfunction of ${\mathcal H}$ satisfying \eqref{eq:EVBOUND}, define 
\begin{align} \label{eq:DEF2}
\tilde{\psi}_h &= e^{-is\xi_0/h^{1/2}}\chi(\frac{t}{h^{1/2-\eta}}) \chi(\frac{s}{h^{1/8-\eta}}) 
\chi\big(\frac{|h^{1/2}D_s-\xi_0|}{h^{3/8-\eta}}\big)  \chi(\frac{4s}{|\partial \Omega|})
u_h\;, \nonumber\\
\psi_h(\sigma,\tau) & = h^{5/16} \tilde{\psi}_h(h^{1/8}\sigma,h^{1/2}\tau)\;.
\end{align}

Calculations will from now on be carried out in the variables $(\sigma, \tau)$. All functions considered will be localized on a scale of order $h^{-\eta}$ in the $(\sigma, \tau)$-variables. This implies (in particular) that they are localized to a tubular neighborhood of size $h^{1/2-\eta}$ near the boundary in the original coordinates $x \in \Omega$. The natural measure in $(\sigma, \tau)$-variables, inherited from $L^2(\Omega,dx)$ by implementing unitarily the change of coordinates, is $(1-h^{1/2}\tau \kappa(h^{-1/8}\sigma))d\sigma d\tau$. However, due to the localization of our functions (and the boundedness of $\kappa$) we can replace this measure by $d\sigma d\tau$ (since $h^{1/2}\tau \kappa(h^{-1/8}\sigma) = {\mathcal O}(h^{1/2-\eta})$ on $| \tau | \leq C h^{-\eta}$), without changing our estimates up to multiplicative $h$-independent constants. Therefore we can (and will !) do all our estimates by choosing the norms in  $L^2({\mathbb R} \times {\mathbb R}_{+}, d\sigma d\tau)$ or in $\mathcal L (L^2({\mathbb R} \times {\mathbb R}_{+}, d\sigma d\tau))$. Thus all $L^2$-norms below refer to $L^2({\mathbb R} \times {\mathbb R}_{+}, d\sigma d\tau)$ and similarly for operator norms.

With these conventions we have, using Theorems~\ref{prop:Agmon-t}, \ref{thm:Agmons} and \ref{prop:LocXi}, for all eigenfunctions $u_h$ corresponding to eigenvalues $\mu(h)$ satisfying \eqref{eq:EVBOUND}, that the corresponding $\psi_h$ (given in \eqref{eq:DEF2}) satisfies (with error terms ${\mathcal O}_{\rm unif}(h^{\infty})$ uniform for eigenfunctions $u_h$ as long as \eqref{eq:EVBOUND} is satisfied),
\begin{align}
\label{eq:alpha2}
\| \psi_h \|_{L^2} & = 1 + {\mathcal O}_{\rm unif}(h^{\infty})\;, & \text{ and } &&
L \psi_h & = \psi_h + {\mathcal O}_{\rm unif}(h^{\infty})\;,
\end{align}
for all 
\begin{align}
\label{eq:Type_L-2}
L = \tilde{\chi}(\frac{\tau}{h^{-\eta}}) \tilde{\chi}(\frac{\sigma}{h^{-\eta}}) \tilde{\chi}(h^{\eta}D_{\sigma})\;,
\end{align}
with $\tilde{\chi}$ satisfying \eqref{eq:locFunc2} and with $\eta$ from \eqref{eq:ChoiceEpsilon2}. Let us fix an $L_0$ as in \eqref{eq:Type_L-2} in the rest of this Section.

Let $H_{\rm harm}$ be the harmonic oscillator on $L^2({\mathbb R})$ defined by
\begin{align}\label{eq:DefHarmOsc}
H_{\rm harm}:= 3C_1 \sqrt{\Theta_0} D_{\sigma}^2 + C_1 \tfrac{k_2 \sigma^2}{2}\;,
\end{align}
(compare with Section~\ref{GrushinUpper}). Clearly, $H_{\rm harm}$ has eigenvalues
\begin{align}
\label{eq:EvHarmOsc2}
e_\ell := C_1 \Theta_0^{1/4} \sqrt{\tfrac{3k_2}{2}} (2\ell + 1)\;,
\end{align}
with $\ell \in {\mathbb N}$. Let $v_{\ell}$ be the corresponding (unique up to scalar multiple) normalized eigenfunction.
For $N \in {\mathbb N}$, 
the value $C_1 \Theta_0^{1/4} \sqrt{6k_2}  N$ is right in the middle between two eigenvalues ($e_{N-1}$ and $e_N$). We
define the vector space $V_N \subset L^2({\mathbb R})$ as the space spanned by eigenfunctions of $H_{\rm harm}$ corresponding to eigenvalues below $C_1 \Theta_0^{1/4} \sqrt{6k_2}  N$, i.e.
\begin{align}
V_N:= \Ran 1_{[0, C_1 \Theta_0^{1/4} \sqrt{6k_2}  N]}(H_{\rm harm}) = \Span \{v_0, \ldots, v_{N-1}\}\;.
\end{align}
Clearly, $\dim V_N = N$.

Similarly, we define $U_N(h) \subset L^2(\Omega)$ as the spectral subspace 
attached to the interval $I_N(h)$, with 
\begin{align}
\label{eq:apubN}
I_N(h) = \big(-\infty, \Theta_0 h - k_{\rm max} C_1 h^{3/2} + C_1 \Theta_0^{1/4} \sqrt{6k_2}  N h^{7/4}\big]\;.
\end{align}
Let $\Pi_{V_N}: L^2({\mathbb R}) \rightarrow V_N$ and 
$\Pi_{U_N}: L^2(\Omega) \rightarrow U_N(h)$ be the orthogonal projections. 
We define a linear map ${\mathcal M}_1^{(N)}(h)$ from $V_N$ to $U_N(h)$ by 
$$
{\mathcal M}_1^{(N)}(h) v_{\ell} = \Pi_{U_N} \phi_{M_0}^{(\ell)}\;,
$$
(where $\phi_{M}^{(n)}$ was defined in \eqref{eq:formquasimode}) and extended by linearity.
The number $M_0$ is chosen fixed, but suffficiently large---the choice $M_0=10$ would suffice.

Furthermore, we define a linear map ${\mathcal M}_2^{(N)}(h)$ from $U_N(h)$ to $V_N$ by
$$
{\mathcal M}_2^{(N)}(h) u_h = \Pi_{V_N} R_0^{-} L_0 \psi_h \;,
$$
where $\psi_h$ is defined from $u_h$ by \eqref{eq:DEF2}.
We will prove the following lemma.

\begin{lemma}
\label{lem:LinAlg}
Let $N \in {\mathbb N}\setminus \{ 0 \}$. Then there exists $h_0 >0$, and $\eta>0$ (as in \eqref{eq:ChoiceEpsilon2} and used in \eqref{eq:DEF2}) such that for all $h<h_0$,  
${\mathcal M}_1^{(N)}(h)$ and ${\mathcal M}_2^{(N)}(h)$ are bijective.
\end{lemma}

We will prove Lemma~\ref{lem:LinAlg} below. First we apply it to finish the proof of Theorem~\ref{thm:GrushinLower2}. 

Lemma~\ref{lem:LinAlg} implies that, for sufficiently small $h$, 
$$
\dim U_N(h) = \dim V_N = N\;.
$$
But Corollary~\ref{cor:Upper} describes $N$ distinct points of $\Spec ({\mathcal H})$ below
$$
\Theta_0 h - k_{\rm max} C_1 h^{3/2} + C_1 \Theta_0^{1/4} \sqrt{\tfrac{3k_2}{2}} 2 N h^{7/4}\;.
$$
This finishes the proof of Theorem~\ref{thm:GrushinLower2}. 
\end{proof}

\begin{proof}[Proof of Lemma~\ref{lem:LinAlg}]~\\
We only need to prove that ${\mathcal M}_1^{(N)}(h)$ and ${\mathcal M}_2^{(N)}(h)$ are both injective.
Injectivity of ${\mathcal M}_1^{(N)}(h)$ is clear from Section~\ref {GrushinUpper}, so we only consider injectivity of ${\mathcal M}_2^{(N)}(h)$.

The key to the proof of injectivity of ${\mathcal M}_2^{(N)}(h)$ is the following lemma.
\begin{lemma}
\label{lem:harmOscAlmost2}
There exists $\eta_0>0$ such that if $\eta<\eta_0$ in \eqref{eq:ChoiceEpsilon2}, then there exists $C>0$ such that for all normalized eigenfunctions $u_h \in U_N(h)$ with corresponding eigenvalue $\mu(h)$, we have
\begin{align}\label{eq:specProblem2}
\big\| \big(
\nu(h) - H_{\rm harm} \big) R_{0}^{-} L_0 \psi_h \big\| \leq
C h^{1/16}\;,
\end{align}
where $\psi_h$ is related to $u_h$ by \eqref{eq:DEF2} and $\nu(h)$ is defined by
\begin{align}\label{eq:mu2}
\nu(h) := h^{-7/4}\big\{ \mu(h)-(\Theta_0 h - k_{\rm max} C_1 h^{3/2})\big\}\;.
\end{align}
\end{lemma}

\begin{proof}[Proof of Lemma~\ref{lem:harmOscAlmost2}]
With $P$ from \eqref{eq:Def_of_P_scaled} and
\begin{align}\label{eq:lambda2}
\lambda(h) = h^{-1}(\mu(h) - \Theta_0 h)\;,
\end{align}
we have (using Theorems~\ref{prop:Agmon-t}, \ref{thm:Agmons}, and \ref{prop:LocXi}), uniformly for normalized eigenfunctions $u_h \in U_N(h)$,
\begin{align}
\label{eq:gamma2}
(P-\lambda(h)) \psi_h &= {\mathcal O}_{\rm unif}(h^{\infty})\;, & \text{ and } && (P-\lambda(h)) L \psi_h = {\mathcal O}_{\rm unif}(h^{\infty})\;.
\end{align}

In the rest of the proof of Lemma~\ref{lem:harmOscAlmost2} we will often have estimates like \eqref{eq:gamma2}. We will generally not repeat the phrase `uniformly for normalized eigenfunctions $u_h \in U_N(h)$', but the estimates are meant to have such uniformity.

Using \eqref{eq:gamma2} and the notation from \eqref{eq:def_matrices}, we get:
\begin{align}
\label{eq:matrix1-2}
\left(\begin{matrix} P-\lambda(h) & R_0^{+} \\ R_{0}^{-} & 0 \end{matrix} \right)
\left(\begin{matrix} L_0\psi_h \\ 0 \end{matrix} \right)
=
\left(\begin{matrix} 0 \\ R_{0}^{-}  L_0 \psi_h  \end{matrix} \right) + {\mathcal O}_{\rm unif}(h^{\infty})\;.
\end{align}
Furthermore, with ${\mathcal E}_0$ from \eqref{eq:def_matrices}, 
\begin{align}
\label{eq:matrix3-2}
{\mathcal E}_0 \left(\begin{matrix} P-\lambda(h) & R_0^{+} \\ R_{0}^{-}  & 0 \end{matrix} \right)\left(\begin{matrix} L_0\psi_h \\ 0 \end{matrix} \right)
&=
\Big\{ L_0 + \left(\begin{matrix}  E_0 (\partial P) L_0 & 0\\ R_0^{-} (\partial P) L_0 & 0 \end{matrix} \right)\Big\}
\left(\begin{matrix} \psi_h \\ 0 \end{matrix} \right)\;.
\end{align}
Here we have introduced
\begin{align}
\partial P = (P - P_0)-\lambda(h)\;.
\end{align}
In order to proceed we need a bound on the matrix in $\{ \cdot \}$ in \eqref{eq:matrix3-2}.
\begin{lemma}
\label{lem:boundedGrushin2}~\\
There exists a constant $C>0$ such that (with $\eta$ from \eqref{eq:ChoiceEpsilon2})
\begin{align} \label{eq:normbd2}
\| E_0 (P - P_0) L_0 \| + \| R_0^{-} (P - P_0) L_0 \| \leq C h^{3/8 - \eta}\;.
\end{align}
More precisely, with $P_1$, $P_2$ and $P_3$ from \eqref{eq:P1s}-\eqref{eq:P3s}, there exists $N_0 \in \mathbb N$ such that, for  $\eta$ satisfying  \eqref{eq:ChoiceEpsilon2},
\begin{align}
\label{eq:Tri2}
\| E_0 (P - P_0) L_0 - E_0 \big( h^{3/8}P_1 + h^{1/2} P_2 + h^{3/4} P_3 \big) L_0\| &\leq C h^{7/8- N_0 \eta}\;, \nonumber\\
\| R_0^{-} (P - P_0) L_0 - R_0^{-} \big( h^{3/8}P_1 + h^{1/2} P_2 + h^{3/4} P_3 \big) L_0\| &\leq C h^{7/8- N_0 \eta}\;.
\end{align}
Furthermore, 
\begin{align}
\label{eq:Sq2}
h^{3/8} \| E_0 P_1 L_0 \| + h^{3/8}\| R_0^{-} P_1 L_0 \| & \leq C h^{\frac 38  - 2\eta}\;, \nonumber\\
h^{1/2} \| E_0 P_2 L_0 \| + h^{1/2}\| R_0^{-} P_2 L_0 \| & \leq C h^{1/2 - 3\eta}\;, \nonumber\\
h^{3/4} \| E_0 P_3 L_0 \| + h^{3/4}\| R_0^{-} P_3 L_0 \| & \leq C h^{3/4-5\eta}\;.
\end{align}
\end{lemma}
~\\

\begin{proof}[Proof of Lemma~\ref{lem:boundedGrushin2}]~\\
With $\tilde{a}$ and $\tilde{a}_2$ from \eqref{eq:as} and omitting the tilda's on the $a$'s, we have by definition
\begin{align}
P & = a^{-1}\big[(\tau + \xi_0) + h^{3/8} D_{\sigma} - \tau (1- a_2)\big] a^{-1} \big[(\tau + \xi_0) + h^{3/8} D_{\sigma} - \tau (1- a_2)\big] \nonumber\\
&\quad\quad+ a^{-1}D_{\tau} a D_{\tau} - \Theta_0\;,
\end{align}
and therefore
\begin{align}
\label{eq:P-P0-2}
 P-P_0 
& = \Big\{ a^{-1}\big[(\tau + \xi_0) + h^{3/8} D_{\sigma} - \tau (1- a_2)\big] a^{-1} \big[(\tau + \xi_0) + h^{3/8} D_{\sigma} \nonumber\\
&\quad\quad\quad\quad - \tau (1- a_2)\big] - (\tau + \xi_0)^2 \Big\}
-i D_{\tau} \frac{\partial_{\tau} a}{a}   + \partial_{\tau} \big( \frac{\partial_{\tau} a}{a}\big)\;.
\end{align}
We will use the property that $L$ localizes to $\{\tau < 2h^{-\eta}, |\sigma| < 2h^{-\eta}\}$ 
and that $E_0\,,\; E_0D_{\tau}\,,\; R_0^{-},$ and $R_0^{-} D_{\tau}$ are bounded.
We introduce $f:=(\tau+\xi_0) - \tau(1-a_2)$ and calculate
\begin{align*}
&a^{-1}\big(f + h^{3/8} D_{\sigma} \big) a^{-1} \big(f + h^{3/8} D_{\sigma}\big) - (\tau + \xi_0)^2\\
&=
\big(f^2/a^2 - (\tau+ \xi_0)^2\big) + i h^{3/8} \frac{\partial_{\sigma}a}{a^3} f 
+ 2a^{-2} f h^{3/8} D_{\sigma} + i \frac{\partial_{\sigma}a}{a^3} h^{3/4} D_{\sigma}  + a^{-2} h^{3/4} D_{\sigma}^2\;.
\end{align*}
Thus,
\begin{align*}
P-P_0 = {\mathcal Q}_1 + {\mathcal Q}_2 + {\mathcal Q}_3\;,
\end{align*}
where
\begin{align*}
{\mathcal Q}_1 &= f^2/a^2 - (\tau+ \xi_0)^2 + i h^{3/8} \frac{\partial_{\sigma}a}{a^3} f + \partial_{\tau} \big( \frac{\partial_{\tau} a}{a}\big)\;,\\
{\mathcal Q}_2 &= -i D_{\tau} \frac{\partial_{\tau} a}{a} \;,\\
{\mathcal Q}_3 &= 2a^{-2} f (h^{3/8} D_{\sigma}) + i \frac{\partial_{\sigma}a}{a^3} (h^{3/4} D_{\sigma}) + a^{-2} (h^{3/4} D_{\sigma}^2)\;.
\end{align*}
Now, on $\{\tau < 2h^{-\eta}, |\sigma| < 2h^{-\eta}\}$, we have
\begin{align}
\label{eq:Taylor2}
a&=1 + {\mathcal O}(h^{1/2-\eta})\;,&
1-a_2 &= {\mathcal O}(h^{1/2-\eta})\;.
\end{align}
Therefore,
\begin{align*}
f^2/a^2 &= (\tau + \xi_0)^2 + {\mathcal O}(h^{1/2-2\eta})\;,&
&f = {\mathcal O}(h^{-\eta})\;, &
\partial_{\sigma}a &= {\mathcal O}(h^{1/2+1/8-\eta})\;, \\
\partial_{\tau} a &= {\mathcal O}(h^{1/2})\;, &
&\partial_{\tau} \big( \frac{\partial_{\tau} a}{a}\big)= {\mathcal O}(h)\;.
\end{align*}
Thus, ${\mathcal Q}_1 = {\mathcal O}(h^{1/2-2\eta})$, so
\begin{align*}
&\| E_0 {\mathcal Q}_1 L_0 \| + \| R_0^{-} {\mathcal Q}_1 L_0 \| = {\mathcal O}(h^{\frac 12 - 2 \eta})\;.
\end{align*}
Furthermore, for all $j \geq 0$,
\begin{align}
\label{eq:stern2}
D_{\sigma}^j  \tilde{\chi}(h^{\eta} \sigma) \tilde{\chi}(h^{\eta} D_{\sigma})
=
\tilde{\chi}(2h^{\eta} \sigma)
D_{\sigma}^j \tilde{\chi}(h^{\eta} \sigma) \tilde{\chi}(h^{\eta} D_{\sigma})\;,
\end{align}
and
\begin{align}\label{eq:zweistern2}
\| D_{\sigma}^j \tilde{\chi}(h^{\eta} \sigma) \tilde{\chi}(h^{\eta} D_{\sigma}) \| \leq 
c_j (h^{- \eta})^j\;.
\end{align}
Using \eqref{eq:stern2} and \eqref{eq:zweistern2}, we get
\begin{align*}
&\| E_0 {\mathcal Q}_2 L_0 \| + \| R_0^{-} {\mathcal Q}_2 L_0 \| = {\mathcal O}(h^{1/2})\;, \\
&\| E_0 {\mathcal Q}_3 L_0 \| + \| R_0^{-} {\mathcal Q}_3 L_0 \| = {\mathcal O}(h^{-\eta} h^{3/8})+
{\mathcal O}(h^{1/2 + 1/8-\eta} h^{-\eta})+{\mathcal O}(h^{3/4-2\eta})
\;.
\end{align*}
This finishes the proof of \eqref{eq:normbd2}.

The more precise estimates, \eqref{eq:Tri2} and \eqref{eq:Sq2}, follow in the same manner, using that on $\{\tau < 2h^{-\eta}, |\sigma| < 2h^{-\eta}\}$,
\begin{align*}
a &= 1 - h^{1/2} \tau\big( \kappa(0) - \tfrac{1}{2} h^{1/4} \sigma^2 \kappa''(0) + {\mathcal O}(h^{3(1/8-\eta)})\big)\;, \\
a_2 &= 1 - h^{1/2} \tau \tfrac{1}{2}\big( \kappa(0) - \tfrac{1}{2} h^{1/4} \sigma^2 \kappa''(0) + {\mathcal O}(h^{3(1/8-\eta)}) \big)\;,
\end{align*}
instead of \eqref{eq:Taylor2}. We omit the details.

This finishes the proof of Lemma~\ref{lem:boundedGrushin2}.
\end{proof}
Combining  \eqref{eq:matrix1-2} and  \eqref{eq:matrix3-2}, we get~:
\begin{equation}\label{w1-2}
L_0 \psi_h = R_0^+ R_0^- L_0 \psi_h - E_0 (\pa P) L_0 \psi_h + \Og_{\rm unif}(h^\infty)\;,
\end{equation}
and
\begin{equation}\label{w2-2}
R_0^- (\pa P) L_0 \psi_h = \Og_{\rm unif}(h^\infty)\;.
\end{equation}
We now introduce an additional localization through an operator $L$
 as in \eqref{eq:Type_L-2}, which is chosen (slightly 'bigger' than $L_0$, i.e.) such that
\begin{align}\label{eq:Lbigger2}
L L_0 = L_0 + {\mathcal O}(h^{\infty})\,.
\end{align}
We observe that \eqref{w2-2} is also valid with $L_0$ replaced by $LL_0$
 and, applying the (uniformly) bounded operator $R_0^-(\partial P) L$ to \eqref{w1-2}, we obtain
\begin{equation} \label{w4-2}
R_0^- (\pa P)L R_0^+ R_0^- L_0 \psi_h -R_0^- (\pa P) L E_0 (\pa P) L_0 \psi_h = \Og_{\rm unif}(h^\infty)\;.
\end{equation}

We again  apply Lemma~\ref{lem:boundedGrushin2}, \eqref{w4-2} and the
 comparison estimates \eqref{w6-2}  and \eqref{w7-2} (to be proved below)  and obtain that, for all $\delta>0$, there exists $\eta_0\in (0,\frac 18)$, such that if $\eta<\eta_0$, then
\begin{align*}
&-R_0^{-} \big( h^{3/8}P_1 + h^{1/2} P_2 + h^{3/4} P_3 - \lambda(h) \big) L R_{0}^{+} R_{0}^{-} L_0 \psi_h \nonumber \\
&\quad\quad\quad+
R_0^{-} \big( h^{3/8}P_1  \big) L 
E_0 \big( h^{3/8}P_1 \big) R_{0}^{+} R_{0}^{-} L_0 \psi_h \nonumber\\
&\quad\quad\quad\quad\quad\quad= {\mathcal O}_{\rm unif}(h^{\infty}) + {\mathcal O}_{\rm unif}(h^{7/8-\delta}) \|R_{0}^{+} R_{0}^{-} L_0 \psi_h \|\;.
\end{align*}

Using \eqref{w2-2}, the rapid decay of the function $u_0$, the support properties (in $\tau$) of $L$ and the pseudo-differential calculus in the $\sigma$ variable for controlling commutators (in order to push $L$ to the right), 
we finally get~:
\begin{align}\label{eq:FiniteGrushin2}
&-R_0^{-} \big( h^{3/8}P_1 + h^{1/2} P_2 + h^{3/4} P_3 - \lambda(h) \big)  R_{0}^{+} R_{0}^{-} L_0 \psi_h \nonumber \\
&\quad\quad\quad+
R_0^{-} \big( h^{3/8}P_1  \big) 
E_0 \big( h^{3/8}P_1 \big) R_{0}^{+} R_{0}^{-} L_0 \psi_h \nonumber\\
&\quad\quad\quad\quad\quad\quad= {\mathcal O}_{\rm unif}(h^{\infty}) + {\mathcal O}_{\rm unif}(h^{7/8-\delta}) \|R_{0}^{+} R_{0}^{-} L_0 \psi_h \|\;.
\end{align}
We get \eqref{eq:specProblem2} from \eqref{eq:FiniteGrushin2} by calculations similar to those leading to the expressions for $E_1$, $E_2$, $E_3$ in Subsection~\ref{ConstTrial}.  We just recall that
\begin{align*}
R_0^- P_1 R_0^+&=0\;,& - R_0^- P_2 R_0^+ &= \kappa(0) C_1\;,
\end{align*}
and
$$
-R_0^- P_3R_0^+ + R_0^-P_1E_0P_1 R_0^+ = - H_{{\rm harm}}\;.
$$
This finishes the proof of Lemma~\ref{lem:harmOscAlmost2}
\end{proof}
We now  compare (as already used above) various norms and observe~:
\begin{lemma}
\label{lem:bignorm2}~\\
Let $N \in {\mathbb N}$.
There exists $c>0$ and $h_0>0$ such that if $\psi_h$ is associated (as in \eqref{eq:DEF2}) to a normalized eigenfunction $u_h$ of ${\mathcal H}$ with $u_h \in U_N(h)$, then for all $h \in (0,h_0]$,
\begin{align}\label{w7-2}
\| R_0^{-} L_0 \psi_h \| - c h^{1/4} \leq \| \psi_h \| \leq \| R_0^{-} L_0 \psi_h \| + c h^{1/4}\;.
\end{align}
\end{lemma}

\begin{proof}~\\
Since clearly $\| R_0^{-}\| = 1$, we get from \eqref{eq:alpha2},
$$
\| R_0^{-} L_0 \psi_h \| \leq \| L_0 \psi_h \| = \| \psi_h \| + {\mathcal O}_{\rm unif}(h^{\infty})\;.
$$
This implies the first inequality in Lemma~\ref{lem:bignorm2}. To get the second inequality, we apply \eqref{w1-2}, Lemma~\ref{lem:boundedGrushin2} and \eqref{eq:alpha2},
 and  get
\begin{equation}\label{w6-2}
\psi_h = R_0^{+} R_0^{-} L_0 \psi_h + {\mathcal O}_{\rm unif}(h^{3/8 - \eta})\;.
\end{equation}
Since $\|R_0^{+}\| =1$ and $\psi_h$ satisfies \eqref{eq:alpha2}, this implies the lemma. 
\end{proof}
Using the self-adjointness of the harmonic oscillators, we get the following 
proposition.
\begin{proposition}
\label{cor:Intervals2}
Let $N \in {\mathbb N}$.
There exist $h_0>0$ and $C>0$ such that if $(\mu(h))_{h\in (0,h_0]}$ is an eigenvalue of ${\mathcal H}$ satisfying with $\mu(h) \in I_N(h)$ (see \eqref{eq:apubN}), then $\nu(h)$ (defined by \eqref{eq:mu2}) satisfies
\begin{align}
\nu(h) \in \cup_{\ell=0}^{N-1} \{e_\ell\} + [-C h^{1/16}, + C  h^{1/16}]\;.
\end{align}
\end{proposition}

\begin{proof}
Using Lemma~\ref{lem:bignorm2} above, Lemma~\ref{lem:harmOscAlmost2} implies that
\begin{align*}
\dist\big( \nu(h), \Spec\{ H_{{\rm harm}}\}\big) = {\mathcal O}_{\rm unif}(h^{1/16})\;.
\end{align*}
\end{proof}
\begin{lemma}
\label{lem:dimU}
Let $N \in {\mathbb N}$.
There exists $h_0>0$ such that if $h\in (0,h_0]$, then $\dim U_N(h) = N$.
\end{lemma}

\begin{proof}
We know from Section~\ref{GrushinUpper} that $\dim U_N(h) \geq N$.
In order to prove Lemma~\ref {lem:dimU} we only have to prove that the eigenspace attached to some  interval 
$\nu(h) \in [e_\ell -C h^{1/16}, e_\ell + C h^{1/16}]$, with $e_{\ell}$ from \eqref{eq:EvHarmOsc2}, and $\ell <N$, is necessarily of dimension $\leq 1$.
If it was not the case, let $u_{1,h}$, $u_{2,h}$ 
be normalized orthogonal eigenfunctions corresponding to eigenvalues $\mu_1(h)$ and 
$\mu_2(h)$ in the interval
\begin{align*}
\Theta_0 h - C_1 k_{{\rm max}} h^{3/2} + h^{7/4} [e_\ell -C h^{1/16}, e_\ell + C h^{1/16}]\;,
\end{align*}
for some $\ell = \ell(h)$.
Let $\psi_{1,h}$, $\psi_{2,h}$ be defined as in \eqref{eq:DEF2} and let $\nu_1$, $\nu_2$ be as in \eqref{eq:mu2}.
Let $e_\ell$ and $v_\ell$ be as in \eqref{eq:EvHarmOsc2} and below.

For $a,b \in {\mathbb C}$ ($a,b$ will depend on $h$) with $|a|^2 + |b|^2 = 1$. We have, using the almost orthonormality of $\psi_{1,h}$, $\psi_{2,h}$ and \eqref{eq:alpha2},
\begin{align}\label{eq:Dim2-2}
1 + {\mathcal O}_{\rm unif}(h^{\infty}) = \| a \psi_{1,h} + b \psi_{2,h}\|^2 \leq 
\| R_0^{-} L_0(a \psi_{1,h}+ b \psi_{2,h})\|^2 + {\mathcal O}_{\rm unif}(h^{1/8-\eta})\;.
\end{align}
With $\ell=\ell(h)$ as above, we may choose $a,b$ such that
\begin{align}\label{eq:Orthog2}
\int_{-\infty}^{\infty} \overline{v_{\ell}(\sigma)} R_0^{-} L_0(a \psi_{1,h}+ b \psi_{2,h})\,d\sigma = 0\;.
\end{align}
Lemma~\ref{lem:harmOscAlmost2} implies that
\begin{align}\label{eq:EF2}
&\quad( e_\ell - H_{{\rm harm}} ) R_{0}^{-} L_0 (a \psi_{1,h}+ b \psi_{2,h}) \nonumber\\
&=
a ( \nu_1(h) - H_{{\rm harm}} ) R_{0}^{-} L_0 \psi_{1,h}
+
b ( \nu_2(h) - H_{{\rm harm}} ) R_{0}^{-} L_0 \psi_{2,h} +
{\mathcal O}_{\rm unif}(h^{1/16}) \nonumber\\
&={\mathcal O}_{\rm unif}(h^{1/16})
\;.
\end{align}
Using \eqref{eq:Orthog2}, \eqref{eq:EF2} implies that
\begin{align}
\| R_{0}^{-} L_0 (a \psi_{1,h} + b \psi_{2,h}) \| = {\mathcal O}_{\rm unif}(h^{1/16})\;,
\end{align}
which is in contradiction to \eqref{eq:Dim2-2}. This finishes the proof of Lemma~\ref{lem:dimU}.
\end{proof}

Thus, for sufficiently small $h$, $U_N(h) = \Span\{ u_h^{(j)} \}_{j=1}^N$.

\begin{lemma}
\label{lem:jToj}
Let $N \in {\mathcal N}$. There exists $h_0>0$  such that
$$
{\mathcal M}_2^{(N)} u^{(j)}_h = v_{j-1}+ {\mathcal O}_{\rm unif}(h^{1/16})\;,
$$
for all $h<h_0$ and all $j\in \{1,\ldots,N\}$.
\end{lemma}

\begin{proof}
By induction it suffices to prove the lemma for $j=N$.
By Lemma~\ref{lem:harmOscAlmost2} and the spectral theorem there exists $\ell(h) \in \{0,\ldots,N-1\}$ such that (with $\psi_N$ being associated to $u^{(N)}_h$ as in \eqref{eq:DEF2})
$$
R_0^{-} L_0 \psi_N - \langle v_{\ell(h)}, R_0^{-} L_0 \psi_N \rangle v_{\ell(h)} = {\mathcal O}_{\rm unif}(h^{1/16})\;.
$$
Suppose $\ell(h_n)<N-1$ for a sequence $\{h_n\}$ with $h_n \searrow 0$. Then $ u^{(N)}_{h_n} \in U_{N-1}(h_n)$ and therefore $\dim U_{N-1}(h_n) \geq N$, in contradiction to Lemma~\ref{lem:dimU}. 
Thus,
$$
R_0^{-} L_0 \psi_N - \langle v_{N-1}, R_0^{-} L_0 \psi_N \rangle v_{N-1} = {\mathcal O}_{\rm unif}(h^{1/16})\;,
$$
and therefore
$$
{\mathcal M}_2^{(N)} u^{(N)}_h =
\langle v_{N-1}, R_0^{-} L_0 \psi_N \rangle v_{N-1} + {\mathcal O}_{\rm unif}(h^{1/16})\;.
$$
Lemma~\ref{lem:jToj} now follows from Lemma~\ref{lem:bignorm2}.
\end{proof}

The injectivity of ${\mathcal M}_2^{(N)}$ clearly follows from Lemma~\ref{lem:jToj}. This finishes the proof of Lemma~\ref{lem:LinAlg}.
\end{proof}

\appendix

\section{On an important family of ordinary differential equations}\label{AppA}
Let us recall for the comfort of the reader the main properties (mainly due to \cite{DaHe} and \cite{BeSt}), concerning
 the Neumann realization
 of $H^{N,\xi}$ in $L^2({\mathbb R}^+)$ associated 
to $D_{x}^{2}+(x+\xi )^2$. We denote by ${\hat \mu}^{(1)}(\xi)$
 the lowest eigenvalue of $H^{N,\xi}$ and by $\varphi_\xi$ the
 corresponding strictly positive normalized eigenfunction. 
More simply we will write $\mu(\xi)$ instead 
of ${\hat \mu}^{(1)}(\xi)$ in this appendix.
It has been proved  that the infimum 
$\inf_{\xi\in {\mathbb R}} \inf \Spec ( H^{N,\xi} )$ is actually
 a minimum. Then one can show that there exists
 $\xi_0 < 0$ such that  $ \mu ( \xi )$ decays monotonically to a
 minimum value $\Theta_0 <1$ and then increases monotonically again.
So it can be proved that~:
\begin{equation}\label{nm6} 
\Theta _0= \inf_{\xi} \big( \inf \; \Spec (H^{N,\xi })\big)
 = \inf \; \Spec (H^{N,\xi_0 })\;, 
\end{equation}
and moreover that:
\begin{equation}\label{Dada}
\Theta_0 = \xi_0^2\;.
\end{equation} 
It is indeed proved in \cite{DaHe}  that 
\begin{equation}\label{lw20} 
\mu' (\xi )=[\mu (\xi )-\xi ^2]\varphi_{\xi }(0) ^{2}\;. 
\end{equation} 
From (\ref{lw20}), we get that 
\begin{equation}\label{lw21} 
\mu''(\xi_0  )=-2\xi_0 \varphi ^{2}_{\xi_0 }(0)>0\;. 
\end{equation} 
We will write $u_0$ instead of $\phi_{\xi_0}$, and define the constant $C_1$ by \eqref{eq:C1}.

Let us now recall some formulas appearing in \cite{BeSt}.
Define $M_k$ to be the $k$'th moment, centered at $-\xi_0$, of the measure  $u_{0}^{2}(x)\,dx$~:
\begin{equation}
M_k=\int _{{\mathbb R} _{+}}(x+\xi_0 )^k u_{0}^{2}(x)\;dx\;.
\end{equation}
These moments were calculated in \cite{BeSt}. 
\begin{lemma}\label{Mom}~\\
The first moments can be expressed by the following formulas~:
\begin{align}
\label{eq:moments}
  M_0 &=1\;, & M_1 &= 0\;,&
  M_2 &=\frac{\Theta _0}{2}\;,&
  M_3 &=\frac{u_{0}^{2}(0)}{6}>0\;. 
\end{align}
\end{lemma} 
We will also need a few other results on the model operator.
\begin{prop}
\label{prop:Iij}~\\
We have the following identities
\begin{align*}
\int_0^{\infty} [2\tau (\tau+\xi_0)^2 -\tau^2(\tau+\xi_0)] u_0^2(\tau)\,d\tau
&=
\frac{u_0^2(0)}{6} = \frac{C_1}{2}> 0\;,\\
i \int_0^{\infty} u_0(\tau) D_{\tau} u_0(\tau)\,d\tau &= - \frac{u_0^2(0)}{2} = - \frac{3}{2}C_1\;.
\end{align*}
\end{prop}

\begin{proof}~\\
The first identity clearly follows from the known moments of $u_0^2$ and \eqref{eq:C1}.
The second identity follows from partial integration.
\end{proof}

\begin{prop}
\label{prop:secondPerturb}~\\
For $z \in {\mathbb R}$, let $E(z)$ be defined as the ground state energy of the Neumann realization of
$$
H(z) = -\frac{d^2}{d\tau^2} + (\tau+\xi_0 + z)^2\;,
$$
on $L^2({\mathbb R}_{+})$. Then $E(z)$ is a smooth function and satisfies
\begin{align}
\label{eq:Edobbeltmaerke}
E''(0) = 2 ( 1 - 4 I_2)\;,
\end{align}
with $I_2$ from \eqref{eq:I2}.

Furthermore,
\begin{align}
\label{eq:EdobbeltmaerkeExplicit}
E''(0) = 6 C_1 \sqrt{\Theta_0}\;.
\end{align}

\end{prop}

\begin{proof}~\\
By analytic perturbation theory, $E(z)$ is analytic and there exists an analytic function ${\mathbb R} \ni z \mapsto \phi(z) \in L^2({\mathbb R}_{+})$ such that
\begin{align}
\| \phi(z) \| &= 1\;, & H(z) \phi(z) &= E(z) \phi(z)\;, & \phi(0) &= u_0\;.
\end{align}
By differentiating the identity $\| \phi(z) \|^2 = 1$ twice with respect to $z$,  we find
\begin{align}
\label{eq:normphimaerke}
2 \Re \langle \phi'(0) \;|\; u_0 \rangle &= 0\;, &
- \| \phi'(0) \|^2 &=  \Re \langle \phi''(0)\;|\; u_0 \rangle\;.
\end{align}
From the equation $H(z) \phi(z) = E(z) \phi(z)$, and the fact that $E(z)$ is minimal  at $z=0$, we get, with $P_0=H(0) - \Theta_0\;$,
$$
P_0 \phi'(0) = -2(\tau+\xi_0) u_0\;,
$$
which implies with $P_0^{-1}$ from \eqref{eq:P0InverseFirst} and \eqref{eq:A14}, since $u_0 \perp (\tau + \xi_0) u_0$ (by \eqref{eq:moments}),
\begin{align}
\label{eq:phimaerke}
\phi'(0) = -2 P_0^{-1} \Big( (\tau+\xi_0) u_0 \Big) + c u_0\;,
\end{align}
for some $c \in i {\mathbb R}$.
Finally, differentiating the relation 
$ E(z) = \langle \phi(z)\;|\; H(z) \phi(z) \rangle$ twice gives us the formula~:
\begin{multline}
\label{eq:Etwice}
E''(0) = 2 \Theta_0 \Re \langle \phi''(0)\;|\; u_0 \rangle + 8 \Re \langle \phi'(0)\;|\; (\tau + \xi_0)u_0 \rangle \\
+ 2 \langle \phi'(0)\;|\; H(0) \phi'(0) \rangle + 2\;.
\end{multline}
Upon inserting \eqref{eq:normphimaerke} and  \eqref{eq:phimaerke} in \eqref{eq:Etwice}, we get \eqref{eq:Edobbeltmaerke}.
The final identity, \eqref{eq:EdobbeltmaerkeExplicit} is a rephrasing of \eqref{lw21} .
\end{proof}
We also have the following easy observation~:
\begin{lemma}
\label{lem:Schwarzbdry-Schwarzdomain}~\\
Let $R_0^{+}$ be the operator from \eqref{eq:DefR0+}.
Suppose $\phi \in {\mathcal S}({\mathbb R})$, then $R_0^+ \phi \in {\mathcal S}({\mathbb R}\times \overline{{\mathbb R}_{+}})\,$.
\end{lemma}

\begin{proof}~\\
This is an easy consequence of the regularity and decay of $u_0\,$. 
\end{proof}

Finally, we will need the following mapping properties of the regularized resolvent.
\begin{lemma}
\label{lem:regresolvent}~\\
Let $P_0$ be the Neumann realization of
$$
-\frac{d^2}{d\tau^2} + (\tau + \xi_0)^2 - \Theta_0\;,
$$
on $L^2({\mathbb R}_+)$.
For $\phi \perp u_0$ we can define $P_0^{-1} \phi$ as the unique solution $f$ to
\begin{align}\label{eq:P0InverseFirst}
P_0 f &= \phi\;, & f \perp u_0\;.
\end{align}
Let $P_0^{-1} \in {\mathcal L}(L^2({\mathbb R}_+))$ be the regularized resolvent~:
\begin{align}\label{eq:A14}
P_0^{-1} \phi = \begin{cases}
0\;,& \phi \parallel u_0 \\ P_0^{-1} \phi,& \phi \perp u_0\;,
\end{cases}
\end{align}
(and extended by linearity).
Then $P_0^{-1}$ is continuous from ${\mathcal S}(\overline{{\mathbb R}_+})$
 into ${\mathcal S}(\overline{{\mathbb R}_+})\,$.\\
Moreover, for any $\alpha \geq 0 $, $ P_0^{-1}$
 is continuous in $L^2(\rz^+\;;\; \exp - \alpha \tau)\,$.
\end{lemma}

\begin{proof}~\\
Using the local regularity up to the boundary of $P_0$, one first gets
 that $P_0^{-1}$ sends ${\mathcal S}(\overline{{\mathbb R}_+})$
 into ${C^\infty}(\overline{{\mathbb R}_+})$. For the control at $\infty$,
 one then observes, after cutting away from $0$, that the problem is reduced
 to the analysis of inverting the harmonic oscillator
 $ -\frac{d^2}{d\tau^2} + (\tau)^2 - \Theta_0$ on $\mathcal S (\rz)$,
 which is a standard result.\\
For the last statement, we can also observe that, for any real $\alpha$, the operator 
$$
\exp -\alpha \sqrt{1+\tau^2}\cdot (-\frac{d^2}{d\tau^2} + (\tau+\xi_0)^2 - \Theta_0)^{-1}\cdot  \exp \alpha \sqrt{1+\tau^2}
$$
extends continuously on $L^2(\rz)$ and $\mathcal S(\rz)$.\\
One can also show by the same technique that
\begin{equation}\label{propu0}
\tau^j u_0^{(k)} (\tau) \in L^2(\rz^+\;;\; \exp - \alpha \tau)\;,
\mbox{ for all } \alpha \geq 0\;,\; \mbox{ and for all } j,k\;.
\end{equation}
With additional work, one could actually get a better decay.
\end{proof}

\section{Coordinates near the boundary.} \label{AppB}
It is convenient in most calculations to straighten out the
boundary by a coordinate transformation. This, quite standard,
procedure, will be defined below.
Let $z_0\in \partial \Omega$ and let $\ell $ be the length of the 
boundary $\partial \Omega$ and 
$I=]-\frac{\ell}{2}, \frac{\ell}{2}]\,.$ 
Let $M\in C^{\infty}(I;\partial \Omega )$ be a 
parametrization of $\partial \Omega $
  such that 
$M(0)=z_0$ and $s$ is the  distance inside $\partial \Omega$  between $M(s)$ and $z_0$.
We denote by $$ T(s):= M'(s)\;, $$  the unit tangent vector of 
$\partial \Omega $ at $\ M(s)$ and the scalar curvature by  $\kappa (s)$, which can
be defined by
$$ T'(s)=\kappa (s)\, \nu(s) \;,$$ where  
$\nu(s)$ is  the interior normal unit vector of 
$\partial \Omega $ at $M(s)$.\\  
Moreover the parametrization is chosen positive~:  
$$ 
{\rm {det}}\; (T(s),\nu(s))=1,\ \forall \; s\; \in \; I\;.
$$
For any $\ z\; \in \; {\overline { \Omega }}\,,$ 
we denote by $\ t(z)$ the standard distance of $\ z$ to 
$\  \partial \Omega :$ 
$$t(z)=\inf _{\omega \in \partial \Omega }\; |z-\omega |\;.$$ 
So, there exists $t_0>0$ and a 
diffeomorphism of class $C^{\infty}$~:
\begin{align}
\label{eq:t0}
\psi :\Omega _{t_0}\; \to \; 
S^{1}_{\ell/(2\pi )} \times (0,t_0) \;,
\end{align}
such that $\psi (z)=w=(s(z),t(z))$ and 
$|z-M(s(z))|=t(z)\;$. \\ 
We have denoted, for small enough $\varepsilon$, by $\Omega _{\varepsilon }$ the tubular neighborhood of $\partial \Omega$~:
$$ \Omega_{\varepsilon}~:=\{ z\in \Omega ;\ \dist (z, \partial \Omega )<\varepsilon \} $$ 
and $S^{1}_{r}$ is the circle of radius $r$ 
is identified  with $[-\pi r,\pi r[\,$.\\ 
So we have the identity 
\begin{equation}\label{br6} 
z=M(s(z))+t(z)\nu(s(z)),\ \forall \; z\; \in \; \Omega _{\varepsilon _0}\;. 
\end{equation} 
From this equality, it is easy to check that 
\begin{equation}\label{br7} 
T(s(z))=[1-t(z)\kappa (s(z))]\nabla  s(z)
\ \ {\rm {and}}\ \ \nu(s(z))=\nabla  t(z)\;. 
\end{equation} 
So for all $ u\; \in \; H^{1}(\Omega )$ such that $ \supp (u)
\subset  \Omega _{\varepsilon_0}\,$, 
\begin{align}\label{br8} 
&  \int _{\omega } |(hD_z-A)u|^2\, dz=\nonumber\\ 
&\qquad \int _K\big[|(hD_t-{\widetilde {A}}_2)v|^2
+(1-t\kappa (s))^{-2}|(hD_s-{\widetilde {A}}_1)v|^2 \big]
(1-t\kappa (s))\, dw
\end{align}
and 
\begin{equation}\label{br08}
\int _{\omega }  |u|^{2}\, dz=\int _K|v|^2(1-t\kappa (s))\, dw\;,
\end{equation} 
with $\ v(w)=u(\psi ^{-1}(w))\,$, 
$K=I\times (0,t_0)\,$, $ w=(s,t)$ and $ dw=ds\,dt\,$.\\  
The magnetic potential  $\ {\widetilde {A}}$ satisfies 
$${\widetilde {A}}_1\, ds+{\widetilde {A}}_2\, dt = A_1\, dx+A_2\, dy\;.$$ 
So 
\begin{equation}\label{br9} 
[\frac{\partial \widetilde {A}_2  }{\partial s}(w)-
\frac{\partial \widetilde {A}_1 }{\partial t}(w)]\; ds\wedge dt 
=B(z)\; dx\wedge dy={\hat B}(w)[1-t\kappa (s)]\; ds\wedge dt \;,
\end{equation} 
with $\psi (z)=w$ and ${\hat B}$ defined as~:
\begin{equation}\label{hatB}
{\hat B}(w)= B(z)\;.
\end{equation}
This gives~: 
\begin{equation}\label{br10} 
\frac{\partial \widetilde {A}_2}{\pa s}(w)-
\frac{\partial \widetilde {A}_1}{\pa t}(w)
=B(\psi^{-1}(w))[1-t\kappa (s)]={\hat B}(t,s) (1-t \kappa(s))\;. 
\end{equation} 
Then we get the identity between   differential operators 
\begin{equation}\label{br11} 
(hD_z-A)^2=a^{-1}[(hD_s-{\widetilde {A}}_1)a^{-1}
(hD_s-{\widetilde {A}}_1)+
(hD_t- {\widetilde {A}}_2)a(hD_t- {\widetilde {A}}_2)]\;,  
\end{equation} 
where $a(w)=1-t\kappa (s)$. \\
The usual Hilbert space $L^2(\Omega _{t_0})$ is transformed 
to $L^2(K;a\,dw)\,$. \\
In the new coordinates and using a gauge transform, we can always assume that the magnetic potential 
has no normal component in a neighborhood of 
$\partial \Omega$~:
\begin{equation}\label{br13} 
 {\widetilde {A}}_2=0\;. 
\end{equation} 
In this case, we have~:
\begin{equation}\label{br13a}
\partial_t  {\widetilde {A}}_1 = - {\hat B}(t,s) (1- t \kappa(s))\;,
\end{equation}
where ${\hat B}$ was introduced in (\ref{hatB}).\\

\centerline{{\bf Acknowledgements}}
The authors wish to thank J.~Sj\"{o}strand for having suggested the
use of a reduction to the boundary for getting an optimal result.
We also acknowledge useful
discussions with A.~Morame on related subjects and V.~Bonnaillie and R.~Frank for
comments on preliminary versions.

\end{document}